\documentclass[twocolumn]{aastex63}
\usepackage{amsmath}

\newcommand{\msun}{${\rm M_{\sun}}$}
\def\ltsima{$\; \buildrel < \over \sim \;$}
\def\simlt{\lower.5ex\hbox{\ltsima}}
\def\gtsima{$\; \buildrel > \over \sim \;$}
\def\simgt{\lower.5ex\hbox{\gtsima}}
\def\km{{\rm\,km}}
\def\kms{{\rm\,km\,s^{-1}}}
\def\km2s2{{\rm\,km^2\,s^{-2}}}

\def\kpc{{\rm\,kpc}}

\def\msun{{\rm\,M_\odot}}


\def\deg{^\circ}

\def\Gyr{{\rm\,Gyr}}

\def\masyr{{\rm\,mas \, yr^{-1}}}

\def\ltsima{$\; \buildrel < \over \sim \;$}
\def\gtsima{$\; \buildrel > \over \sim \;$}

\def\Gaia{{\it Gaia}}
\def\enlink{{\texttt{ENLINK}}}
\def\EJ{{($\textit{\textbf{J}}, E$)}}
\def\J{{$\textit{\textbf{J}}$}}

\def\nss{{\rm\,41}}
\def\nGCs{{\rm\,170}}
\def\nDGs{{\rm\,46}}
\def\nobjects{{\rm\,257}}
\def\Pth{{\,P_{\rm Threshold}}}

\def\Pg{{\,P_{\rm Group}}}
\def\nmergers{{\rm\,6}}
\def\nMergerObjects{{\rm\,62}}
\def\fMergerObjects{{\rm\,25\%}}

\def\Pontus{{\it Pontus}}

\def\Sgr{{\it Sagittarius}}
\def\Helmi{{\it Helmi}}
\def\Cetus{{\it Cetus}}
\def\GSE{{\it Gaia-Sausage/Enceladus}}
\def\lms1{{\it LMS-1/Wukong}}
\def\asi{{\it Arjuna/Sequoia/I'itoi}}
\def\Kraken{{\it Kraken/Koala}}

\submitjournal{ApJ}

\shorttitle{Detecting the mergers of the Milky Way halo}
\shortauthors{Malhan et al.}
\graphicspath{{./}{figures/}}

\begin{document}

\title{The Global Dynamical Atlas of the Milky Way mergers:\\ constraints from Gaia EDR3 based orbits of globular clusters, stellar streams and satellite galaxies}

\correspondingauthor{Khyati Malhan}
\email{kmalhan07@gmail.com}

\author[0000-0002-8318-433X]{Khyati Malhan}
\affiliation{Humboldt Fellow and IAU Gruber Fellow}
\affiliation{Max-Planck-Institut f\"ur Astronomie, K\"onigstuhl 17, D-69117, Heidelberg, Germany}

\author[0000-0002-3292-9709]{Rodrigo A. Ibata}
\affiliation{Universit\'e de Strasbourg, CNRS, Observatoire astronomique de Strasbourg, UMR 7550, F-67000 Strasbourg, France}

\author[0000-0002-0920-809X]{Sanjib Sharma}
\affiliation{Sydney Institute for Astronomy, School of Physics, The University of Sydney, NSW 2006, Australia}
\affiliation{ARC Centre of Excellence for All Sky Astrophysics in Three Dimensions (ASTRO-3D)}

\author[0000-0003-3180-9825]{Benoit Famaey}
\affiliation{Universit\'e de Strasbourg, CNRS, Observatoire astronomique de Strasbourg, UMR 7550, F-67000 Strasbourg, France}

\author[0000-0001-8200-810X]{Michele Bellazzini}
\affiliation{INAF  -  Osservatorio  di  Astrofisica  e  Scienza  dello  Spazio,via Gobetti 93/3, 40129 Bologna, Italy}

\author[0000-0002-7667-0081]{Raymond G. Carlberg}
\affiliation{Department of Astronomy \& Astrophysics, University of Toronto, Toronto, ON M5S 3H4, Canada}

\author[0000-0001-9269-8167]{Richard D'Souza}
\affiliation{Vatican Observatory, Specola Vaticana, V-00120, Vatican City State}

\author[0000-0002-8129-5415]{Zhen Yuan}
\affiliation{Universit\'e de Strasbourg, CNRS, Observatoire astronomique de Strasbourg, UMR 7550, F-67000 Strasbourg, France}

\author[0000-0002-1349-202X]{Nicolas F. Martin}
\affiliation{Universit\'e de Strasbourg, CNRS, Observatoire astronomique de Strasbourg, UMR 7550, F-67000 Strasbourg, France}
\affiliation{Max-Planck-Institut f\"ur Astronomie, K\"onigstuhl 17, D-69117, Heidelberg, Germany}

\author[0000-0002-2468-5521]{Guillaume F. Thomas}
\affiliation{Universidad de La Laguna, Dpto. Astrof\'isica E-38206 La Laguna, Tenerife, Spain}

%
\begin{abstract}
The Milky Way halo was predominantly formed by the merging of numerous progenitor galaxies. However, our knowledge of this process is still incomplete, especially in regard to the total number of mergers, their global dynamical properties and their contribution to the stellar population of the Galactic halo. Here, we uncover the Milky Way mergers by detecting groupings of globular clusters, stellar streams and satellite galaxies in action ($\textit{\textbf{J}}$) space. While actions fully characterize the orbits, we additionally use the redundant information on their energy ($E$) to enhance the contrast between groupings. For this endeavour, we use {\it Gaia} EDR3 based measurements of $170$ globular clusters, $41$ streams and $46$ satellites to derive their $\textit{\textbf{J}}$ and $E$. To detect groups, we use the \texttt{ENLINK} software, coupled with a statistical procedure that accounts for the observed phase-space uncertainties of these objects. We detect a total of $N=6$ groups, including the previously known mergers {\it Sagittarius}, {\it Cetus}, {\it Gaia-Sausage/Enceladus}, {\it LMS-1/Wukong}, {\it Arjuna/Sequoia/I'itoi} and one new merger that we call {\it Pontus}. All of these mergers, together, comprise $62$ objects ($\approx 25\%$ of our sample). We discuss their members, orbital properties and metallicity distributions. We find that the three most metal-poor streams of our Galaxy -- ``C-19'' ([Fe/H]$=-3.4$~dex), ``Sylgr'' ([Fe/H]$=-2.9$~dex) and ``Phoenix'' ([Fe/H]$=-2.7$~dex) -- are associated with {\it LMS-1/Wukong}; showing it to be the most metal-poor merger. The global dynamical atlas of Milky Way mergers that we present here provides a present-day reference for galaxy formation models.
\end{abstract}

\section{Introduction}\label{sec:Introduction}

The stellar halo of the Milky Way was predominantly formed by the merging of numerous progenitor galaxies \citep{Ibata1994, Helmi_1999, Chiba2000, Majewski2003, Bell2008, Newberg_2009_Cetus, Nissen2010, Belokurov2018, Helmi2018, Myeong2019, Matsuno2019, Koppelman2019, Yuan2020a, Naidu2020}, and this observation appears consistent with the $\Lambda$CDM based models of galaxy formation (e.g., \citealt{Bullock_2005, Pillepich2018}). However, challenging questions remain, for instance: how many progenitor galaxies actually merged with our Galaxy? What were the initial physical properties of these merging galaxies, including their stellar and dark matter masses, their stellar population, and their chemical composition (e.g., their [Fe/H] distribution function)? Which objects among the observed population of globular clusters, stellar streams and satellite galaxies in the Galactic halo were accreted inside these mergers? Answering these questions is important to understand the hierarchical build up of our Galaxy, and thereby inform galaxy formation models. 

It was recently proposed that a significant fraction of the Milky Way's stellar halo ($\sim95\%$ of the stellar population) resulted from the merging of $\approx 9-10$ progenitor galaxies. This scenario is suggested by \cite{Naidu2020}, who identified these mergers by selecting ``overdensities'' in the chemo-dynamical space of $\sim 5700$ giant stars (these giants lie within $50\kpc$ from the Galactic center). Many of their selections were based on the knowledge of the previously known mergers. Here, our motivation is also to find the mergers of our Galaxy, but using a different approach from theirs. First, our objective is to be able to {\it detect} these mergers using the data (and not {\it select} them) while being agnostic about the previously claimed mergers of our Galaxy. Secondly, we aim for a procedure that is possibly reproducible in cosmological simulations. Finally, we use a very different sample of halo objects, comprising only of globular clusters, stellar streams and satellite galaxies.

The Milky Way halo harbours a large population of globular clusters \citep{Harris2010, Vasiliev_GCs_2021}, stellar streams \citep{Ibata_2021, Li_2021_12streams} and satellite galaxies \citep{McConnachie_2020_GaiaEDR3, Battaglia_2021}, and these objects represent the most ancient and metal-poor structures of our Galaxy (e.g., \citealt{Harris2010, Kirby_2013, Helmi_2020})\footnote{While globular clusters and satellite galaxies represent two very different categories of stellar systems, streams do not represent a third category as they are produced from either globular clusters or satellite galaxies. Streams differ from the other two objects only in terms of their dynamical evolution; in the sense that streams are much more dynamically evolved.}. A majority of halo streams are the tidal remnants of either globular clusters or very low-mass satellites (\citealt{Ibata_2021, Li_2021_12streams}, see Section~\ref{subsec:data}). For all these halo objects, a significant fraction of their population is expected to have been brought into the Galactic halo inside massive progenitor galaxies (e.g., \citealt{Deason_2015, Kruijssen_2019, Carlberg_2020}). This implies that these objects, that today form part of our Galactic halo, can be used to trace their progenitor galaxies (e.g., \citealt{Massari_2019, Malhan_cocoon_2019, Bonaca2021}). Consequently, this knowledge also has direct implications on the long-standing question -- which of the globular clusters (or streams) were initially formed within the stellar halo (and represent an {\it in-situ} population) and which were initially formed in different progenitors that only later merged into the Milky Way (and represent an {\it ex-situ} population). Therefore, using these halo objects as tracers of mergers is also important to understand their own origin and birth sites. 

In this regard, using these halo objects also provides a powerful means to detect even the most metal-poor mergers of the Milky Way. This is important, for instance, to understand the origin of the metal-poor population of the stellar halo (e.g., \citealt{Komiya_2010, Sestito_2021}) and also to constrain the formation scenarios of the metal-deficient globular clusters inside high-redshift galaxies (e.g., \citealt{Forbes_2018_GC}). In the context of stellar halo, we currently lack the knowledge of the origin of the `metallicity floor' for globular clusters; that has recently been pushed down from [Fe/H]=$-2.5$~dex \citep{Harris2010} to [Fe/H]=$-3.4$~dex \citep{Martin2022_C19}. In fact, the stellar halo harbours several very metal-poor globular clusters (e.g., \citealt{Simpson_2018}) and also streams (e.g., \citealt{Roederer_2019, Wan_2020_Phoenix, Martin2022_C19}). These observations raise the question: did these metal-poor objects originally form in the Milky Way itself or were they accreted inside the merging galaxies? Moreover, such globular cluster-merger associations also allow us to understand the cluster formation processes inside those proto-galaxies that had formed in the early universe (e.g., \citealt{Freeman2002, Frebel2015}).

Our underlying strategy to identify the Milky Way's mergers is as follows. 
For each halo object we compute their three actions \J\ and then detect those `groups' that clump together tightly in the \J\ space\footnote{Conceptually, \J\ represents the amplitude of an object's orbit along different directions. For instance, in cylindrical coordinates, \J$\equiv (J_R, J_\phi,J_z)$, where $J_\phi$ represents the z-component of angular momentum ($\equiv L_z$), and $J_R$ and $J_z$ describe the extent of oscillations in cylindrical radius and $z$ directions, respectively.}. 
However, to enhance the contrast between groups, we additionally use the redundant information on their energy $E$ as this allows us to separate the groups even more confidently. 

The motivation behind this strategy can be explained as follows. First, imagine a progenitor galaxy (that is yet to be merged with the Milky Way) containing its own population of globular clusters, satellite galaxies and streams\footnote{The population of streams can originate from the tidal stripping of the member globular clusters and/or satellites inside the progenitor galaxy (e.g., \citealt{Carlberg2018}).}, along with its population of stars. Upon merging with the Milky Way, the progenitor galaxy will get tidally disrupted and deposit its contents into the Galactic halo. If the tidal disruption occurs slowly, the stars of the merging galaxy will themselves form a vast stellar stream in the Galactic halo (e.g., this is the case for the \Sgr\ merger, \citealt{Ibata_2020_Sgr, Vasiliev_Sgr_2020}). However, if the disruption occurs rapidly, then the stars will quickly get phase-mixed and no clear clear signature of the stream will be visible (e.g., this is expected for the \GSE\ merger, \citealt{Belokurov2018, Helmi2018}). In either case, the member objects of the progenitor galaxy (i.e., its member globular clusters, satellites and streams), that are now inside the Galactic halo, will possess very similar values of actions \J. This is because the dynamical quantities \J\ are conserved for a very long time, if the potential of the primary galaxy changes adiabatically. The Milky Way's potential likely evolved adiabatically (e.g., \citealt{Cardone_2005}), and therefore, those objects that merge inside the same progenitor galaxy are expected to remain tightly clumped in the \J\ space of the Milky Way; even long after they have been tidally removed from their progenitor. While $E$ is not by itself an adiabatic invariant, objects that merge together are expected to occupy a small subset of the energy space. Hence, even though actions fully characterize the orbits, $E$ is useful as a redundant `weight' to enhance the contrast between different groups. Moreover, since the mass of the merging galaxies ($M_{\rm halo}\simlt 10^{9-11}\msun$, e.g., \citealt{Robertson_2005, DSouza_2018}) are typically much smaller than that of our Galaxy ($M_{\rm halo}\sim 10^{12}\msun$, e.g., \citealt{Karukes_2020}), the merged objects are expected to occupy only a small volume of the \EJ\ space. Therefore, detecting tightly-clumped groups of halo objects in \J\ and $E$ space potentially provides a powerful means to detect the past mergers that contributed to the Milky Way's halo. 

This strategy for detecting mergers has now become feasible in the era of ESA/\Gaia\ mission \citep{Prusti2016}, because the precision of this astrometric dataset allows one to compute reasonably accurate \EJ\ values for a very large population of halo objects. Particularly the excellent \Gaia\ EDR3 dataset \citep{GaiaEDR3_Brown_2020, Lindegren_2021_GaiaEDR3} has provided the means to obtain very precise phase-space measurements for an enormously large number of globular clusters (e.g., \citealt{Vasiliev_GCs_2021}), stellar streams (e.g., \citealt{Ibata_2021, Li_2021_12streams}) and satellite galaxies (e.g., \citealt{McConnachie_2020_GaiaEDR3}), and we use these measurements in the present study. 

Before proceeding further, we note that some recent studies have also analysed energies and angular momenta of globular clusters (e.g., \citealt{Massari_2019}) and streams (e.g., \citealt{Bonaca2021}). However, the objective of those studies was largely to {\it associate} these objects with the previously known mergers of the Milky Way\footnote{An exception is the study by \cite{Myeong2019}, who used the \Gaia\ DR2 measurements of globular clusters and hypothesized the ``Sequoia'' merger.}. Here, our objective is fundamentally different, namely -- to {\it detect} the Milky Way's mergers by being completely agnostic about the previously hypothesized groupings of mergers and accretions. 

This paper is arranged as follows. In Section~\ref{sec:Computing_EJ}, we describe the data used for the halo objects and explain our method to compute their actions and energy values. In Section~\ref{sec:ENLINK}, we present our procedure to detect the mergers by finding ``groups of objects'' in \EJ\ space. In Section~\ref{sec:analyse_mergers}, we analyse the detected mergers for their member objects, their dynamical properties and their [Fe/H] distribution function. In Section~\ref{sec:merger_candidate}, we discuss the properties of a specific candidate merger. Additionally, in Section~\ref{sec:Streams_progs}, we find several physical connections between streams and
other objects (based on the similarity of their orbits and [Fe/H]). Finally, we discuss our findings and conclude in Section~\ref{sec:Disc_Conc}.

\begin{figure*}
\begin{center}
\vspace{-0.25cm}
\includegraphics[width=1.02\hsize]{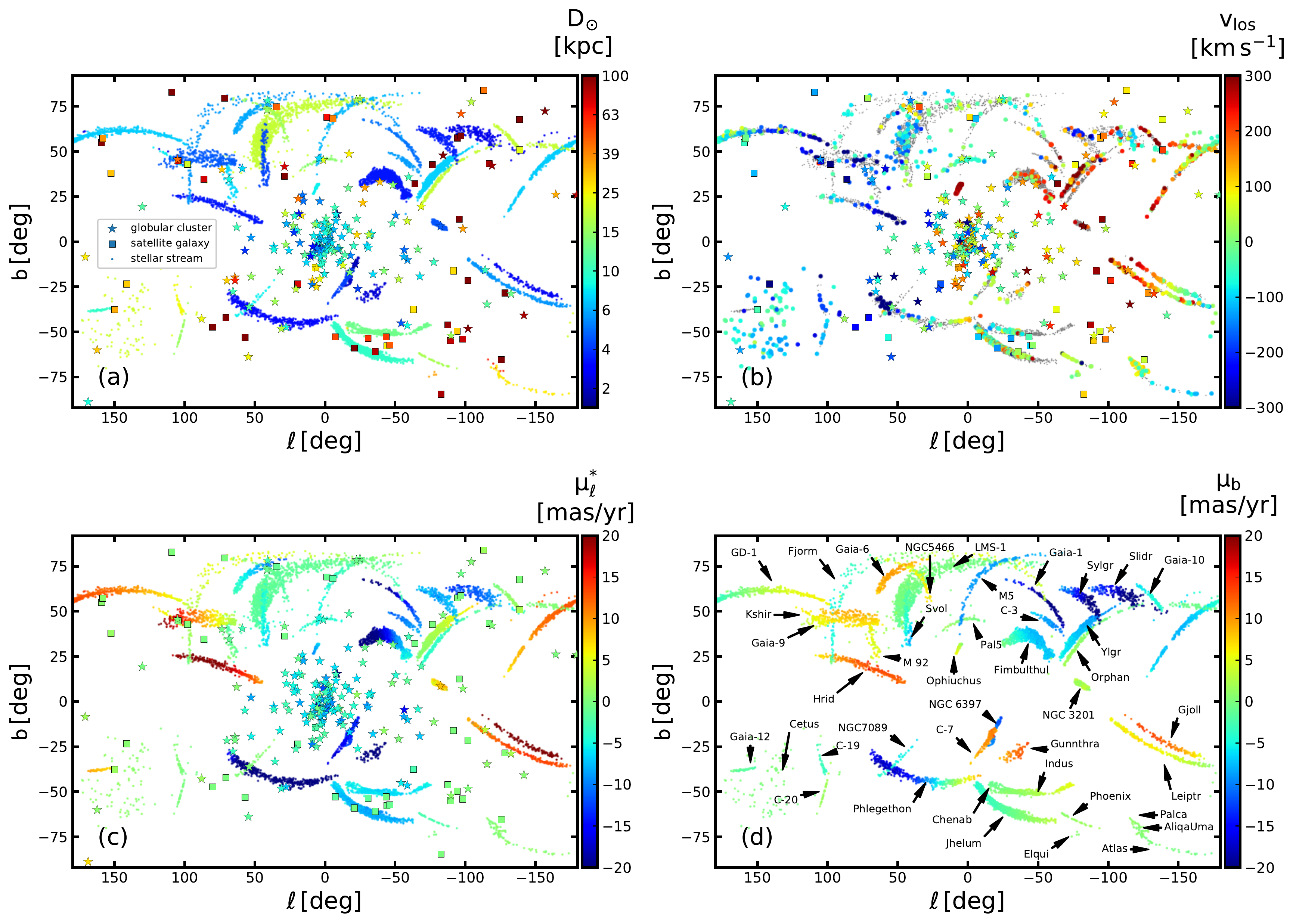}
\end{center}
\vspace{-0.5cm}
\caption{The Galactic maps showing phase-space measurements of $n=\nobjects$ halo objects used in our study, namely $\nGCs$ globular clusters (denoted by `star' markers), $\nss$ stellar streams (denoted by `dot' markers) and $\nDGs$ satellite galaxies (denoted by `square' markers). From panel `a' to `d', these objects are colored by their heliocentric distances ($D_{\odot}$), line-of-sight velocities ($v_{\rm los}$), proper motion in the Galactic longitude direction ($\mu^{*}_{\ell}$) and proper motion in the Galactic latitude direction direction ($\mu_{b}$), respectively. In panel `d', we only show streams with their names, and do not plot other objects to avoid crowding.}
\label{fig:Fig_all_substructures_sky}
\end{figure*}
\section{Computing actions and energy of globular clusters, stellar streams and satellite galaxies}\label{sec:Computing_EJ}

To compute \EJ\ of an object, we require (1) data of the complete 6D phase-space measurements of that object; i.e., its 2D sky position ($\alpha,\delta$), heliocentric distance ($D_{\odot}$) or parallax ($\varpi$), 2D proper motion ($\mu^{*}_{\alpha}\equiv \mu_{\alpha}{\rm cos}\delta,\,\mu_{\delta}$) and line-of-sight velocity ($v_{\rm los}$), and (2) a Galactic potential model that suitably represents the Milky Way. Below, Section~\ref{subsec:data} describes the phase-space measurements of $n=\nobjects$ objects and Section~\ref{subsec:compute_EJ} details the adopted Galactic potential model and our procedure to compute the \EJ\ quantities.

\subsection{Data}\label{subsec:data}

For globular clusters, we obtain their phase-space measurements from the \cite{Vasiliev_GCs_2021} catalog. This catalog provides, for $\nGCs$ globular clusters, their phase-space measurements and we use the observed heliocentric coordinates (i.e., $\alpha$, $\delta$, $D_{\odot}$, $\mu^{*}_{\alpha}$, $\mu_{\delta}$, $v_{\rm los}$) along with the associated uncertainties. \cite{Vasiliev_GCs_2021} derives the 4D astrometric measurement ($\alpha$, $\delta$, $\mu^{*}_{\alpha}$ , $\mu_{\delta}$) of globular clusters using the \Gaia\ EDR3 dataset, while the parameters $D_{\odot}$ and $v_{\rm los}$ are based on a combination of \Gaia\ EDR3 and other surveys. 

For satellite galaxies, we obtain their phase-space measurements from the \cite{McConnachie_2020_GaiaEDR3} catalog. This catalog provides data in a similar heliocentric coordinates format as described above, but for the satellite galaxies. From this catalog, we use only those objects that lie within a distance of $D_{\odot}<250\kpc$ (equivalent to the virial radius of the Milky Way, \citealt{CorreaMagnus2021}), yielding a sample of $44$ objects. In \cite{McConnachie_2020_GaiaEDR3}, the uncertainties on each component of proper motion are only the observational uncertainties, and therefore, we add in quadrature a systematic uncertainty of $0.033\masyr$ to each component of proper motion (A. McConnachie, private communication). While inspecting this catalog, we found that it lacks the proper motion measurements of two other satellites of the Milky Way, namely Bootes~III \citep{Grillmair_BootesIII_Styx} and the Sagittarius dSph \citep{Ibata1994}. For Bootes~III, we obtain its \Gaia\ DR2 based proper motion from \cite{Carlin_2018}. For Sagittarius, we use the \cite{Vasiliev_Sgr_2020} catalog that provides \Gaia\ DR2 based proper motions for this dwarf. From this, we compute the median and uncertainty for the Sagittarius dSph as ($\mu^{*}_{\alpha}$, $\mu_{\delta}$) = $(-2.67\pm0.45,-1.40\pm0.40)\masyr$. Our final sample comprises of $\nDGs$ satellite galaxies.

For stellar streams, we acquire their phase-space measurements primarily from the \cite{Ibata_2021} catalog, but we also use some other public stream catalogs (as described below). We first use the the \cite{Ibata_2021} catalog that contains those streams detected in \Gaia\ DR2 and EDR3 datasets using the \texttt{STREAMFINDER} algorithm \citep{Malhan_SF_2018, Malhan_Ghostly_2018, Ibata_norse_2019}. In this catalog, all the stream stars possess the 5D astrometric measurements on ($\alpha$, $\delta$, $\varpi$, $\mu^{*}_{\alpha}$, $\mu_{\delta}$), along with their observational uncertainties, as listed in the EDR3 catalog. However, most of these stream stars lack spectroscopic $v_{\rm los}$ measurements; this is because (to date) \Gaia\ has provided $v_{\rm los}$ for only very bright stars with $G\simlt 12$~mag. Therefore, to obtain the missing $v_{\rm los}$ measurements, we use various available spectroscopic surveys and also rely on the data from our own follow-up spectroscopic campaigns. These spectroscopic measurements are already presented in \cite{Ibata_2021} for the streams ``Pal~5'' (originally discovered by \citealt{Odenkirchen2001}), ``GD-1'' \citep{Grillmair2006GD1}, ``Orphan'' \citep{Belokurov_2007_Orphan, Grillmair_2006_Orphan}, ``Atlas'' \citep{Shipp_2018}, ``Gaia-1'' \citep{Malhan_Ghostly_2018}, ``Phlegethon'' \citep{Ibata_Phlegethon2018}, ``Slidr'' \citep{Ibata_norse_2019}, ``Ylgr'' \citep{Ibata_norse_2019}, ``Leiptr'' \citep{Ibata_norse_2019}, ``Sv\"ol'' \citep{Ibata_norse_2019}, ``Gj\"oll'' (\citealt{Ibata_norse_2019}, stream of NGC~3201, \citealt{Hansen_2020_3201_Gjoll, Palau_2021_NGC3201_Gjoll}), ``Fj\"orm'' (\citealt{Ibata_norse_2019}, stream of NGC~4590/M~68, \citealt{Palau2019_M68_Fjorm}), ``Sylgr'' (\citealt{Ibata_norse_2019}, low-metallicity stream with [Fe/H]=$-2.92$~dex, \citealt{Roederer_2019}), ``Fimbulthul'' (stream of the $\omega$-Centauri cluster, \citealt{Ibata_2019_wcen}), ``Kshir'' \citep{Malhan_Kshir2019}, ``M~92'' \citep{Thomas_2020_M92, Sollima_2020}, ``Hr\'id'', ``C-7'', ``C-3'', ``Gunnthr\`a'' and  ``NGC~6397''. This spectroscopic campaign suggests that $\simgt85\%$ of the \cite{Ibata_2021} sample stars are bonafide stream members.

\cite{Ibata_2021} also detected other streams, namely ``Indus'' \citep{Shipp_2018}, ``Jhelum'' \citep{Shipp_2018}, ``NGC~5466'' \citep{Grillmair_2006_NGC5466, Belokurov_2006_NGC5466}, ``M~5'' \citep{Grillmair2019_M5stream}, ``Phoenix'' (\citealt{Balbinot2016_Phoenix}, low-metallicity globular cluster stream with [Fe/H]=$-2.7$~dex, \citealt{Wan_2020_Phoenix}), ``Gaia-6'', ``Gaia-9'', ``Gaia-10'', ``Gaia-12'',  ``NGC~7089''. For these streams, we obtain their $v_{\rm los}$ measurements in this study by cross-matching their stars to various public spectroscopic catalogs, namely SDSS/Segue \citep{Yanny2009}, LAMOST DR7 \citep{Zhao_2012_LAMOST}, APOGEE DR16 \citep{Majewski_APOGEE_2017}, S5 DR1 \citep{Li_S5_DR1_2019} and our own spectroscopic data (that we have collected from our follow-up campaigns, \citealt{Ibata_2021}).

Finally, we include additional streams into our analysis from some of the public stream catalogs. From \cite{Malhan_2021_LMS1} we take the data for the ``LMS-1'' stream (a recently discovered dwarf galaxy stream, \citealt{Yuan2020}). We use the \cite{Yuan_2022_Cetus} catalog for the streams ``Palca'' \citep{Shipp_2018}, ``C-20'' \citep{Ibata_2021} and ``Cetus'' \citep{Newberg_2009_Cetus}. For ``Ophiuchus'' \citep{Bernard_2014_Ophiuchus}, we use those stars provided in the \cite{Caldwell_2020} catalog that possess membership probabilities$>0.8$. We also include those streams provided in the S5 DR1 survey \citep{Li_S5_DR1_2019} but were not detected by \cite{Ibata_2021}. These include ``Elqui'', ``AliqaUma'' and ``Chenab''. Lastly, we also include into our analysis the ``C-19'' stream (the most metal poor globular cluster stream known to date with [Fe/H]$\approx-3.4$~dex, \citealt{Martin2022_C19}). For all these streams, we use the \Gaia\ EDR3 astrometry.

In summary, we use a total of $\nss$ stellar streams for this study. This stream sample comprises $n=9192$ \Gaia\ EDR3 stars of which $1485$ possess spectroscopic $v_{\rm los}$ measurements. The parallaxes of all the stream stars are corrected for the global parallax zero-point in \Gaia\ EDR3 using \cite{Lindegren_2021_GaiaEDR3} value. 

Figure~\ref{fig:Fig_all_substructures_sky} shows phase-space measurements of all the $n=\nobjects$ objects considered in our study. In this plot, the distance of a given stream corresponds to the inverse of the uncertainty-weighted average mean parallax value of its member stars.

\begin{figure*}
\begin{center}
\vspace{-0.3cm}
\vbox{
\includegraphics[width=0.88\hsize]{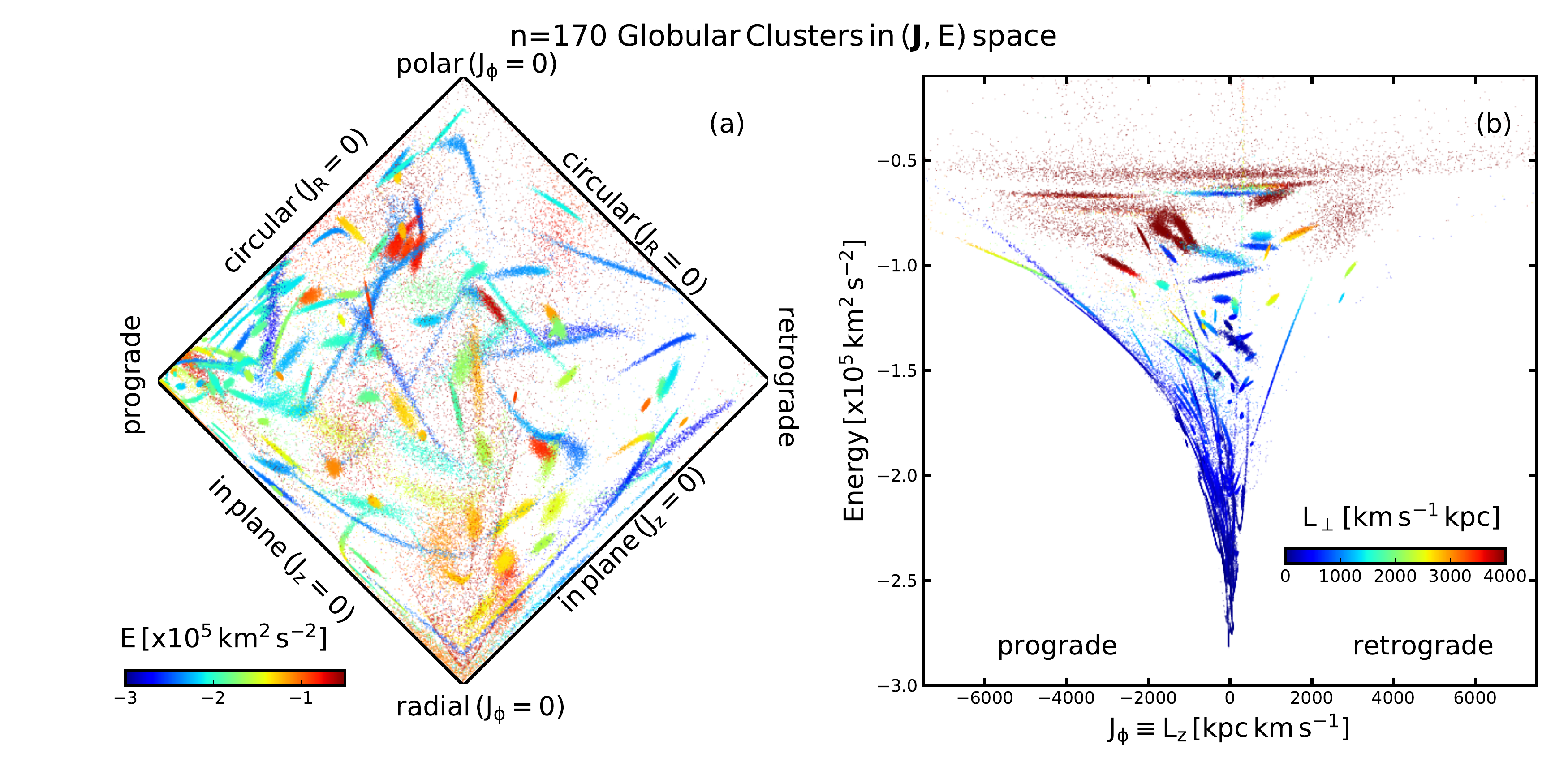}
\includegraphics[width=0.88\hsize]{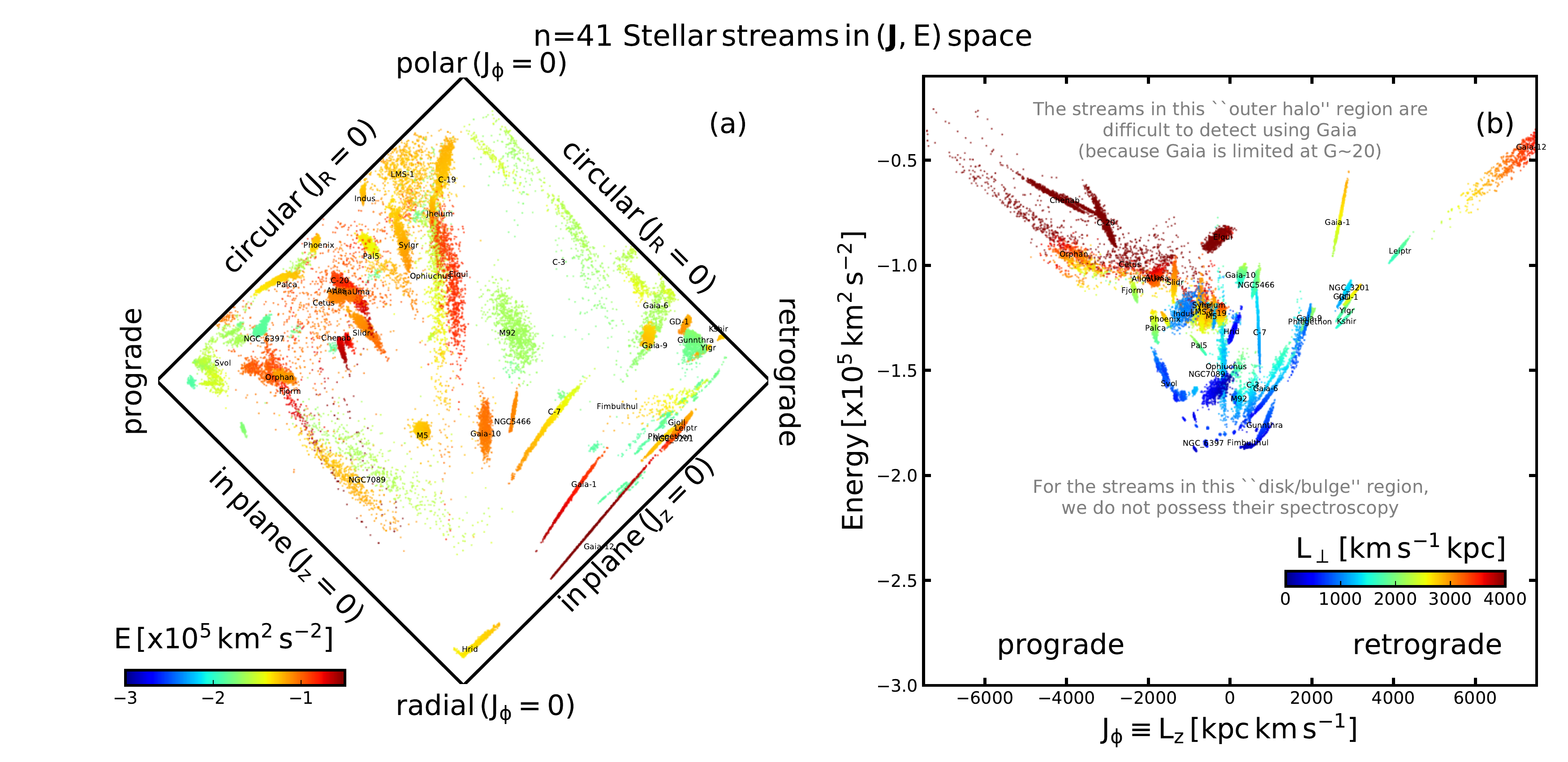}
\includegraphics[width=0.88\hsize]{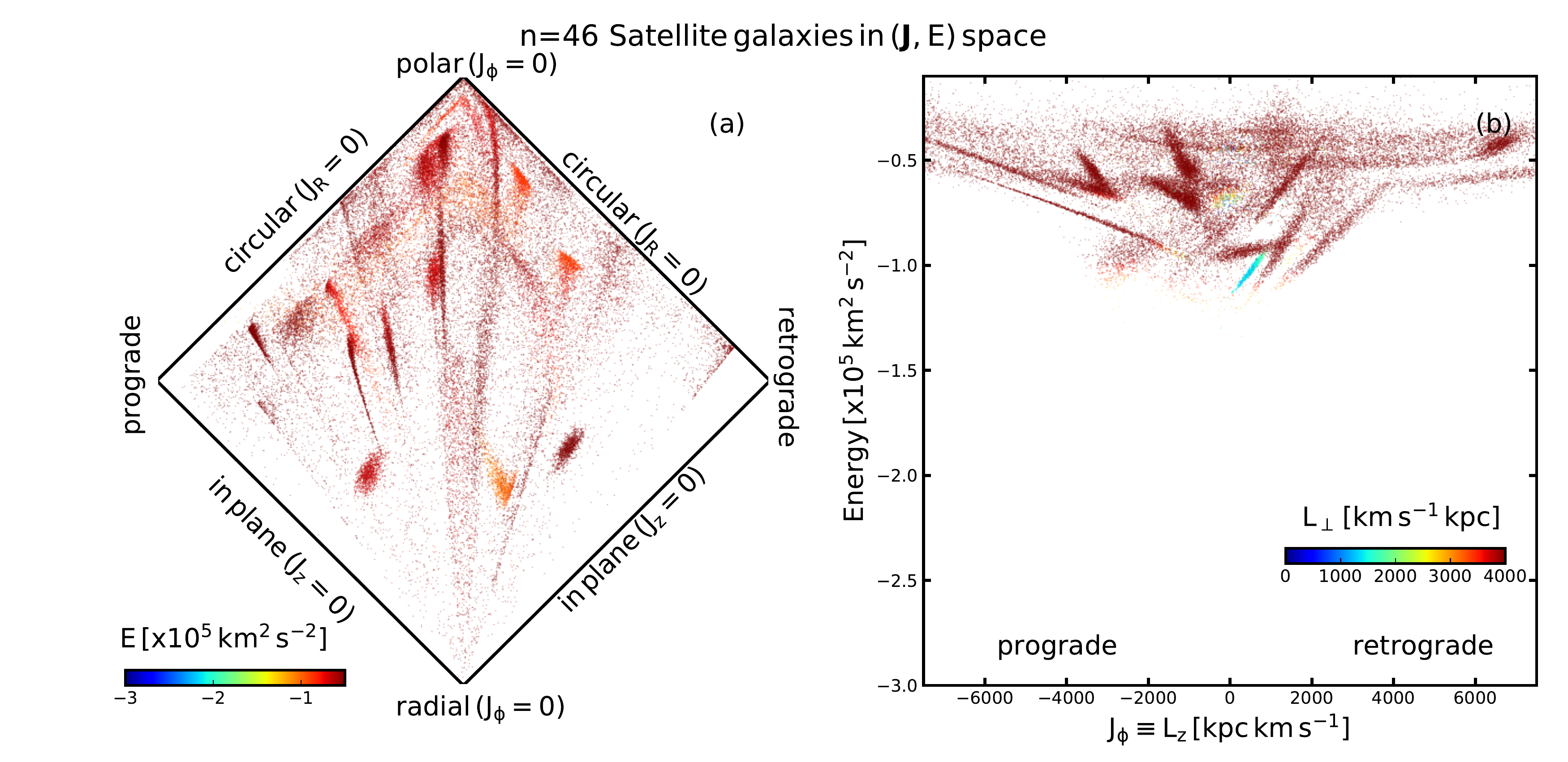}
}
\end{center}
\vspace{-0.5cm}
\caption{Action-Energy \EJ\ space of the Milky Way showing globular clusters (top panels), stellar streams (middle panels) and satellite galaxies (bottom panels). Each object can be seen as a `cloud' of $1000$ Monte Carlo representations of its orbit (see Section~\ref{subsec:compute_EJ}). In each row, the left panel corresponds to the projected action-space map, where the horizontal axis is $J_{\phi}/J_{\rm tot}$ and the vertical axis is ($J_z$-$J_R$)/$J_{\rm tot}$ with $J_{\rm tot}=J_R+J_z+|J_\phi|$. In these panels, the points are colored by the total energy of their orbits ($E$). The right panels show the $z-$component of the angular momentum ($J_\phi \equiv L_z$) vs $E$, and the points are colored by their orthogonal component of the angular momenta ($L_\perp=\sqrt{L^2_x+L^2_y}$).}
\label{fig:Fig_EJ_of_objects}
\end{figure*}
\subsection{Computing actions and energy of the halo objects}\label{subsec:compute_EJ}

To compute orbits of the objects in our sample, we adopt the Galactic potential model of \cite{McMillan2017}. This is a static and axisymmetric model comprising a bulge, disk components and an NFW halo. For this potential model, the total galactic mass within the galactocentric distance $r_{\rm gal }<20\kpc$ is $2.2\times10^{11}\msun$, $r_{\rm gal }<50\kpc$ is $4.9\times10^{11}\msun$ and $r_{\rm gal }<100\kpc$ is $8.1\times10^{11}\msun$. Another model that is often used to represent the Galactic potential is \texttt{MWPotential2014} of \cite{Bovy2015} and this model (on average) is $\sim 1.5$ times lighter than the \cite{McMillan2017} model. For our study, we prefer the \cite{McMillan2017} model because (1) the predicted velocity curve of this model is more consistent with the measurements of the Milky Way (e.g., \citealt{Bovy2020, Nitschai_2021}) and (2) we find that all the halo objects in this mass model possess $E<0$ (i.e., their orbits are bound), however, in the case of \texttt{MWPotential2014} we infer that $34$ clusters and all the satellite galaxies possess $E>=0$ (i.e., their orbits are unbound). To set the \cite{McMillan2017} potential model, and to compute \EJ\ and other orbital parameters, we make use of the \texttt{galpy} module \citep{Bovy2015}. Moreover, to transform the heliocentric phase-space measurements of the objects into Galactocentric frame (that is required for computing orbits), we adopt the Sun's Galactocentric distance from \cite{Gravity2018} and the Sun's galactic velocity from \cite{Reid2014_Sun} and \cite{Schornich2010_Sun}.

To compute \EJ\ values of globular clusters, we do the following. For a given globular cluster, we sample $1000$ orbits using the mean and the uncertainty on its phase-space measurement. For that particular cluster, this provides an \EJ\ distribution of $1000$ data points and this distribution represents the uncertainty in the derived \EJ\ value for that cluster. Note that this \EJ\ uncertainty, for a given cluster, reflects its uncertainty on the phase-space measurement. This orbit-sampling procedure is repeated for all the globular clusters, and for each cluster we retain their respective \EJ\ distribution. This \EJ\ distribution is a vital information and we subsequently use this while detecting the mergers (as shown in Section~\ref{sec:ENLINK}). The resulting \EJ\ distribution of all the globular clusters is shown in Figure~\ref{fig:Fig_EJ_of_objects}, where each object is effectively represented by a distribution of $1000$ points. 

We analyse actions in cylindrical coordinates, i.e., in the \J$\equiv(J_R,J_\phi,J_z)$ system, where $J_\phi$ corresponds to the z-component of angular momentum (i.e., $J_\phi\equiv L_z$) and negative $J_\phi$ represents prograde motion (i.e., rotational motion in the direction of the Galactic disk). Similarly, components $J_R$ and $J_z$ describe the extent of oscillations in cylindrical radius and $z$ directions, respectively. Figure~\ref{fig:Fig_EJ_of_objects} shows these globular clusters in (1) the ``projected action space'', represented by a diagram of $J_\phi/J_{\rm total}$ vs $(J_z-J_R)/J_{\rm total}$; where $J_{\rm total}=J_R + J_z + |J_\phi|$, and (2) the $J_\phi$ vs. $E$ space. The reason for using the projected action space is that this plot is effective in separating objects that lie along circular, radial and in-plane orbits, and it is considered to be superior to other commonly used kinematic spaces (e.g., \citealt{Lane_2021}). We also use the orthogonal component of the angular momentum $L_\perp = \sqrt{L^2_x+L^2_y}$ for representation. Note that even though $L_\perp$ is not fully conserved in an axisymmetric potential, it still serves as a useful quantity for orbital characterization (e.g., \citealt{Bonaca2021}). Along with retrieving the \EJ\ values, we also retrieve other orbital parameters (e.g., $r_{\rm apo}, r_{\rm peri}$, eccentricity -- these values are used at a later stage for the analysis of the detected mergers). 

To compute \EJ\ values of satellite galaxies, we use exactly the same orbit-sampling procedure as described above for globular clusters. The corresponding \EJ\ distribution is shown in Figure~\ref{fig:Fig_EJ_of_objects}. 

To compute \EJ\ values of stellar streams, we follow the orbit-fitting procedure; this approach is more sophisticated than the above described orbit-sampling procedure and more suitable for stellar streams. That is, we obtain \EJ\ solutions of a given stream by fitting orbits to the phase-space measurements of all its member stars (e.g., \citealt{Koposov_2010}). This procedure ensures that the resulting orbit solution provides a reasonable representation of the entire stream structure and also that the resulting \EJ\ values are precise\footnote{By employing the orbit-fitting procedure, we are assuming that the entire phase-space structure of a stream can be well represented by an orbit. Although streams do not strictly delineate orbits \citep{Sanders_2013}, our assumption is still reasonable as far as the scope of this study is concerned.}. We use this method only for narrow and dynamically cold streams (that make up most of our stream sample), but for the other broad and dynamically hot streams we rely on the orbit-sampling procedure (see further below). To carry out the orbit-fitting of streams, we follow the same procedure as described in \cite{Malhan_2021_LMS1}. Briefly, we survey the parameter space using our own Metropolis-Hastings based MCMC algorithm, where the log-likelihood of each member star $i$ is defined as
\begin{equation}\label{eq:Loglikehood}
\ln \mathcal{L}_i = -\ln ((2\pi)^{5/2}\sigma_{\rm sky} \sigma_{\varpi} \sigma_{\mu_{\alpha}} \sigma_{\mu_{\delta}} \sigma_{v_{\rm los}}) +\ln N -\ln D,
\end{equation}
where 
\begin{equation}
\begin{aligned}
N &= \prod_{j=1}^5 (1-e^{-R^2_j/2}) \, , 
\quad\text{\,}\quad 
D = \prod_{j=1}^5 R^2_j \, , \\
R_1^2 &= \dfrac{\theta^2_{\rm sky}}{\sigma^2_{\rm sky}} \, , 
\quad\text{\,\,\,\,\,\,\,\,\,\,\,\,\,\,\,\,}\quad 
R_2^2 = \dfrac{(\varpi^{\rm d} - \varpi^{\rm o})^2}{\sigma^2_{\varpi}} \, , \\
R_3^2 &= \dfrac{(\mu^{\rm d}_{\rm \alpha} - \mu^{\rm o}_{\rm \alpha})^2}{\sigma^2_{\mu_{\alpha}}} \, , 
\quad\text{\,}\quad 
R_4^2 = \dfrac{(\mu^{\rm d}_{\rm \delta} - \mu^{\rm o}_{\rm \delta})^2}{\sigma^2_{\mu_{\delta}}} \, , \\
R_5^2 &= \dfrac{(v^{\rm d}_{\rm los} - v^{\rm o}_{\rm los})^2}{\sigma^2_{v_{\rm los}}} \, .\\
\end{aligned}
\end{equation}

Here, $\theta_{\rm sky}$ is the on-sky angular difference between the orbit and the data point, $\varpi^{\rm d}, \mu^{\rm d}_{\rm \alpha}, \mu^{\rm d}_{\rm \delta}$  and $v^{\rm d}_{\rm los}$ are the measured data parallax, proper motion and los velocity, with the corresponding orbital model values marked with ``$o$''. The Gaussian dispersions $\sigma_{\rm sky}, \sigma_{\varpi}, \sigma_{\mu_{\alpha}}, \sigma_{\mu_{\delta}}, \sigma_{v_{\rm los}}$ are the sum in quadrature of the intrinsic dispersion of the model and the observational uncertainty of each data point. The reason for particularly adopting this ``conservative formulation'' of the log-likelihood function \citep{sivia1996data} is to lower the contribution from outliers that could be contaminating the stream data. Furthermore, in a given stream, those stars that lack spectroscopic measurements, we set them all to $v_{\rm los}=0\kms$ with a $10^4\kms$ Gaussian uncertainty. While undertaking this orbit-fitting procedure for a given stream, we chose to anchor the orbit solutions at fixed R.A. value (that was approximately half-way along the stream), while leaving all the other parameters to be varied. We do this because without setting an anchor, the solution would have wandered over the full length of the stream. The success of such a procedure in fitting streams has been demonstrated before \citep{Malhan_GalPot_2019, Malhan_2021_LMS1}. This procedure works well for most of the streams, as the final MCMC chains are converged and the resulting best fit orbits provide good representations to the phase-space structures of all these streams.

The above orbit-fitting procedure was carried out for the majority of streams, however, for a subset of them we considered it better to instead adopt the orbit-sampling procedure. This subset includes LMS-1, Orphan, Fimbulthul, Cetus, Svol, NGC~6397, Ophiuchus, C-3, Gaia-6, Chenab. The orbit-sampling procedure means that we no longer use equation~\ref{eq:Loglikehood} (that ensures that the resulting orbit provides a reasonable fit to the entire stream structure), but instead, we simply sample orbits using directly the phase-space measurements of the individual member stars (this does not guarantee an orbit-fit to the entire stream structure). The reason for adopting this scheme for LMS-1, Cetus (that are dwarf galaxy streams) and Fimbulthul (that is the stream of the massive $\omega$~Cen cluster) is that these are dynamically hot and physically broad streams, and the aforementioned orbit-fitting procedure would have underestimated their dispersions in the derived \EJ\ quantities. Similarly, Ophiuchus also appears to possess a broader dispersion in the $v_{\rm los}$ space ($\sim10-15\kms$, see Figure~10 of \citealt{Caldwell_2020}). For Orphan, that is a stream with a ``twisted'' shape (due to perturbation by the LMC, \citealt{Erkal_2019}), we deemed it better to sample its orbits (\citealt{Li_2021_12streams} also adopt a similar procedure to compute the orbit of Orphan). For the remaining streams, although they did appear narrow and linear in ($\alpha$, $\delta$) and ($\mu^{*}_{\alpha}$, $\mu_{\delta}$) space, it was difficult to visualise this linearity in the $v_{\rm los}$ space. This was primarily because these streams lack enough spectroscopic measurements so that a clear stream signal can be visible in the $v_{\rm los}$ space. Therefore, it was difficult to apply the orbit-fitting procedure for them and we resort to the orbit-sampling procedure. For all of these streams, the sampling in $\alpha$, $\delta$, $\mu^{*}_{\alpha}$, $\mu_{\delta}$ and $v_{\rm los}$ was performed directly using the measurements and the associated uncertainties. However, to sample over the distance parameter in a given stream, we computed the average distance (and the uncertainty) using the uncertainty-weighted average mean parallax of the member stars.  

The above orbit-fitting and orbit-sampling schemes generate the MCMC chains for the orbital parameters of all the $\nss$ streams, and for each stream we randomly sample $1000$ steps (this we do after rejecting the burn-in phase). These sampled values are shown in Figure~\ref{fig:Fig_EJ_of_objects}. Note that for most of the streams, their \EJ\ dispersions are much smaller than those of globular clusters and satellite galaxies. This is because the orbits of streams are much more precisely constrained (since we employ the above orbit-fitting procedure). The derived orbital properties of our streams are provided in Tables~\ref{tab:table_stream_helio} and \ref{tab:table_stream_EJ}.

\begin{table*}
\centering
\caption{Constrained heliocentric parameters of stellar streams. For each stream, the following values represent the posterior distribution at the stream's ``anchor'' point (i.e., at a fixed R.A. value). This anchor is defined during the orbit-fitting procedure.}
\label{tab:table_stream_helio}
\begin{tabular}{|l|c|c|c|c|c|c|c|c|}

\hline
\hline
Stream & $\#$ of \Gaia & $\#$ of & R.A. & Decl. & $D_\odot$ & $\mu^{*}_{\alpha}$ & $\mu_\delta$ & $v_{\rm los}$\\
& sources & $v_{\rm los}$ sources & (deg) & (deg) & (kpc) & ($\masyr$) & ($\masyr$) & ($\kms$)\\
\hline
\hline
& & & & & & & &\\

Gjoll  &  102  &  35  &  82.1  & $ -13.95 ^{+ 0.35 }_{- 0.36 }$  & $ 3.26 ^{+ 0.03 }_{- 0.03 }$  & $ 23.58 ^{+ 0.08 }_{- 0.09 }$  & $ -23.7 ^{+ 0.06 }_{- 0.05 }$  & $ 78.73 ^{+ 2.36 }_{- 1.84 }$ \\

Leiptr  &  237  &  67  &  89.11  & $ -28.37 ^{+ 0.2 }_{- 0.27 }$  & $ 7.39 ^{+ 0.07 }_{- 0.07 }$  & $ 10.59 ^{+ 0.03 }_{- 0.04 }$  & $ -9.9 ^{+ 0.03 }_{- 0.04 }$  & $ 194.22 ^{+ 2.23 }_{- 1.86 }$ \\

Hrid  &  233  &  24  &  280.51  & $ 33.3 ^{+ 0.75 }_{- 0.6 }$  & $ 2.75 ^{+ 0.1 }_{- 0.07 }$  & $ -5.88 ^{+ 0.11 }_{- 0.08 }$  & $ 20.08 ^{+ 0.21 }_{- 0.19 }$  & $ -238.77 ^{+ 3.3 }_{- 5.52 }$ \\

Pal5  &  48  &  29  &  229.65  & $ 0.26 ^{+ 0.1 }_{- 0.13 }$  & $ 20.16 ^{+ 0.24 }_{- 0.33 }$  & $ -2.75 ^{+ 0.03 }_{- 0.02 }$  & $ -2.68 ^{+ 0.02 }_{- 0.02 }$  & $ -57.03 ^{+ 1.08 }_{- 1.04 }$ \\

Gaia-1  &  106  &  8  &  190.96  & $ -9.16 ^{+ 0.15 }_{- 0.1 }$  & $ 5.57 ^{+ 0.16 }_{- 0.1 }$  & $ -14.39 ^{+ 0.04 }_{- 0.04 }$  & $ -19.72 ^{+ 0.03 }_{- 0.04 }$  & $ 214.91 ^{+ 3.5 }_{- 2.16 }$ \\

Ylgr  &  699  &  32  &  173.82  & $ -22.31 ^{+ 0.22 }_{- 0.3 }$  & $ 9.72 ^{+ 0.16 }_{- 0.14 }$  & $ -0.44 ^{+ 0.04 }_{- 0.03 }$  & $ -7.65 ^{+ 0.05 }_{- 0.04 }$  & $ 317.86 ^{+ 2.83 }_{- 3.05 }$ \\

Fjorm  &  182  &  28  &  251.89  & $ 65.38 ^{+ 0.24 }_{- 0.22 }$  & $ 6.42 ^{+ 0.16 }_{- 0.14 }$  & $ 3.92 ^{+ 0.07 }_{- 0.08 }$  & $ 3.1 ^{+ 0.06 }_{- 0.06 }$  & $ -25.37 ^{+ 1.89 }_{- 2.19 }$ \\

Kshir  &  55  &  16  &  205.88  & $ 67.25 ^{+ 0.13 }_{- 0.17 }$  & $ 9.57 ^{+ 0.08 }_{- 0.08 }$  & $ -7.67 ^{+ 0.04 }_{- 0.04 }$  & $ -3.92 ^{+ 0.04 }_{- 0.05 }$  & $ -249.88 ^{+ 2.62 }_{- 2.92 }$ \\

Gunnthra  &  61  &  8  &  284.22  & $ -73.49 ^{+ 0.23 }_{- 0.14 }$  & $ 2.83 ^{+ 0.12 }_{- 0.13 }$  & $ -15.83 ^{+ 0.11 }_{- 0.13 }$  & $ -24.04 ^{+ 0.15 }_{- 0.17 }$  & $ 132.26 ^{+ 6.23 }_{- 4.97 }$ \\

Slidr  &  181  &  29  &  160.05  & $ 10.22 ^{+ 0.43 }_{- 0.41 }$  & $ 2.99 ^{+ 0.11 }_{- 0.09 }$  & $ -24.6 ^{+ 0.08 }_{- 0.08 }$  & $ -6.65 ^{+ 0.06 }_{- 0.06 }$  & $ -87.98 ^{+ 3.44 }_{- 3.17 }$ \\

M92  &  84  &  9  &  259.89  & $ 43.08 ^{+ 0.2 }_{- 0.2 }$  & $ 8.94 ^{+ 0.2 }_{- 0.18 }$  & $ -5.15 ^{+ 0.05 }_{- 0.05 }$  & $ -0.63 ^{+ 0.06 }_{- 0.04 }$  & $ -140.66 ^{+ 6.28 }_{- 7.53 }$ \\

NGC~3201  &  388  &  4  &  152.46  & $ -46.32 ^{+ 0.11 }_{- 0.08 }$  & $ 4.99 ^{+ 0.01 }_{- 0.02 }$  & $ 8.87 ^{+ 0.02 }_{- 0.02 }$  & $ -2.22 ^{+ 0.02 }_{- 0.02 }$  & $ 489.63 ^{+ 3.36 }_{- 3.82 }$ \\

Atlas  &  46  &  10  &  25.04  & $ -29.81 ^{+ 0.1 }_{- 0.1 }$  & $ 19.93 ^{+ 0.76 }_{- 0.75 }$  & $ 0.04 ^{+ 0.02 }_{- 0.02 }$  & $ -0.89 ^{+ 0.02 }_{- 0.02 }$  & $ -85.65 ^{+ 1.48 }_{- 1.58 }$ \\

C-7  &  120  &  10  &  287.15  & $ -50.17 ^{+ 0.16 }_{- 0.14 }$  & $ 6.77 ^{+ 0.28 }_{- 0.21 }$  & $ -13.79 ^{+ 0.07 }_{- 0.06 }$  & $ -12.38 ^{+ 0.06 }_{- 0.07 }$  & $ 55.05 ^{+ 1.5 }_{- 2.51 }$ \\

Palca  &  24  &  24  &  36.57  & $ -36.15 ^{+ 0.33 }_{- 0.31 }$  & $ 12.31 ^{+ 1.68 }_{- 1.44 }$  & $ 0.9 ^{+ 0.02 }_{- 0.02 }$  & $ -0.23 ^{+ 0.04 }_{- 0.04 }$  & $ 106.32 ^{+ 2.62 }_{- 2.54 }$ \\

Sylgr  &  165  &  19  &  179.68  & $ -2.44 ^{+ 0.27 }_{- 0.4 }$  & $ 3.77 ^{+ 0.07 }_{- 0.11 }$  & $ -13.98 ^{+ 0.12 }_{- 0.14 }$  & $ -12.9 ^{+ 0.09 }_{- 0.1 }$  & $ -184.8 ^{+ 15.48 }_{- 8.15 }$ \\

Gaia-9  &  286  &  15  &  233.27  & $ 60.42 ^{+ 0.04 }_{- 0.11 }$  & $ 4.68 ^{+ 0.08 }_{- 0.09 }$  & $ -12.49 ^{+ 0.11 }_{- 0.12 }$  & $ 6.37 ^{+ 0.14 }_{- 0.08 }$  & $ -359.86 ^{+ 4.65 }_{- 4.11 }$ \\

Gaia-10  &  90  &  9  &  161.47  & $ 15.17 ^{+ 0.14 }_{- 0.14 }$  & $ 13.32 ^{+ 0.34 }_{- 0.28 }$  & $ -4.14 ^{+ 0.05 }_{- 0.05 }$  & $ -3.15 ^{+ 0.04 }_{- 0.04 }$  & $ 289.64 ^{+ 2.75 }_{- 3.32 }$ \\

Gaia-12  &  38  &  1  &  41.05  & $ 16.45 ^{+ 0.13 }_{- 0.13 }$  & $ 15.71 ^{+ 1.29 }_{- 1.03 }$  & $ 5.84 ^{+ 0.05 }_{- 0.05 }$  & $ -5.66 ^{+ 0.07 }_{- 0.06 }$  & $ -303.83 ^{+ 22.55 }_{- 15.07 }$ \\

Indus  &  454  &  45  &  340.12  & $ -60.58 ^{+ 0.1 }_{- 0.1 }$  & $ 14.96 ^{+ 0.19 }_{- 0.16 }$  & $ 3.59 ^{+ 0.03 }_{- 0.03 }$  & $ -4.89 ^{+ 0.02 }_{- 0.03 }$  & $ -49.15 ^{+ 2.45 }_{- 3.68 }$ \\

Jhelum  &  972  &  246  &  351.95  & $ -51.74 ^{+ 0.08 }_{- 0.08 }$  & $ 11.39 ^{+ 0.13 }_{- 0.15 }$  & $ 7.23 ^{+ 0.04 }_{- 0.04 }$  & $ -4.37 ^{+ 0.04 }_{- 0.03 }$  & $ -1.29 ^{+ 2.6 }_{- 3.11 }$ \\

Phoenix  &  35  &  19  &  23.96  & $ -50.01 ^{+ 0.24 }_{- 0.24 }$  & $ 16.8 ^{+ 0.33 }_{- 0.36 }$  & $ 2.72 ^{+ 0.03 }_{- 0.03 }$  & $ -0.07 ^{+ 0.03 }_{- 0.03 }$  & $ 45.92 ^{+ 1.63 }_{- 1.58 }$ \\

NGC5466  &  62  &  4  &  214.41  & $ 26.84 ^{+ 0.12 }_{- 0.11 }$  & $ 14.09 ^{+ 0.27 }_{- 0.25 }$  & $ -5.64 ^{+ 0.03 }_{- 0.03 }$  & $ -0.72 ^{+ 0.03 }_{- 0.02 }$  & $ 95.04 ^{+ 7.4 }_{- 5.91 }$ \\

M5  &  139  &  5  &  206.96  & $ 13.5 ^{+ 0.15 }_{- 0.14 }$  & $ 7.44 ^{+ 0.12 }_{- 0.11 }$  & $ 3.5 ^{+ 0.03 }_{- 0.04 }$  & $ -8.76 ^{+ 0.04 }_{- 0.04 }$  & $ -42.97 ^{+ 3.33 }_{- 3.83 }$ \\

C-20  &  34  &  9  &  359.81  & $ 8.63 ^{+ 0.16 }_{- 0.16 }$  & $ 18.11 ^{+ 1.45 }_{- 1.39 }$  & $ -0.58 ^{+ 0.03 }_{- 0.03 }$  & $ 1.44 ^{+ 0.02 }_{- 0.02 }$  & $ -116.87 ^{+ 1.46 }_{- 1.44 }$ \\

C-19  &  34  &  8  &  355.28  & $ 28.82 ^{+ 0.63 }_{- 1.17 }$  & $ 18.04 ^{+ 0.55 }_{- 0.53 }$  & $ 1.25 ^{+ 0.03 }_{- 0.03 }$  & $ -2.74 ^{+ 0.03 }_{- 0.05 }$  & $ -193.48 ^{+ 2.61 }_{- 2.52 }$ \\

Elqui  &  4  &  4  &  19.77  & $ -42.36 ^{+ 0.3 }_{- 0.29 }$  & $ 51.41 ^{+ 4.64 }_{- 7.04 }$  & $ 0.33 ^{+ 0.02 }_{- 0.03 }$  & $ -0.49 ^{+ 0.02 }_{- 0.02 }$  & $ 15.86 ^{+ 8.82 }_{- 20.38 }$ \\

AliqaUma  &  5  &  5  &  34.08  & $ -33.97 ^{+ 0.31 }_{- 0.34 }$  & $ 21.48 ^{+ 2.32 }_{- 1.2 }$  & $ 0.24 ^{+ 0.02 }_{- 0.03 }$  & $ -0.79 ^{+ 0.03 }_{- 0.03 }$  & $ -42.33 ^{+ 2.29 }_{- 2.23 }$ \\

Phlegethon  &  365  &  41  &  319.89  & $ -32.07 ^{+ 0.43 }_{- 0.37 }$  & $ 3.29 ^{+ 0.05 }_{- 0.05 }$  & $ -3.97 ^{+ 0.09 }_{- 0.09 }$  & $ -37.66 ^{+ 0.08 }_{- 0.09 }$  & $ 15.9 ^{+ 4.97 }_{- 6.12 }$ \\

GD-1  &  811  &  216  &  160.02  & $ 45.9 ^{+ 0.25 }_{- 0.19 }$  & $ 8.06 ^{+ 0.07 }_{- 0.07 }$  & $ -6.75 ^{+ 0.04 }_{- 0.03 }$  & $ -10.88 ^{+ 0.04 }_{- 0.05 }$  & $ -101.83 ^{+ 2.05 }_{- 2.47 }$ \\

& & & & & & & &\\
\hline
\hline
\end{tabular}
\tablecomments{From left to right the columns provide stream's name, number of \Gaia\ EDR3 sources in the stream, number of sources with spectroscopic line-of-sight velocities ($v_{\rm los}$), right ascension (R.A., that acts as the anchor point in our orbit-fitting procedure), declination (Decl.), heliocentric distance ($D_{\odot}$), proper motions ($\mu^{*}_{\alpha},\mu_\delta$) and $v_{\rm los}$. The quoted values are medians of the sampled posterior distributions and the corresponding uncertainties represent their $16$ and $84$ percentiles. Only those streams are listed for which orbit-fitting procedure was employed (see Section~\ref{subsec:compute_EJ}).}
\end{table*}
\begin{table*}
\centering
\caption{Actions, energies, orbital parameters and metallicities of stellar streams.}
\label{tab:table_stream_EJ}
\begin{tabular}{|l|l|c|c|c|c|c|c|}

\hline
\hline
Stream & $(J_R,J_\phi,J_z)$ & Energy & $r_{\rm peri}$ & $r_{\rm apo}$ & $z_{\rm max}$ & ecc. & [Fe/H]\\
& ($\kpc\kms$) & ($\times10^2\km2s2$) & (kpc) & (kpc) & (kpc) & & (dex) \\
\hline
\hline
& & & & & & & \\

LMS-1  & $( 255 ^{+ 239 }_{- 149 }, -627 ^{+ 183 }_{- 232 }, 2514 ^{+ 383 }_{- 263 })$  & $ -1227 ^{+ 65 }_{- 39 }$  & $ 10.8 ^{+ 2.5 }_{- 1.8 }$  & $ 20.6 ^{+ 3.7 }_{- 1.9 }$  & $ 20.2 ^{+ 3.6 }_{- 2.0 }$  & $ 0.32 ^{+ 0.11 }_{- 0.12 }$  & $ -2.1\pm0.4 $ \\

Gjoll  & $( 783 ^{+ 73 }_{- 65 }, 2782 ^{+ 60 }_{- 60 }, 274 ^{+ 7 }_{- 6 })$  & $ -1152 ^{+ 19 }_{- 18 }$  & $ 8.5 ^{+ 0.1 }_{- 0.1 }$  & $ 27.4 ^{+ 1.4 }_{- 1.2 }$  & $ 10.8 ^{+ 0.5 }_{- 0.5 }$  & $ 0.52 ^{+ 0.01 }_{- 0.01 }$  & $ -1.78 $ \\

Leiptr  & $( 1455 ^{+ 133 }_{- 119 }, 4128 ^{+ 77 }_{- 73 }, 378 ^{+ 11 }_{- 10 })$  & $ -933 ^{+ 19 }_{- 18 }$  & $ 12.3 ^{+ 0.1 }_{- 0.1 }$  & $ 45.1 ^{+ 2.3 }_{- 2.1 }$  & $ 17.6 ^{+ 0.9 }_{- 0.9 }$  & $ 0.57 ^{+ 0.01 }_{- 0.01 }$  & $ - $ \\

Hrid  & $( 1642 ^{+ 110 }_{- 79 }, 78 ^{+ 54 }_{- 47 }, 83 ^{+ 6 }_{- 5 })$  & $ -1319 ^{+ 30 }_{- 22 }$  & $ 1.1 ^{+ 0.0 }_{- 0.0 }$  & $ 22.0 ^{+ 1.4 }_{- 1.0 }$  & $ 7.0 ^{+ 0.9 }_{- 0.5 }$  & $ 0.9 ^{+ 0.0 }_{- 0.0 }$  & $ -1.1 $ \\

Pal5  & $( 282 ^{+ 19 }_{- 17 }, -744 ^{+ 61 }_{- 44 }, 1357 ^{+ 42 }_{- 51 })$  & $ -1385 ^{+ 11 }_{- 15 }$  & $ 6.9 ^{+ 0.3 }_{- 0.4 }$  & $ 15.8 ^{+ 0.2 }_{- 0.3 }$  & $ 14.7 ^{+ 0.2 }_{- 0.3 }$  & $ 0.39 ^{+ 0.02 }_{- 0.02 }$  & $ -1.35\pm0.06 $ \\

Orphan  & $( 959 ^{+ 978 }_{- 271 }, -3885 ^{+ 405 }_{- 1017 }, 1199 ^{+ 484 }_{- 213 })$  & $ -949 ^{+ 175 }_{- 64 }$  & $ 15.6 ^{+ 3.8 }_{- 2.1 }$  & $ 41.2 ^{+ 23.6 }_{- 6.2 }$  & $ 26.4 ^{+ 16.9 }_{- 4.8 }$  & $ 0.48 ^{+ 0.09 }_{- 0.06 }$  & $ -1.85\pm0.53 $ \\

Gaia-1  & $( 3638 ^{+ 1307 }_{- 632 }, 2678 ^{+ 89 }_{- 57 }, 997 ^{+ 93 }_{- 58 })$  & $ -794 ^{+ 90 }_{- 55 }$  & $ 8.2 ^{+ 0.1 }_{- 0.0 }$  & $ 67.6 ^{+ 18.3 }_{- 8.9 }$  & $ 45.7 ^{+ 13.6 }_{- 6.5 }$  & $ 0.78 ^{+ 0.04 }_{- 0.03 }$  & $ -1.36 $ \\

Fimbulthul  & $( 202 ^{+ 109 }_{- 78 }, 427 ^{+ 244 }_{- 588 }, 95 ^{+ 197 }_{- 44 })$  & $ -1847 ^{+ 73 }_{- 16 }$  & $ 2.4 ^{+ 0.8 }_{- 0.7 }$  & $ 7.2 ^{+ 0.4 }_{- 0.3 }$  & $ 2.4 ^{+ 3.5 }_{- 0.7 }$  & $ 0.51 ^{+ 0.12 }_{- 0.13 }$  & $ -1.36\,to\,-1.8 $ \\

Ylgr  & $( 205 ^{+ 40 }_{- 35 }, 2766 ^{+ 68 }_{- 66 }, 556 ^{+ 30 }_{- 25 })$  & $ -1219 ^{+ 20 }_{- 19 }$  & $ 11.5 ^{+ 0.1 }_{- 0.1 }$  & $ 20.7 ^{+ 1.2 }_{- 1.1 }$  & $ 11.2 ^{+ 0.8 }_{- 0.7 }$  & $ 0.29 ^{+ 0.02 }_{- 0.02 }$  & $ -1.87 $ \\

Fjorm  & $( 831 ^{+ 60 }_{- 62 }, -2332 ^{+ 24 }_{- 24 }, 877 ^{+ 43 }_{- 40 })$  & $ -1123 ^{+ 15 }_{- 15 }$  & $ 9.1 ^{+ 0.1 }_{- 0.1 }$  & $ 29.1 ^{+ 1.1 }_{- 1.1 }$  & $ 19.7 ^{+ 1.0 }_{- 1.0 }$  & $ 0.52 ^{+ 0.01 }_{- 0.01 }$  & $ -2.2 $ \\

Kshir  & $( 18 ^{+ 6 }_{- 4 }, 2755 ^{+ 60 }_{- 53 }, 491 ^{+ 14 }_{- 13 })$  & $ -1268 ^{+ 12 }_{- 11 }$  & $ 13.4 ^{+ 0.2 }_{- 0.2 }$  & $ 16.0 ^{+ 0.6 }_{- 0.5 }$  & $ 8.2 ^{+ 0.3 }_{- 0.3 }$  & $ 0.09 ^{+ 0.01 }_{- 0.01 }$  & $ -1.78 $ \\

Cetus  & $( 815 ^{+ 513 }_{- 317 }, -2416 ^{+ 841 }_{- 1064 }, 2287 ^{+ 1282 }_{- 954 })$  & $ -1000 ^{+ 124 }_{- 64 }$  & $ 14.7 ^{+ 7.2 }_{- 4.5 }$  & $ 35.9 ^{+ 9.9 }_{- 3.7 }$  & $ 30.2 ^{+ 10.9 }_{- 4.9 }$  & $ 0.45 ^{+ 0.14 }_{- 0.1 }$  & $ -2.0 $ \\

Svol  & $( 97 ^{+ 94 }_{- 32 }, -1501 ^{+ 384 }_{- 248 }, 224 ^{+ 107 }_{- 54 })$  & $ -1566 ^{+ 89 }_{- 61 }$  & $ 5.9 ^{+ 0.6 }_{- 0.6 }$  & $ 10.0 ^{+ 2.8 }_{- 1.0 }$  & $ 5.0 ^{+ 0.9 }_{- 0.8 }$  & $ 0.28 ^{+ 0.07 }_{- 0.05 }$  & $ -1.98\pm 0.10 $ \\

Gunnthra  & $( 69 ^{+ 14 }_{- 7 }, 852 ^{+ 67 }_{- 77 }, 218 ^{+ 30 }_{- 34 })$  & $ -1765 ^{+ 31 }_{- 28 }$  & $ 4.2 ^{+ 0.3 }_{- 0.4 }$  & $ 7.2 ^{+ 0.3 }_{- 0.2 }$  & $ 3.8 ^{+ 0.4 }_{- 0.4 }$  & $ 0.27 ^{+ 0.03 }_{- 0.02 }$  & $ - $ \\

Slidr  & $( 1076 ^{+ 217 }_{- 149 }, -1358 ^{+ 23 }_{- 23 }, 1831 ^{+ 126 }_{- 102 })$  & $ -1086 ^{+ 41 }_{- 32 }$  & $ 8.7 ^{+ 0.1 }_{- 0.1 }$  & $ 32.3 ^{+ 3.5 }_{- 2.5 }$  & $ 29.1 ^{+ 3.4 }_{- 2.4 }$  & $ 0.58 ^{+ 0.03 }_{- 0.02 }$  & $ -1.8 $ \\

M92  & $( 361 ^{+ 9 }_{- 9 }, 181 ^{+ 39 }_{- 41 }, 544 ^{+ 83 }_{- 70 })$  & $ -1639 ^{+ 12 }_{- 10 }$  & $ 3.0 ^{+ 0.2 }_{- 0.1 }$  & $ 10.7 ^{+ 0.2 }_{- 0.2 }$  & $ 9.9 ^{+ 0.5 }_{- 0.6 }$  & $ 0.56 ^{+ 0.01 }_{- 0.01 }$  & $ -2.16 \pm 0.05 $ \\

NGC~6397  & $( 75 ^{+ 5 }_{- 5 }, -586 ^{+ 15 }_{- 29 }, 222 ^{+ 23 }_{- 14 })$  & $ -1851 ^{+ 11 }_{- 6 }$  & $ 3.4 ^{+ 0.1 }_{- 0.1 }$  & $ 6.4 ^{+ 0.1 }_{- 0.1 }$  & $ 3.7 ^{+ 0.2 }_{- 0.1 }$  & $ 0.3 ^{+ 0.01 }_{- 0.01 }$  & $ - $ \\

NGC~3201  & $( 975 ^{+ 48 }_{- 48 }, 2860 ^{+ 32 }_{- 33 }, 296 ^{+ 5 }_{- 5 })$  & $ -1110 ^{+ 10 }_{- 10 }$  & $ 8.5 ^{+ 0.0 }_{- 0.0 }$  & $ 30.5 ^{+ 0.8 }_{- 0.8 }$  & $ 12.3 ^{+ 0.3 }_{- 0.3 }$  & $ 0.56 ^{+ 0.01 }_{- 0.01 }$  & $ - $ \\

Ophiuchus  & $( 507 ^{+ 387 }_{- 202 }, -160 ^{+ 34 }_{- 41 }, 1192 ^{+ 91 }_{- 98 })$  & $ -1490 ^{+ 130 }_{- 84 }$  & $ 3.9 ^{+ 0.3 }_{- 0.4 }$  & $ 14.2 ^{+ 4.9 }_{- 2.7 }$  & $ 14.1 ^{+ 4.9 }_{- 2.6 }$  & $ 0.58 ^{+ 0.11 }_{- 0.1 }$  & $ -1.80\pm0.09 $ \\

Atlas  & $( 757 ^{+ 39 }_{- 35 }, -1817 ^{+ 34 }_{- 33 }, 2093 ^{+ 100 }_{- 86 })$  & $ -1061 ^{+ 12 }_{- 11 }$  & $ 11.7 ^{+ 0.3 }_{- 0.3 }$  & $ 32.4 ^{+ 1.0 }_{- 0.9 }$  & $ 28.6 ^{+ 1.1 }_{- 1.0 }$  & $ 0.47 ^{+ 0.01 }_{- 0.01 }$  & $ -2.22\pm0.03 $ \\

C-7  & $( 1059 ^{+ 397 }_{- 215 }, 706 ^{+ 17 }_{- 25 }, 728 ^{+ 107 }_{- 69 })$  & $ -1319 ^{+ 100 }_{- 65 }$  & $ 3.5 ^{+ 0.0 }_{- 0.0 }$  & $ 21.0 ^{+ 5.2 }_{- 2.8 }$  & $ 18.1 ^{+ 5.1 }_{- 2.8 }$  & $ 0.72 ^{+ 0.05 }_{- 0.04 }$  & $ - $ \\

C-3  & $( 142 ^{+ 538 }_{- 64 }, 468 ^{+ 1110 }_{- 1185 }, 872 ^{+ 910 }_{- 510 })$  & $ -1571 ^{+ 338 }_{- 111 }$  & $ 5.7 ^{+ 2.0 }_{- 1.2 }$  & $ 10.0 ^{+ 13.2 }_{- 1.4 }$  & $ 8.7 ^{+ 12.5 }_{- 1.3 }$  & $ 0.35 ^{+ 0.18 }_{- 0.1 }$  & $ - $ \\

Palca  & $( 91 ^{+ 37 }_{- 24 }, -1830 ^{+ 28 }_{- 29 }, 1076 ^{+ 138 }_{- 128 })$  & $ -1300 ^{+ 30 }_{- 28 }$  & $ 10.8 ^{+ 0.4 }_{- 0.3 }$  & $ 16.5 ^{+ 1.5 }_{- 1.3 }$  & $ 12.7 ^{+ 1.5 }_{- 1.4 }$  & $ 0.21 ^{+ 0.03 }_{- 0.03 }$  & $ -2.02\pm0.23 $ \\

Sylgr  & $( 602 ^{+ 141 }_{- 202 }, -702 ^{+ 28 }_{- 24 }, 2220 ^{+ 94 }_{- 153 })$  & $ -1192 ^{+ 36 }_{- 61 }$  & $ 8.7 ^{+ 0.0 }_{- 0.0 }$  & $ 24.6 ^{+ 2.4 }_{- 3.7 }$  & $ 23.8 ^{+ 2.4 }_{- 3.7 }$  & $ 0.48 ^{+ 0.03 }_{- 0.06 }$  & $ -2.92\pm0.06 $ \\

Gaia-6  & $( 125 ^{+ 51 }_{- 71 }, 907 ^{+ 342 }_{- 229 }, 557 ^{+ 288 }_{- 204 })$  & $ -1593 ^{+ 129 }_{- 70 }$  & $ 6.0 ^{+ 1.4 }_{- 1.8 }$  & $ 9.5 ^{+ 3.1 }_{- 0.3 }$  & $ 6.9 ^{+ 3.6 }_{- 0.9 }$  & $ 0.3 ^{+ 0.08 }_{- 0.1 }$  & $ -1.16 $ \\

Gaia-9  & $( 393 ^{+ 37 }_{- 38 }, 1928 ^{+ 47 }_{- 61 }, 852 ^{+ 22 }_{- 20 })$  & $ -1255 ^{+ 15 }_{- 18 }$  & $ 8.7 ^{+ 0.1 }_{- 0.1 }$  & $ 20.8 ^{+ 0.8 }_{- 0.9 }$  & $ 14.7 ^{+ 0.6 }_{- 0.6 }$  & $ 0.41 ^{+ 0.01 }_{- 0.01 }$  & $ -1.94 $ \\

Gaia-10  & $( 2189 ^{+ 66 }_{- 64 }, 287 ^{+ 43 }_{- 43 }, 1542 ^{+ 155 }_{- 140 })$  & $ -1051 ^{+ 20 }_{- 18 }$  & $ 4.3 ^{+ 0.4 }_{- 0.3 }$  & $ 37.7 ^{+ 1.6 }_{- 1.4 }$  & $ 37.2 ^{+ 1.6 }_{- 1.5 }$  & $ 0.8 ^{+ 0.01 }_{- 0.01 }$  & $ -1.4 $ \\

Gaia-12  & $( 9834 ^{+ 5378 }_{- 4378 }, 7340 ^{+ 608 }_{- 801 }, 794 ^{+ 80 }_{- 107 })$  & $ -433 ^{+ 98 }_{- 142 }$  & $ 18.5 ^{+ 0.9 }_{- 1.2 }$  & $ 194.3 ^{+ 96.8 }_{- 75.0 }$  & $ 83.0 ^{+ 42.9 }_{- 32.5 }$  & $ 0.83 ^{+ 0.05 }_{- 0.08 }$  & $ -2.6 $ \\

Indus  & $( 99 ^{+ 25 }_{- 19 }, -1121 ^{+ 35 }_{- 36 }, 2211 ^{+ 61 }_{- 49 })$  & $ -1232 ^{+ 17 }_{- 15 }$  & $ 12.6 ^{+ 0.2 }_{- 0.1 }$  & $ 18.9 ^{+ 1.0 }_{- 0.9 }$  & $ 17.8 ^{+ 0.9 }_{- 0.8 }$  & $ 0.2 ^{+ 0.02 }_{- 0.02 }$  & $ -1.96\pm0.41 $ \\

Jhelum  & $( 594 ^{+ 49 }_{- 56 }, -356 ^{+ 19 }_{- 17 }, 2557 ^{+ 62 }_{- 72 })$  & $ -1193 ^{+ 17 }_{- 22 }$  & $ 8.7 ^{+ 0.2 }_{- 0.2 }$  & $ 24.5 ^{+ 1.1 }_{- 1.3 }$  & $ 24.3 ^{+ 1.1 }_{- 1.3 }$  & $ 0.48 ^{+ 0.01 }_{- 0.01 }$  & $ -1.83\pm0.34 $ \\

Phoenix  & $( 107 ^{+ 11 }_{- 10 }, -1563 ^{+ 35 }_{- 32 }, 1578 ^{+ 56 }_{- 62 })$  & $ -1259 ^{+ 10 }_{- 12 }$  & $ 11.7 ^{+ 0.4 }_{- 0.5 }$  & $ 18.1 ^{+ 0.3 }_{- 0.3 }$  & $ 15.6 ^{+ 0.3 }_{- 0.3 }$  & $ 0.22 ^{+ 0.01 }_{- 0.01 }$  & $ -2.70\pm0.06 $ \\

NGC5466  & $( 1769 ^{+ 144 }_{- 114 }, 619 ^{+ 38 }_{- 35 }, 1373 ^{+ 93 }_{- 80 })$  & $ -1098 ^{+ 29 }_{- 24 }$  & $ 4.8 ^{+ 0.3 }_{- 0.2 }$  & $ 33.7 ^{+ 2.3 }_{- 1.8 }$  & $ 31.8 ^{+ 2.2 }_{- 1.7 }$  & $ 0.75 ^{+ 0.01 }_{- 0.01 }$  & $ - $ \\

M5  & $( 1366 ^{+ 69 }_{- 57 }, -353 ^{+ 20 }_{- 18 }, 931 ^{+ 46 }_{- 40 })$  & $ -1246 ^{+ 17 }_{- 14 }$  & $ 3.4 ^{+ 0.1 }_{- 0.1 }$  & $ 24.8 ^{+ 1.0 }_{- 0.8 }$  & $ 23.6 ^{+ 1.1 }_{- 0.9 }$  & $ 0.76 ^{+ 0.01 }_{- 0.01 }$  & $ -1.34\pm 0.05 $ \\

C-20  & $( 1329 ^{+ 526 }_{- 350 }, -3042 ^{+ 131 }_{- 167 }, 3823 ^{+ 503 }_{- 484 })$  & $ -800 ^{+ 65 }_{- 59 }$  & $ 20.8 ^{+ 1.3 }_{- 1.3 }$  & $ 58.5 ^{+ 12.0 }_{- 8.8 }$  & $ 52.4 ^{+ 11.3 }_{- 8.5 }$  & $ 0.47 ^{+ 0.05 }_{- 0.04 }$  & $ -2.44 $ \\

NGC7089  & $( 800 ^{+ 713 }_{- 270 }, -638 ^{+ 414 }_{- 555 }, 359 ^{+ 99 }_{- 123 })$  & $ -1504 ^{+ 307 }_{- 104 }$  & $ 2.9 ^{+ 1.2 }_{- 0.7 }$  & $ 14.7 ^{+ 12.7 }_{- 3.0 }$  & $ 10.9 ^{+ 7.5 }_{- 3.8 }$  & $ 0.71 ^{+ 0.06 }_{- 0.06 }$  & $ - $ \\

C-19  & $( 383 ^{+ 53 }_{- 47 }, -210 ^{+ 48 }_{- 46 }, 2712 ^{+ 258 }_{- 253 })$  & $ -1232 ^{+ 21 }_{- 21 }$  & $ 9.3 ^{+ 1.0 }_{- 1.0 }$  & $ 21.6 ^{+ 0.5 }_{- 0.5 }$  & $ 21.6 ^{+ 0.5 }_{- 0.6 }$  & $ 0.4 ^{+ 0.04 }_{- 0.03 }$  & $ -3.38\pm0.06 $ \\

Elqui  & $( 2072 ^{+ 543 }_{- 602 }, -273 ^{+ 166 }_{- 191 }, 4324 ^{+ 321 }_{- 352 })$  & $ -868 ^{+ 27 }_{- 35 }$  & $ 12.1 ^{+ 1.8 }_{- 2.0 }$  & $ 54.0 ^{+ 4.6 }_{- 6.2 }$  & $ 53.9 ^{+ 4.6 }_{- 6.3 }$  & $ 0.64 ^{+ 0.07 }_{- 0.08 }$  & $ -2.22\pm0.37 $ \\

Chenab  & $( 2469 ^{+ 463 }_{- 286 }, -4062 ^{+ 601 }_{- 509 }, 3735 ^{+ 341 }_{- 338 })$  & $ -690 ^{+ 52 }_{- 53 }$  & $ 22.0 ^{+ 2.2 }_{- 2.7 }$  & $ 81.0 ^{+ 12.8 }_{- 10.4 }$  & $ 69.1 ^{+ 10.6 }_{- 8.2 }$  & $ 0.58 ^{+ 0.01 }_{- 0.01 }$  & $ -1.78\pm0.34 $ \\

AliqaUma  & $( 738 ^{+ 138 }_{- 71 }, -1838 ^{+ 96 }_{- 64 }, 2025 ^{+ 223 }_{- 126 })$  & $ -1067 ^{+ 30 }_{- 18 }$  & $ 11.6 ^{+ 0.4 }_{- 0.3 }$  & $ 31.9 ^{+ 2.6 }_{- 1.5 }$  & $ 27.9 ^{+ 3.0 }_{- 1.6 }$  & $ 0.47 ^{+ 0.03 }_{- 0.02 }$  & $ -2.30\pm0.06 $ \\

Phlegethon  & $( 815 ^{+ 120 }_{- 93 }, 1882 ^{+ 39 }_{- 37 }, 231 ^{+ 11 }_{- 10 })$  & $ -1272 ^{+ 32 }_{- 28 }$  & $ 5.5 ^{+ 0.0 }_{- 0.0 }$  & $ 22.1 ^{+ 1.8 }_{- 1.4 }$  & $ 9.4 ^{+ 0.9 }_{- 0.7 }$  & $ 0.6 ^{+ 0.02 }_{- 0.02 }$  & $ -1.96\pm0.05 $ \\

GD-1  & $( 164 ^{+ 35 }_{- 28 }, 2952 ^{+ 66 }_{- 61 }, 938 ^{+ 23 }_{- 22 })$  & $ -1153 ^{+ 15 }_{- 14 }$  & $ 14.1 ^{+ 0.1 }_{- 0.1 }$  & $ 23.0 ^{+ 1.1 }_{- 1.0 }$  & $ 14.8 ^{+ 0.7 }_{- 0.7 }$  & $ 0.24 ^{+ 0.02 }_{- 0.02 }$  & $ -2.24\pm0.21 $ \\

\hline
\hline
\end{tabular}
\tablecomments{From left to right the columns provide the stream's name, action components (\J), energy ($E$), pericentric distance ($r_{\rm peri}$), apocentric distance ($r_{\rm apo}$), maximum height of the orbit from the Galactic plane ($z_{\rm max}$), eccentricity (ecc) and [Fe/H] measurements (most of these are spectroscopic and a few are photometric). The \EJ\ and other orbital parameters are derived in this study. The quoted orbital parameter values are the medians of the sampled posterior distributions and the corresponding uncertainties reflect their $16$ and $84$ percentiles. The [Fe/H] values of Gaia-6, Gaia-9, Gaia-10 and Gaia-12 correspond to the median of the spectroscopic sample that we obtained in this study. The other streams with spectroscopic [Fe/H] include LMS-1 (its value is taken from \citealt{Malhan_2021_LMS1}), Gj\"oll, Ylgr, Slidr, Fj\"orm \citep{Ibata_norse_2019}, Jhelum, Chenab, Elqui, Ophiuchus, Orphan, Palca, Indus \citep{Li_2021_12streams}, Fimbulthul \citep{Ibata_2019_wcen}, Gaia-1, C-2 and Hr\'id \citep{MalhanSun2020}, Cetus \citep{Yam_2013}, Sylgr \citep{Roederer_2019}, GD-1 \citep{Malhan_GalPot_2019}, Kshir \citep{Malhan_Kshir2019}, C-20 \citep{Yuan_2022_Cetus}, Pal~5 \citep{Ishigaki_2016}, Atlas, AliqaUma \citep{Li_2021_AAS}. Streams with photometric [Fe/H] include Phlegethon, Sv\"ol, M~92, M~5 (their values are taken from \citealt{Martin2022_Pr_SF}).}
\end{table*}
\subsection{A qualitative analysis of the orbits}\label{subsec:analyse_orbits}

As a passing analysis, we qualitatively examine some basic orbital properties of globular clusters, satellite galaxies and stellar streams. The knowledge gained from this analysis allows us to put our final results in some context. 

For globular clusters, we find that $\sim70\%$ of them move along prograde orbits (i.e., their $J_\phi<0$), $18\%$ move along polar orbits (i.e., their orbital planes are inclined almost perpendicularly to the Galactic disk plane, with $\phi>=75\deg$), $\sim 12\%$ have their orbits nearly confined to the Galactic plane (i.e., their $\phi<=20\deg$), $11\%$ have disk-like orbits (i.e., both prograde and in-plane), $1\%$ have in-plane and retrograde orbits and $10\%$ have highly eccentric orbits (with ecc$>0.8$). This excess of prograde globular clusters could be indicating that the Galactic halo itself initially had an excess of prograde clusters or it may owe to the possible spinning of the dark matter halo (e.g., \citealt{Obreja_2021}).

For satellites, we find that $\sim60\%$ of them move along prograde orbits, $\simlt10\%$ have highly eccentric orbits and $\sim45\%$ move along polar orbits (most of these `polar' satellites belong to the `Vast Plane of Satellites' structure, see \citealt{Pawlowski_2021}). None of the satellites move in the disk plane; this could be because satellites on co-planar orbits are expected to be destroyed quickly compared to those on polar orbits (e.g., \citealt{Penarrubia_2002}). The satellites possess quite high energies and angular momenta compared to the globular clusters (and also stellar streams, as we note below). The high $E$ values of satellites suggest that many of them are not ancient inhabitants of the Milky Way, but have only recently arrived into our Galaxy (perhaps $\simlt 4\Gyr$ ago, e.g., \citealt{Hammer_2021}). 

For stellar streams, we find that $55\%$ of them move along prograde orbits, $22\%$ move along polar orbits and $5\%$ possess highly eccentric orbits. Some of these polar streams are LMS-1, C-19, Sylgr, Jhelum, Elqui, Gaia-10, Ophiuchus, Hr\`id. None of the streams orbit in the disk-plane. Our inference on the prograde distribution of streams is somewhat consistent with the study of \cite{Panithanpaisal_2021}, who analysed FIRE~2 cosmological simulations and found that Milky Way-mass galaxies should have an even distribution of streams on prograde and retrograde orbits. 

As a final passing analysis, and not to deviate too much from the prime objective of the paper, we quickly compare the distribution of the orbital phase and eccentricity of all the halo objects (as shown in Figure~\ref{fig:Fig_phase} of Appendix~\ref{appendix:phase}). First, we observe a pile-up of objects at the pericenter and at the apocenter, and this is more prevalent for globular clusters and stellar streams, and not so much for satellite galaxies. Particularly, in the case of streams, we note that more objects are piled-up at the pericenter than at the apocenter. This effect points towards our inefficiency in detecting those streams that, at the present day, could be close to their apocenters (at distances $D_{\odot}\simgt30\kpc$). This inefficiency, in part, is also because of \Gaia's limiting magnitude at $G\sim21$. Our result is different than that of \cite{Li_2021_12streams}, who find that more streams (in their sample of $12$ streams) are piled-up at the apocentre. Secondly, we find that most of the objects (be it clusters, streams or satellites) have eccentricities $e\approx0.5$, and it is rare for the objects to possess very radial orbits ($e\approx1$) or very circular orbits ($e\approx0$). This last inference, in regard to streams, is consistent with that of \cite{Li_2021_12streams}. 

In summary, we now possess \EJ\ information for a total of $n=\nobjects$ halo objects of the Milky Way (as shown in Figure~\ref{fig:Fig_EJ_of_objects}). In the next section, we process this entire \EJ\ data to detect groups of objects (i.e., mergers). Therefore, at this stage, it is important to clarify that some of the objects are being counted twice in our dataset. These objects include those systems that have counterparts both in the globular cluster catalog and the stream catalog. For instance, a subset of these objects include Pal~5, NGC~3201, $\omega$~Centauri and M~5. One possible way to proceed would be to remove their counterparts from either of the catalogs. However, there could be many other streams in our catalog that could be physically associated to other globular clusters (e.g., see Section~\ref{sec:Streams_progs}) or even to other streams (e.g., Orphan--Chenab, \citealt{Koposov_2019}, Palca--Cetus, \citealt{Chang_2020_Cetus}, AliqaUma--Atlas, \citealt{Li_2021_AAS}), and it is a difficult task to separate these plausible associations. We therefore consider it to be less biased to proceed with all of the detected structures. Prior associations will be discussed in our final grouping analysis.

\section{Detecting groups of objects in the \EJ\ space}\label{sec:ENLINK}

To search for the Milky Way mergers, we essentially process the data shown in Figure~\ref{fig:Fig_EJ_of_objects} and detect groups of objects that tightly clump together in the \EJ\ space. For detecting these groups, we employ the \enlink\ software \citep{Sharma_2009_ENLINK} and couple it with a statistical procedure that accounts for the uncertainties in the \EJ\ values of every object. Below, we first briefly describe the working of \enlink\ and then our procedure to detect groups.

\subsection{Description of \enlink}

\enlink\ is a density-based hierarchical group finding algorithm that detects groups of any shape and density in a multidimensional dataset. This software employs non-parametric methods to find groups, i.e., it makes no assumptions about the number of groups being identified or their form. These functionalities of \enlink\ are particularly useful for our study because {\it a priori} we neither know the number of groups (i.e., number of mergers) that are present in the \EJ\ dataset, nor the shapes of these groups (since objects that accrete inside the same merging galaxy can realise extended/irregular ellipsoidal shapes in the \EJ\ space, e.g., \citealt{Wu_2021}). 

To detect groups in the dataset, \enlink\ does not use the typical Euclidean metric, but builds a locally adaptive Mahalanobis (LAM) metric. The importance of this metric can be explained as follows. Generally speaking, the task of finding groups in a given dataset ultimately boils down to computing ``distances'' between different data points. Then, those data points that lie at smaller distances from each other form part of the same group. In a scenario where correlation between different dimensions of the dataset are zero or negligible, one can simply adopt the Euclidean metric to compute these distances. In this case, the distance between two data points $\rm{{\bf x}_i}$ and $\rm{{\bf x}_j}$ is given by ${\rm s^2({\bf x}_i,{\bf x}_j)=({\bf x}_i - {\bf x}_j)^T.({\bf x}_i - {\bf x}_j)}$; where $\rm{\bf x}$ is a 1D matrix whose length equals the dimension of the dataset. However, in real datasets, correlations between different dimensions are non-zero. Particularly in our case, one expects significant correlation in the space constructed with \J\ and $E$ dimensions. Therefore, to find groups in such a correlated dataset, one effectively requires a multivariate equivalent of the Euclidean distance. This is the importance of LAM that \enlink\ employs, because the Mahalonobis distance is the distance between a point and a distribution (and not between two data points). At its heart, \enlink\ uses the LAM metric, where the distance between two data points (under descrete approximation) is defined as 
\begin{equation}
{\rm s^2 ({\bf x_i,x_j}) =|\Sigma({\bf x_i,x_j})|^{1/d}.({\bf x_i} - {\bf x_j})^{T}. \Sigma^{-1}({\bf x_i,x_j}).({\bf x_i} - {\bf x_j})}\,,
\end{equation}
where `d' is the dimension of data, $\Sigma$ is the covariance matrix,  ${\rm \Sigma({\bf x_i,x_j})}$ = $0.5 [{\rm \Sigma ({\bf x_i})}$ + ${\rm \Sigma({\bf x_j})}]$ and ${\rm \Sigma^{-1}({\bf x_i,x_j})}$ = $0.5 [{\rm \Sigma^{-1} ({\bf x_i})}$ + ${\rm \Sigma^{-1}({\bf x_j})}]$. 

The above formula can be intuitively understood as follows. Consider the term ${\rm ({\bf x_i} -{\bf  x_j})^T}.\Sigma^{-1}$. Here, ${\rm ({\bf x_i} -{\bf  x_j})}$ is the distance between two data points. This is then multiplied by the inverse of the covariance matrix $\Sigma$ (or divided by the covariance matrix). So, this is essentially a multivariate equivalent of the regular standardization y = (x - $\mu$)/$\sigma$. The effect of dividing by covariance is that if the $\rm{\bf x}$ values in the dataset are strongly correlated, then the covariance will be high and dividing by this large covariance will reduce the distance. On the other hand, if the $\rm{\bf x}$ are not correlated, then the covariance is small and the distance is not reduced by much. The overall workings and implementation of \enlink\ is detailed in \cite{Sharma_2009_ENLINK} and this software has also been previously applied to various datasets (e.g., \citealt{Sharma_2010_ENLINK_2MASS, Wu_2021}).

\begin{figure*}
\begin{center}
\vspace{-0.30cm}
\includegraphics[width=0.95\hsize]{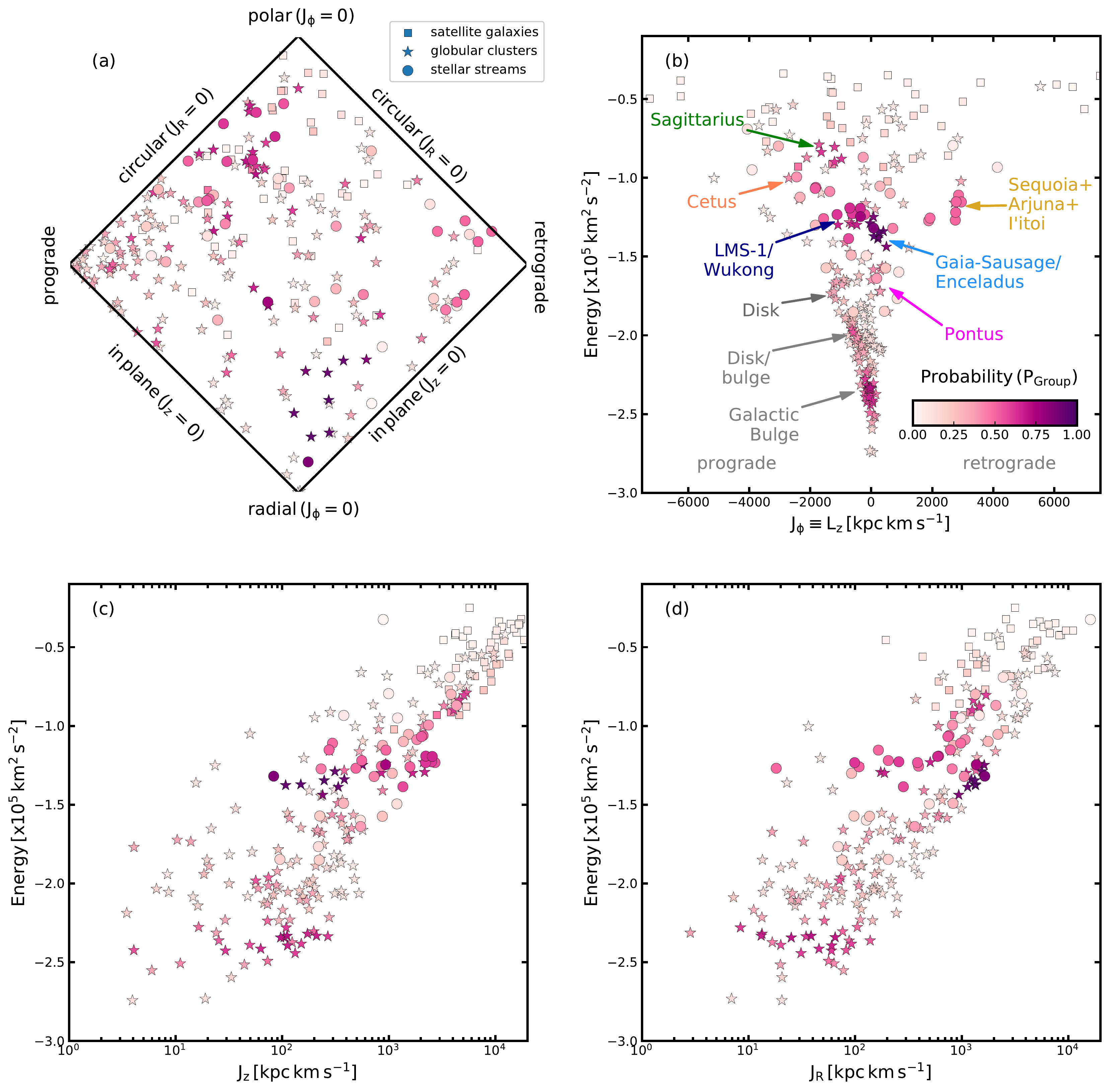}
\end{center}
\vspace{-0.5cm}
\caption{\EJ\ distribution of the halo objects as a function of their group-probability $\Pg$ (see Section~\ref{sec:ENLINK}). In this plot, each object is represented by the median of its \EJ\ distribution (that is shown in Figure~\ref{fig:Fig_EJ_of_objects}). The globular clusters are denoted by `stars', streams by `circles' and satellite galaxies by `squares'. Note that objects with higher $\Pg$ values lie in the denser regions of this \EJ\ space. Such objects with high $\Pg$ values, that also clump together in the \EJ\ space, form part of the same group. In panel `b', we label all the high-significance groups.}
\label{fig:Fig_EJ_probability}
\end{figure*}
\subsection{Applying \enlink}\label{subsec:app_enlink}

To detect groups, we work in the 4-dimensional space of $\rm{\bf x_i}\equiv$ $(J_{R,i}, J_{\phi,i}, J_{z,i}, E_i)$, where $i$ represents a given halo object and the units of \J\ and $E$ are $\kpc\kms$ and $\km2s2$, respectively. The reason for working with both \J\ and $E$ quantities is that their combined information allowed us to detect several groups (as we show below). Initially, we operated \enlink\ only in the 3-dimensional space of \J. However, this resulted in the detection of the \Sgr\ group \citep{Ibata1994, Bellazzini_2020} (although with unusual membership of objects), the \asi\ group \citep{Naidu2020, Bonaca2021} and $1-2$ other very low-significance groups. At first, this may seem odd that \enlink\ requires the additional (redundant) $E$ information to find high-significance groups; since \J\ fully characterize the orbits and the parameter $E$ brings no additional dynamical information. However, this oddity relates to the uncertainties on \J\ and $E$. For instance, the relative uncertainties on $(J_{R,i}, J_{\phi,i}, J_{z,i})$ for all the objects in our sample (on average) are ($12\%,17\%,9\%$), while the relative uncertainty on $E$ is only $2\%$. Therefore, \enlink\ prefers these precise values of $E$, in addition to \J\, as this helps it to easily distinguish between different groups.

The \enlink\ parameters that we use are \texttt{neighbors}, \texttt{min\_cluster\_size}, \texttt{min\_peak\_height}, \texttt{cluster\_separation\_method}, \texttt{density\_method} and \texttt{gmetric}. \texttt{neighbors} is the `smoothning' that is used to compute a local density for each data point, since \enlink\ first estimates the density and then finds groups in the density field. To search for groups in a $d$-dimensional dataset, \enlink\ requires \texttt{neighbours}$\geq (d+1)$. In our case, $d=4$ ($3$ components of \J\ and $E$) and therefore we set \texttt{neighbors}=5. Secondly, we set \texttt{min\_cluster\_size}=5. This is because it is difficult to find groups smaller than the smoothing length (i.e., we satisfy the \texttt{min\_cluster\_size}$\geq$\texttt{neighbours} condition of \enlink). \texttt{min\_peak\_height} can be thought of as the signal-to-noise ratio of the detected groups, and we set \texttt{min\_peak\_height}=3.0. For the parameters \texttt{cluster\_separation\_method} and \texttt{gmetric} we adopt the default values (i.e., 0). Further, we set \texttt{density\_method}=\texttt{sbr} as this uses an adaptive metric to detect groups. We also tried different metric definitions, but these gave very similar results that we obtained from the above parameter setting\footnote{For example, instead of using the adaptive metric, we defined a constant metric using the uncertainties on \EJ\ by setting \texttt{gmetric}=2 and using the \texttt{custom\_metric} parameter. We made this test because since we are dealing with a very low number of data points (only $\nobjects$ points), we wanted to ensure that the detected groups are robust and are not noise driven. However, in this case we found similar results as with the original \enlink\ setting.}. Our experimentation with various parameter settings makes us confident that we are detecting robust groups. 

Before unleashing \enlink\ onto the \EJ\ dataset, we couple it with a statistical procedure that accounts for the dispersion in the \EJ\ values of the objects (these dispersions are visible in Figure~\ref{fig:Fig_EJ_of_objects}). This is important because \enlink\ itself does not account for the dispersion associated with each data point. This statistical procedure can be explained as follows. Fundamentally, we want to compute a ``group-probability'' ($\Pg$) for each halo object, such that this probability is higher for those objects that belong to the groups detected by \enlink. To compute this $\Pg$ value, we undertake an iterative procedure. 

In the first iteration, each halo object is represented by a single \EJ\ value that is sampled from its MCMC chain (we obtained these MCMC chains in Section~\ref{subsec:compute_EJ}). At this stage, the total number of \EJ\ data points equals the total number of objects (i.e., \nobjects). After this, we process this \EJ\ data using \enlink. An attribute that \enlink\ returns is a 1D array \texttt{labels}. \texttt{labels} has the same length as the number of input data points and it stores the grouping information. That is, all the elements in \texttt{labels} possess integer values in the range $1$ to $n$, where $n$ is the total number of groups detected by \enlink, and elements that form part of the same group receive the same values. Furthermore, elements for which \texttt{labels}$=1$ correspond to those objects that form part of the largest group. For all the objects with \texttt{labels}$\geq2$, we explicitly set their probability of group membership at iteration $i$ to be $P_{{\rm Group}, i}=1$. Among objects with \texttt{labels}$=1$, we accept only those objects that possess \texttt{density}$>99$~percentile and set their $P_{{\rm Group}, i}=1$, while the remaining low \texttt{density} objects are set as $P_{{\rm Group}, i}=0$\footnote{The reason that we make such a distinction for the objects in the \texttt{labels}$=1$ group is that -- a majority of objects in this largest group are those that that could not be associated with any ``well defined group'' by \enlink\ (these represent the background objects). However, even in this group, some of the high density objects may still represent a real merger. Therefore, in order to consider these potential objects of interest, we accept only those objects that satisfy the the threshold \texttt{density} criteria.}. In the next iteration, a new set of \EJ\ values is sampled and the the above procedure is repeated. Note that in this new iteration, the input \EJ\ data has changed, and therefore the same object can now belong to a different group, thus receive a different \texttt{labels} value and a different $P_{{\rm Group}, i}$ value. We iterate this procedure $1000$ times. This produces, for each halo object, a one-dimensional array (of length $1000$) that contains a combination of $0$s or $1$s. For each halo object, we take the average of this array and this we interpret as the group-probability $\Pg$ of that object. The $\Pg$ parameter can be defined as the probability of an object belonging to a group in the \EJ\ space. Indeed, those halo objects that lie in denser regions of the \EJ\ space -- i.e., objects whose \EJ\ distributions overlap significantly -- will possess higher $\Pg$ values.

Figure~\ref{fig:Fig_EJ_probability} shows the \EJ\ distribution of the halo objects as a function of the computed $\Pg$ values. In this figure, each object is represented by the median of its \EJ\ distribution. It can be seen that different objects possess different $\Pg$ values. We also note that objects with high $\Pg$ values lie in denser regions of the \EJ\ space; suggesting that our procedure of detecting groups has worked as desired. In Figure~\ref{fig:Fig_EJ_probability}, one can already visually identify many possible groupings -- comprising those objects that possess high $\Pg$ values and which appear well-separated from other groups. 

\begin{figure}
\begin{center}
\vspace{-0.3cm}
\includegraphics[width=\hsize]{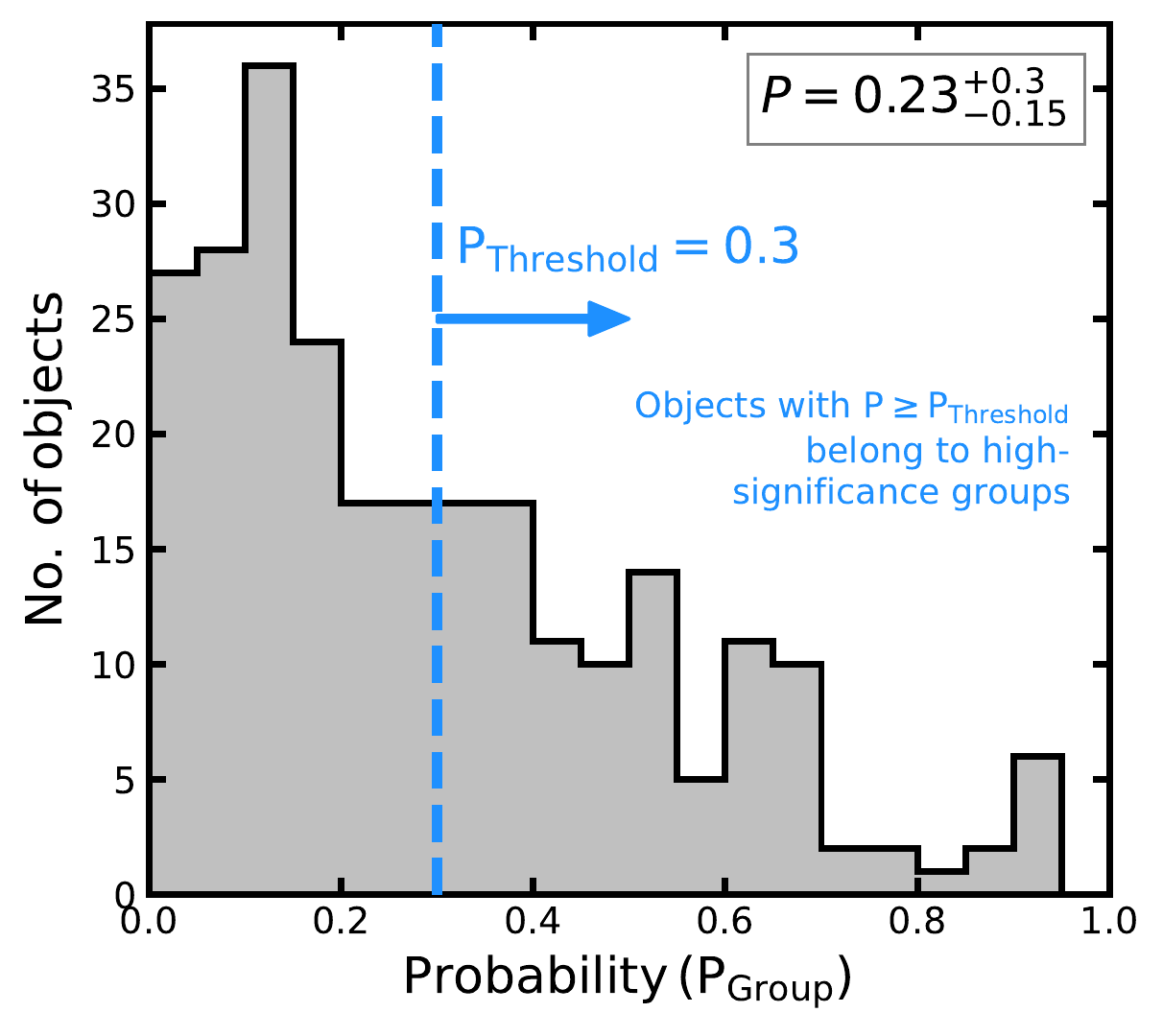}
\end{center}
\vspace{-0.6cm}
\caption{PDF of the group-probability $\Pg$ of all the halo objects (see Section~\ref{subsec:detect_high_sig_groups}). The vertical line represents the $\Pth$ value and all the objects with $\Pg\geq\Pth$ belong to high-significance groups. The value quoted in the top-right corner is the median of the distribution and the corresponding uncertainties reflect the $16$ and $84$ percentiles.}
\label{fig:Fig_ProbDF}
\end{figure}
\subsection{Detecting high-significance groups}\label{subsec:detect_high_sig_groups}

Due to the relatively large \EJ\ uncertainties, the \enlink\ algorithm's output of proposed groupings varies considerably over the $1000$ random iterations described above. This means that the proposed groups cannot be immediately used to identify the Milky Way's mergers.

Therefore, we proceed by first defining a threshold value $\Pth$, such that objects with $\Pg \geq \Pth$ belong to high-significance groups, and this corresponds to a likely detection. To find a suitable $\Pth$ value, we follow a pragmatic approach. We repeat the above analysis of computing the $\Pg$ values of all the halo objects, except this time we use a `randomised' version of our real \EJ\ data. This randomised data is artificially created, where each object is first assigned a random orbital pole and then its new \EJ\ values are computed. This randomised \EJ\ data is shown in Figure~\ref{fig:Fig_EJ_of_random_objects} in Appendix~\ref{appendix:EJ_random}. Such a randomisation procedure erases any plausible correlations between the objects in the \EJ\ space. For the resulting PDF of the new $\Pg$ values (that is shown in  Figure~\ref{fig:Fig_PDFrandom}), its $90$ percentile limit motivates setting a threshold at $\Pth=0.3$ for a $2\sigma$ detection. This procedure may seem convoluted, but it is required by the astrometric uncertainties which project in a complicated, non-linear way into \EJ\ space (hence the usual techniques of error propagation would not have been appropriate). This method of finding the $\Pth$ value is detailed in Appendix~\ref{appendix:EJ_random}. Consequently, for the real $\Pg$ values (shown in Figure~\ref{fig:Fig_ProbDF}), all those objects that possess $\Pg \geq \Pth$ are considered as high-significance groups.

\begin{figure*}
\begin{center}
\vspace{-0.3cm}
\includegraphics[width=\hsize]{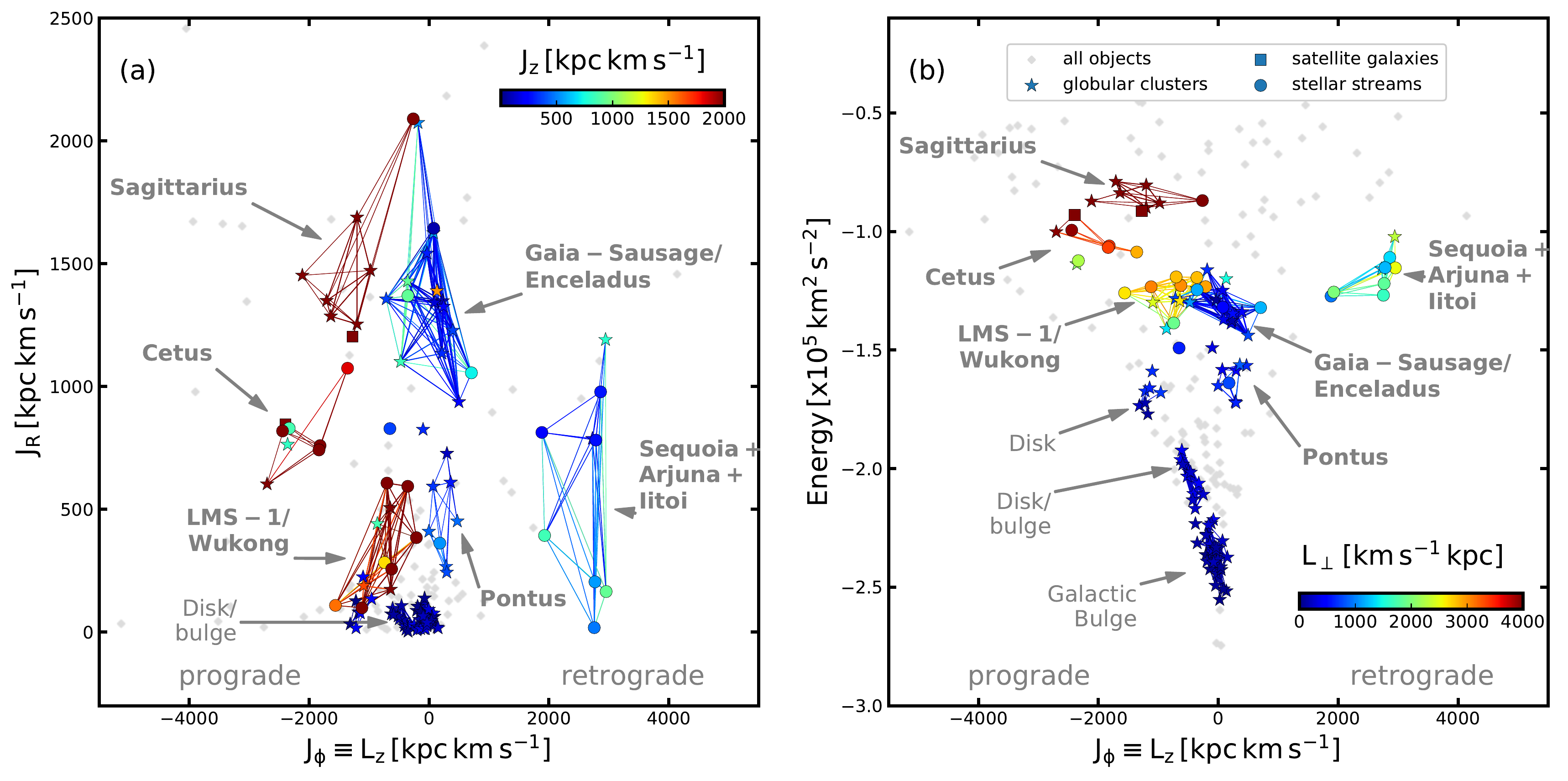}
\end{center}
\vspace{-0.5cm}
\caption{\EJ\ distribution of the groups detected in our study. The plot shows several independent groups that comprise those objects with high probabilities (i.e., $\Pg\geq\Pth$, see Section~\ref{subsec:detect_high_sig_groups}). The left panel shows $J_\phi$ vs. $J_R$ and the objects are colored by their $J_z$ values. The right panel shows $J_\phi$ vs. $E$, colored by $L_\perp$. The gray points are all the remaining objects with $\Pg<\Pth$. The straight lines between any two objects indicate the frequency of these objects being members of the same group -- thicker the line, higher is this frequency. These lines are colored using the same scheme described above. Such a representation automatically reveals several independent groups.}
\label{fig:Fig_probability_cut}
\end{figure*}

The selection $\Pg \geq \Pth$ yields $108$ objects ($42\%$ of the total $n=\nobjects$ objects), and these are shown in Figure~\ref{fig:Fig_probability_cut}. These objects include $81$ globular clusters, $25$ stellar streams and $2$ satellite galaxies. This figure also shows different objects being linked by straight-lines. This ``link'', between given two objects, represents the frequency with which these objects were classified as members of the same group (as per the procedure described in Section~\ref{subsec:app_enlink}). Thicker links imply higher frequency. In Figure~\ref{fig:Fig_probability_cut}, these links are pruned removing those cases where two objects resulted in the same group in less than (approx.) one third ($300/1000$) of the realizations. Due to this pruning, a couple of objects can be seen without any links, even though they satisfy the condition $\Pg \geq \Pth$, and it is therefore difficult to associate them with one unique group. The power of Figure~\ref{fig:Fig_probability_cut} is that -- such a representation automatically reveals the detection of several independent groups. 

Figure~\ref{fig:Fig_probability_cut} shows that we have detected $9$ high-significance groups and the properties of these groups are discussed below.

\begin{figure*}
\begin{center}
\vspace{-0.0cm}
\includegraphics[width=\hsize]{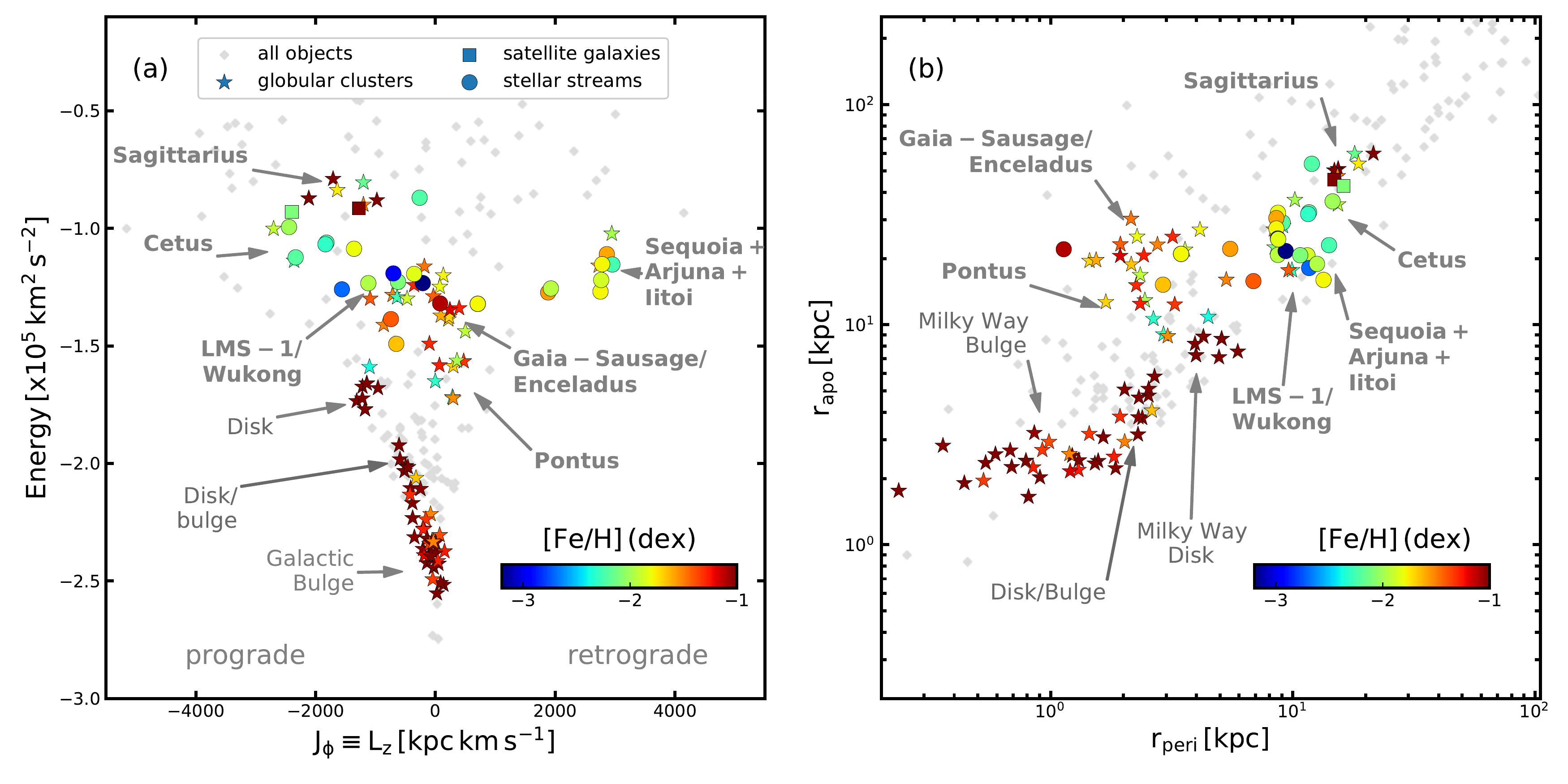}
\end{center}
\vspace{-0.5cm}
\caption{The orbital parameters of the halo objects as a function of their [Fe/H] values. We consider [Fe/H] of only those objects that possess $\Pg\geq\Pth$, and the remaining objects are shown as gray points. The left plot shows $J_\phi$ vs $E$ and the right plot shows $r_{\rm peri}$ vs $r_{\rm apo}$. The \lms1\ group has a minimum of [Fe/H]$=-3.4$~dex.}
\label{fig:Fig_orbits_FeH}
\end{figure*}
\section{Analysing the detected groups}\label{sec:analyse_mergers}

We detect a total of $9$ distinct groups at $\geq 2\sigma$ significance. Among these, we interpret $N=\nmergers$ groups as the mergers of the Milky Way because the remaining three actually contain the in-situ population of the Milky Way (see below). The merger groups comprise $\nMergerObjects$ halo objects ($\fMergerObjects$ of the total \nobjects\ objects considered in our study), including $35$ globular clusters, $25$ streams and $2$ satellite galaxies. For each of the merger groups, we analyse the objects' memberships (that is also summarised in Table~\ref{tab:merger_summary}), their \EJ\ properties (that results from Figure~\ref{fig:Fig_probability_cut}), orbital parameters as a function of [Fe/H] (see Figure~\ref{fig:Fig_orbits_FeH}), [Fe/H] distribution function (MDF, see Figure~\ref{fig:Fig_MDF}), other orbital parameters (see Figure~\ref{fig:Fig_rper_rapo}) and also estimate the masses of the corresponding progenitor galaxies. 

For each group discussed below, we also make comparison between our object membership and those proposed in previous studies. Therefore, to also facilitate this comparison visually, we provide  Figure~\ref{fig:Fig_literature} in Appendix~\ref{appendix:literature}. This figure is constructed by adopting the object-merger associations from other studies (specifically from \citealt{Massari_2019, Myeong2019, Kruijssen_2020_Kraken, Bonaca2021}). We explicitly note that our results are not based on Figure~\ref{fig:Fig_literature}, and we use it purely for comparison with our Figure~\ref{fig:Fig_probability_cut}. 

\subsection{Group~1: The \Sgr\ merger}\label{subsec:Sgr}

The first group we detect is a high-energy and prograde over-density in \EJ\ space. Its member objects possess dynamical properties in the range $E\sim[ -0.91 , -0.79 ]\times10^5\km2s2$, $J_R\sim [ 1205 , 2090 ]\kms\kpc$, $J_\phi\sim [ -2115 , -265 ]\kms\kpc$,  $J_z\sim [ 3285 , 5350 ]\kms\kpc$, $L_\perp\sim [ 4565 , 6835 ]\kms\kpc$, eccentricity$\sim [ 0.5 , 0.6 ]$, $ r_{\rm peri}\sim [ 11 , 22 ]\kpc$, $ r_{\rm apo}\sim [ 45 , 60 ]\kpc$, $ \phi\sim [ 66 \deg, 86 \deg]$; where $\phi$ defines how `polar' the merger group is. This highly polar group represents the previously known \Sgr\ merger \citep{Ibata1994, Majewski2003,Bellazzini_2020}. 

We find that $8$ objects belong to this group: $6$ globular clusters (namely, Pal~12, Whiting~1, Terzan~7, Terzan~8, Arp~2, NGC~6715/M~54), $1$ stream (namely, Elqui) and $1$ satellite (namely, the Sagittarius dSph itself). Our globular cluster member list is similar to those previously reported by other studies (e.g., \citealt{Massari_2019, Bellazzini_2020, Forbes_2020}). We note that our \Sgr\ group lacks NGC~2419 as its member, but previous studies have advocated for this association based on that fact that this cluster too lies within the phase-space distribution of the Sagittarius stream (e.g., \citealt{Sohn_2018, Bellazzini_2020}). A possible reason that our analysis does not identify a strong association between NGC~2419 and \Sgr\ could be due to this cluster's large \EJ\ uncertainties that arise due to its large observational uncertainties (since it is a very distant cluster, $D_{\odot}\approx 83\kpc$). On the other hand, our stream-\Sgr\ association is completely different than that of \cite{Bonaca2021}. \cite{Bonaca2021} found $5$ stream-\Sgr\ associations by comparing the ($J_\phi, E$) values of their stream sample to the ($J_\phi, E$) distribution of the mergers previously found by \cite{Naidu_2021}, but their stream member list does not include Elqui\footnote{Our stream sample contains all the streams that \cite{Bonaca2021} associated with \Sgr, except for ``Turranburra''.}. In fact, we find that most of their \Sgr\ stream members actually belong to the \Cetus\ group (see below). Moreover, given that the Elqui stream is produced from a low-mass dwarf galaxy \citep{Li_2021_12streams}, this further suggests that Elqui was likely the satellite dwarf galaxy of the progenitor \Sgr\ galaxy (i.e., of the Sagittarius dSph galaxy itself).   

\begin{figure}
\begin{center}
\vspace{-0.25cm}
\includegraphics[width=0.95\hsize]{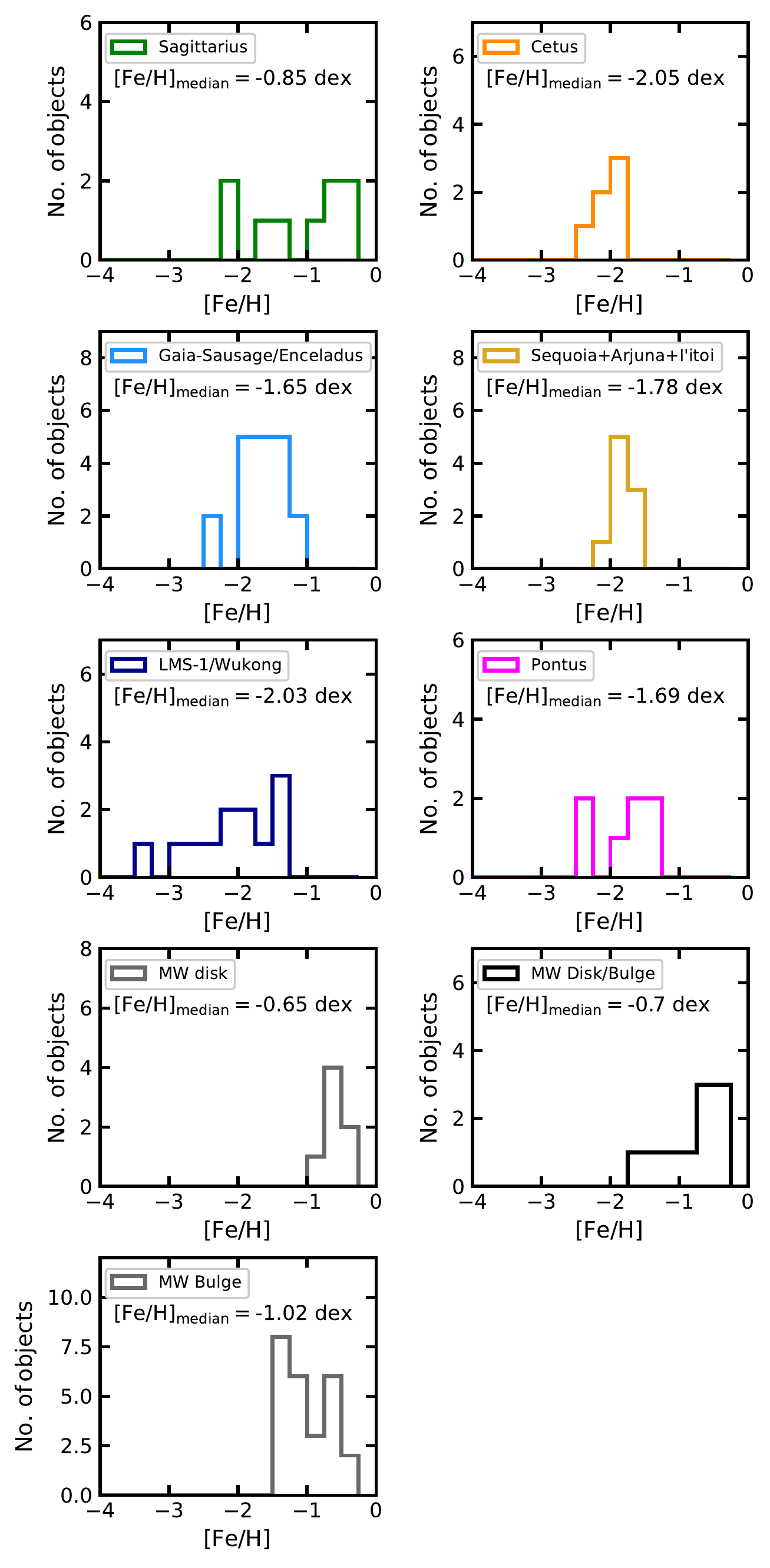}
\end{center}
\vspace{-0.67cm}
\caption{The metallicity distribution function (MDF) of different groups detected in our study. This MDF is constructed using the [Fe/H] measurements of globular clusters and streams that belong to different groups. The \lms1\ group has a minimum of [Fe/H]$=-3.4$~dex.}
\label{fig:Fig_MDF}
\end{figure}

We use the above listed member objects of \Sgr\ and analyse their [Fe/H]. The [Fe/H] measurements of streams are taken from Table~\ref{tab:table_stream_EJ} and for globular clusters we rely on the \cite{Harris2010} catalog. Figure~\ref{fig:Fig_orbits_FeH} shows the orbital properties of the objects as a function of their [Fe/H]. One can notice that the member objects of \Sgr\ possess varied metallicities, and this is consistent with  previous studies (e.g., \citealt{Massari_2019, Bellazzini_2020}). To quantify this [Fe/H] distribution, we also construct the MDF shown in Figure~\ref{fig:Fig_MDF}. This MDF has a median of [Fe/H]=$-0.85$~dex and spans a wide range from $-2.22$~dex to $-0.32$~dex. 

For the progenitor \Sgr\ galaxy, we determine its halo mass ($M_{\rm halo}$) and stellar mass ($M_{*}$) as follows. We first determine $M_{\rm halo}$ using the globular-cluster-to-halo-mass relation \citep{Hudson2014} and then convert this $M_{\rm halo}$ to $M_{*}$ using the stellar-to-halo-mass relation \citep{Read2017}. To this end, we use the masses of the individual globular clusters from \cite{Baumgardt2019}. The combined masses of the clusters provide $M_{\rm halo}\sim 4.5\times10^{10}\msun$ and this further implies $M_{*}\sim 13\times10^7\msun$. These mass values are similar to those found by previous studies (e.g., \citealt{Gibbons_2017, Niederste-Ostholt_2012}). 

Note that such a method provides a very rough estimate of the mass values and is not very accurate. This is because: (1) Both the \cite{Hudson2014} and \cite{Read2017} relations have some scatter that we do not account for. (2) Such a method makes a strong assumption that the present day observations of globular-cluster-to-halo-mass relation and stellar-to-halo-mass relation do not evolve with redshift. Since our estimates are not corrected for redshift, they provide an overestimate of the actual progenitor mass (at merging time). (3) On the other hand, such a mass estimation technique uses the knowledge of only member globular clusters and not member stellar streams (some of which could be produced from globular cluster themselves). Therefore, this may underestimate the actual progenitor mass. (4) Such a method does not account for other objects that in principle could belong to the merger groups, but were actually not identified by our study. For instance, the globular cluster AM~4 has also been previously linked to the \Sgr\ group \citep{Forbes_2020}, but we do not identify it here. In view of these limitations, we note that this method provides an approximate value on the mass of the progenitor galaxy (at the time of merging). 

\subsection{Group~2: The \Cetus\ merger}\label{subsec:Cetus}

This group is the most prograde among all the detected groups, and possesses dynamical properties in the range $E\sim[ -1.09 , -0.93 ]\times10^5\km2s2$, $J_R\sim [ 605 , 1075 ]\kms\kpc$, $J_\phi\sim [ -2700 , -1360 ]\kms\kpc$, $J_z\sim [ 1835 , 2820 ]\kms\kpc$, $L_\perp\sim [ 2905 , 4635 ]\kms\kpc$, eccentricity$\sim [ 0.4 , 0.6 ]$, $ r_{\rm peri}\sim [ 8 , 16 ]\kpc$, $ r_{\rm apo}\sim [ 31 , 43 ]\kpc$, $ \phi\sim [ 50 \deg, 65 \deg]$. It corresponds to the previously known \Cetus\ merger \citep{Newberg_2009_Cetus, Yuan_Cetus_2019}. Inspecting Figure~\ref{fig:Fig_probability_cut}, it can be seen that \Cetus\ is situated in the vicinity of the \Sgr\ group. However, these two groups overall possess quite different $J_R$ and $J_\phi$ components, different orbital properties and they can also be distinguished on the basis of their [Fe/H] properties (the \Cetus\ members are overall more metal-poor than the \Sgr\ members). 

We find that $6$ objects belong to this group: $4$ streams (namely, Cetus itself, Slidr, Atlas, AliqaUma), $1$ cluster (namely, NGC~5824) and $1$ satellite galaxy (Willman~1). Among the stream member list, AliqaUma and Atlas were recently associated with the Cetus stream by \cite{Li_2021_12streams}. On the other hand, \cite{Bonaca2021} associated most of these streams with the \Sgr\ group. \cite{Bonaca2021} found three other streams to be associated with \Cetus, but these streams are not present in our data sample\footnote{These streams are Willka Yaku, Triangulum and Turbio.}. Furthermore, we could not find the streams C~20 and Palca as members of this group, but their associations have been suggested by previous studies (e.g., \citealt{Chang_2020_Cetus,Li_2021_12streams, Yuan_2022_Cetus}). As for the globular cluster-\Cetus\ association, NGC~5824 has been previously linked with the Cetus stream by various studies on the basis that this cluster lies within the phase-space distribution of the Cetus stream (e.g., \citealt{Newberg_2009_Cetus, Yuan_Cetus_2019, Chang_2020_Cetus}). However, other studies indicate that NGC~5824 is associated with the \Sgr\ group \citep{Massari_2019, Forbes_2020}. 

Surprisingly, some of the previous studies do not mention the \Cetus\ group in their analysis. For instance, \cite{Massari_2019} made a selection in $L_z - L_\perp$ space to identify the \Sgr\ globular clusters and found that this integral-of-motion space also contains NGC~5824; so they assigned it to the \Sgr\ group. On the other hand, \cite{Forbes_2020} identify their merger groups by combining the orbit information of globular clusters from \cite{Massari_2019} and the ages and [Fe/H] from \cite{Kruijssen_2019}. Also, they guide their analysis by the previously known cluster-merger memberships from \cite{Massari_2019}. A possible reason that these studies could not identify \Cetus\ is because they were analysing only globular clusters, and the \Cetus\ group (likely) contains only one such object -- NGC~5824 itself. However, we are able to detect \Cetus\ because we have combined the globular cluster information with that of streams and satellites; and \Cetus\ group clearly contains many streams. As for the satellite-\Cetus\ association, it is for the first time that Willman~1 has been associated with this group (to the best of our knowledge). It could be that Willman~1, which is an ultra-faint dwarf galaxy \citep{Willman_2005}, is actually the remnant of the progenitor \Cetus\ galaxy (in other words, the remnant of the Cetus stream). This scenario is also supported by the fact that the [Fe/H] of Willman~1 ($\approx-2.1$~dex, \citealt{McConnachie_2020_GaiaEDR3}) is very similar to that of the Cetus stream (see Table~\ref{tab:table_stream_EJ}).

In Figure~\ref{fig:Fig_probability_cut}, one can see two additional objects that lie close to the \Cetus\ group, namely the globular cluster NGC~4590/M~68 and the stream Fj\"orm (which is the stream produced from NGC~4590/M~68). These two objects have very similar ($J_R,J_\phi,E$) values as that of the \Cetus\ group but possess lower $J_z$ values, rendering this association rather tentative. We note that NGC~4590 was previously associated with the \Helmi\ substructure by \cite{Massari_2019, Forbes_2020, Kruijssen_2020_Kraken} and with the {\it Canis Major} progenitor galaxy \citep{Martin_2004} by \cite{Kruijssen_2019}. On the other hand, Fj\"orm was previously linked with \Sgr\ by \cite{Bonaca2021}. 

From Figure~\ref{fig:Fig_orbits_FeH}, it can be seen that all the \Cetus\ member objects possess similar [Fe/H] values. We find that the MDF of this group has a median at [Fe/H]=$-2.05$~dex and spans a very narrow range from $-2.3$~dex to $-1.8$~dex. Using the mass of NGC~5824, we estimate the mass of the progenitor \Cetus\ galaxy as $M_{\rm halo}\sim 2\times10^{10}\msun$, and this in turn provides $M_{*}\sim 3\times10^7\msun$. Note that these mass values likely represent a severe underestimation of the true (infall) mass of the progenitor \Cetus\ galaxy, because this group contains many streams whose masses we have not accounted for (because we do not possess that information).

\begin{figure*}
\begin{center}
\vspace{-0.2cm}
\includegraphics[width=\hsize]{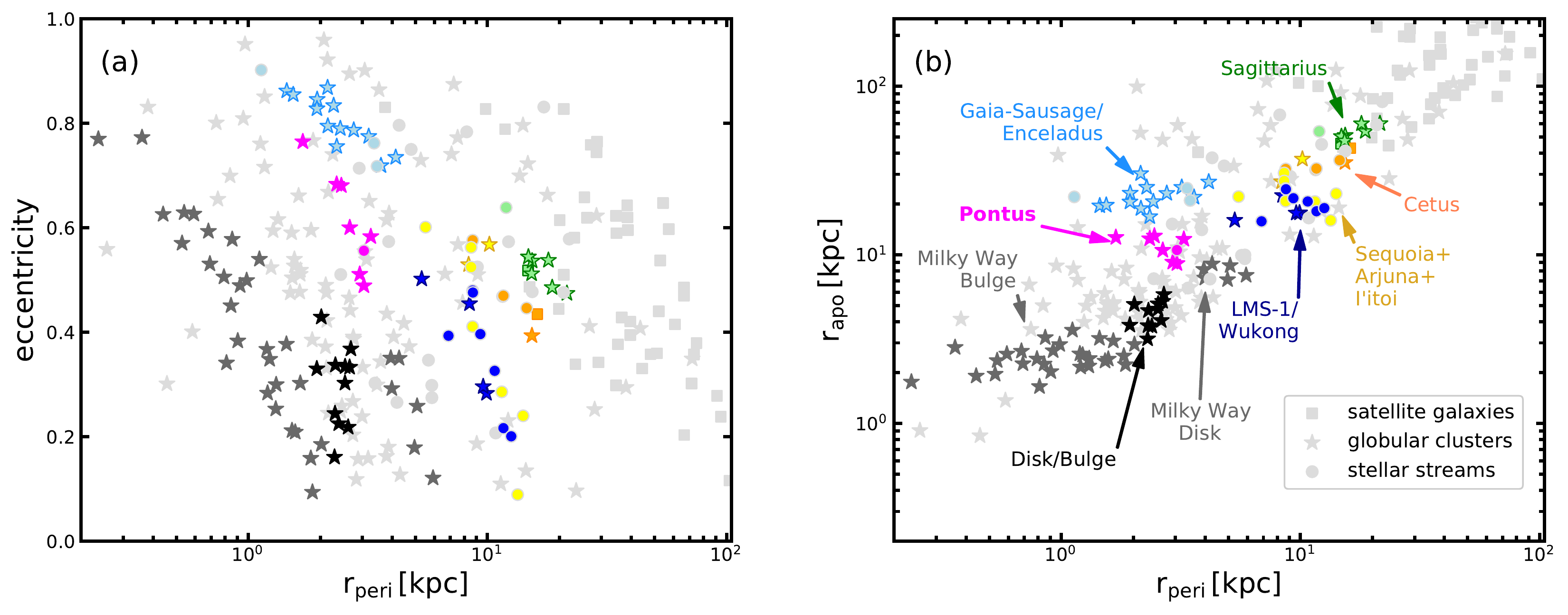}
\end{center}
\vspace{-0.4cm}
\caption{Comparing the orbital properties of those objects that belong to different groups. The left panel shows $r_{\rm peri}$ vs. eccentricity and the right panel shows $r_{\rm peri}$ vs. $r_{\rm apo}$. We use same color for all those objects that belong to the same group. The gray points represent all the objects in our sample. The objects that form part of the same group are clumped in this space (in addition to being clumped in the \EJ\ space, see Figure~\ref{fig:Fig_probability_cut}).}
\label{fig:Fig_rper_rapo}
\end{figure*}
\subsection{Group~3: The \GSE\ merger}\label{subsec:GSE}

This group represents the largest of all the mergers that we detect here. The member objects of this group possess relatively low values in $|J_\phi|$ and $L_\perp$, implying that they lie along radial orbits. This group possesses dynamical properties in the range $E\sim[ -1.44 , -1.16 ]\times10^5\km2s2$, $J_R\sim [ 935 , 2075 ]\kms\kpc$, $J_\phi\sim [ -715 , 705 ]\kms\kpc$, $J_z\sim [ 85 , 1505 ]\kms\kpc$, $L_\perp\sim [ 85 , 1520 ]\kms\kpc$, eccentricity$\sim [ 0.7 , 0.9 ]$, $ r_{\rm peri}\sim [ 1 , 4 ]\kpc$, $ r_{\rm apo}\sim [ 16 , 30 ]\kpc$, $ \phi\sim [ 27 \deg, 85 \deg]$. This group represents the \GSE\ merger \citep{Belokurov2018, Helmi2018, Myeong_2018_SausageGCs, Massari_2019}. 

We find that $16$ objects belong to this group: $3$ streams (namely, C-7, M~5, Hr\`id) and $13$ globular clusters (namely, NGC~7492, NGC~6229, NGC~6584, NGC~5634, NGC~5904/M~5, NGC~2298, NGC~4147, NGC~1261, NGC~6981/M~72, NGC~7089/M~2, IC~1257, NGC~1904/M~79, NGC~1851). There exist two additional objects close to this group, namely the globular cluster NGC~6864/M~75 and the stream NGC~7089, but their association is not very strong (because of their slightly lower $J_R$ values). 

\begin{figure*}
\begin{center}
\hbox{
\includegraphics[width=0.5\hsize]{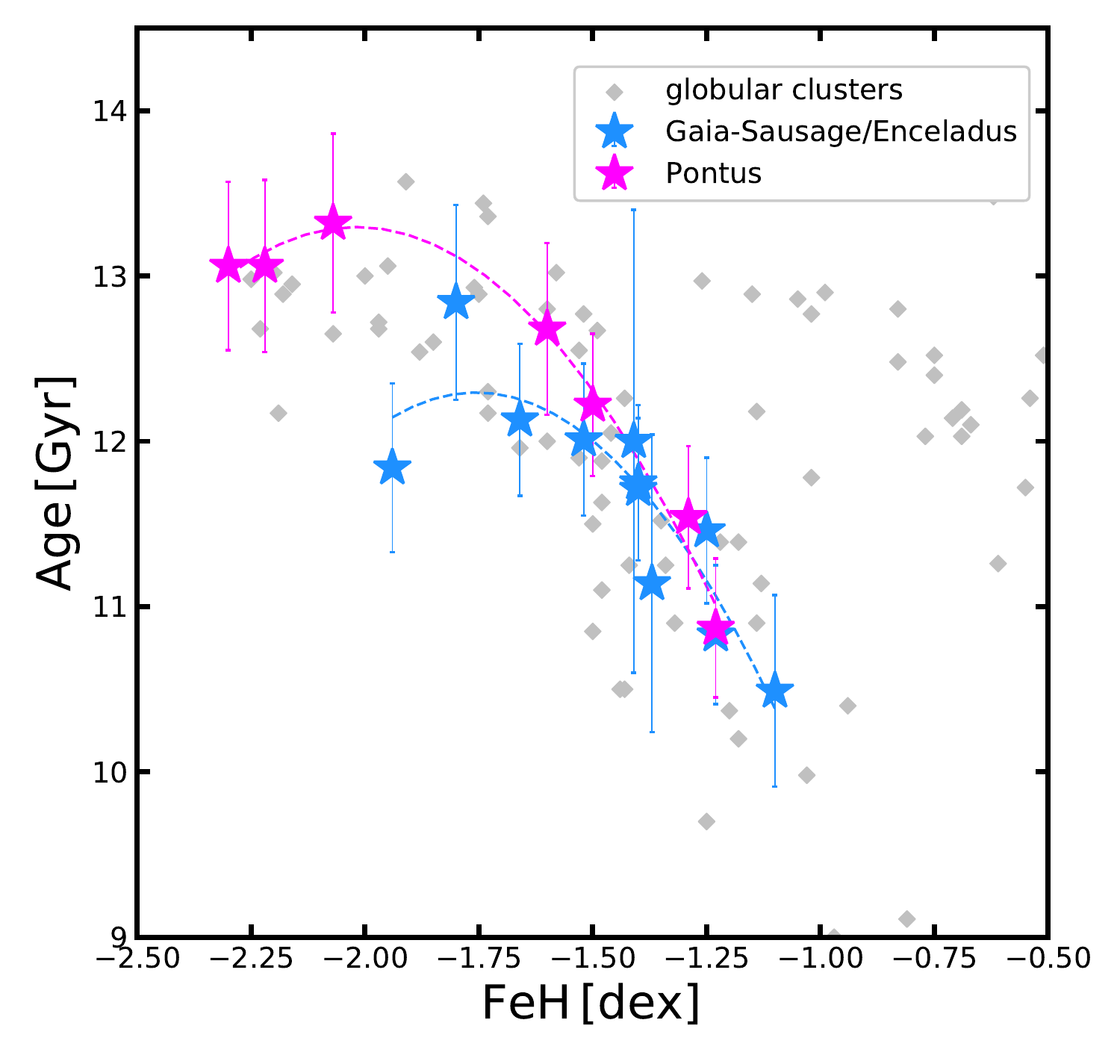}
\includegraphics[width=0.5\hsize]{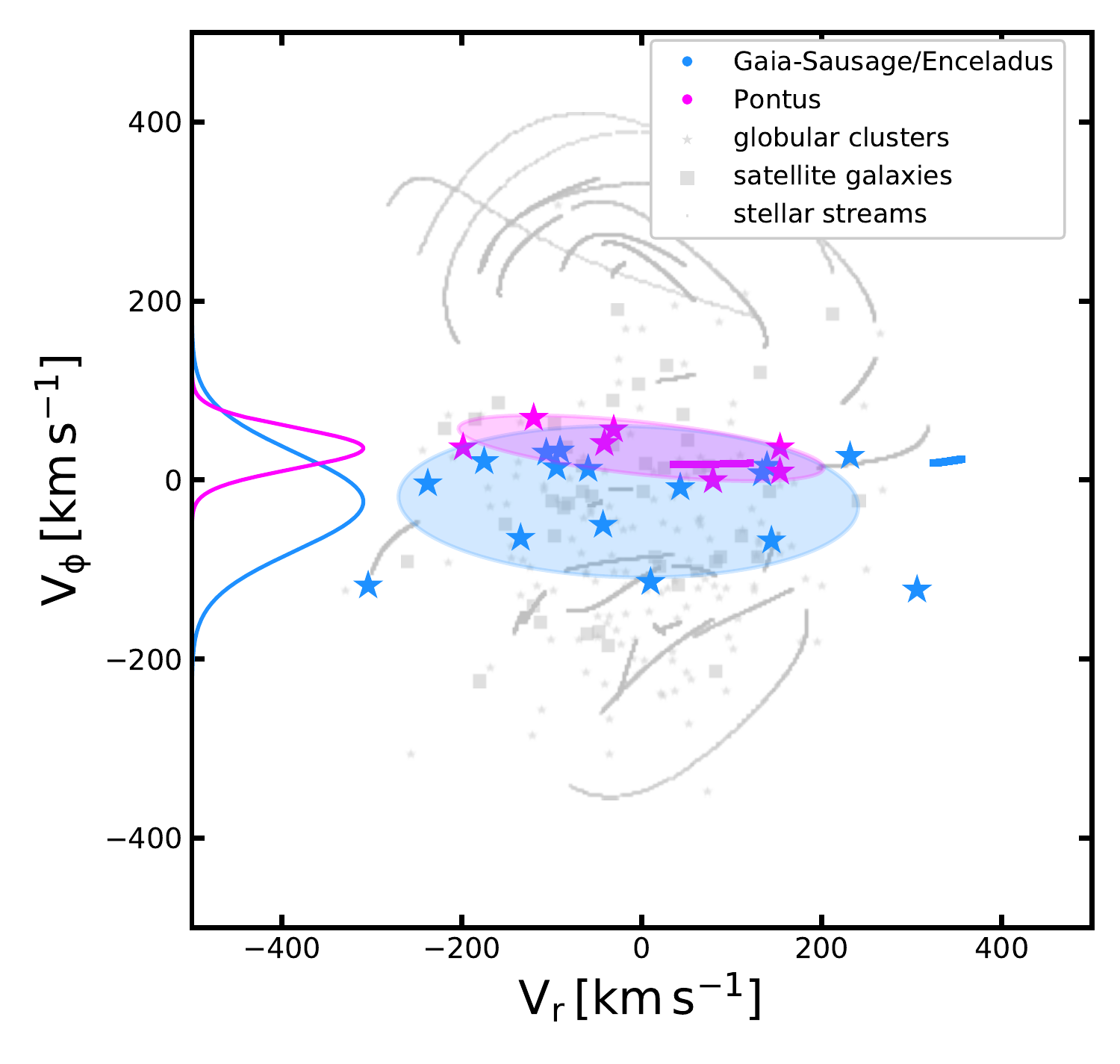}
}
\end{center}
\vspace{-1.0cm}
\caption{Comparing properties of those objects that belong to the groups \GSE\ and \Pontus. The left panel shows the age-metallicity relationship of the globular clusters belonging to these groups; the [Fe/H] and age values are taken from \cite{Kruijssen_2019}. The right panel shows the velocity behaviour of these objects in spherical polar coordinates, namely radial $V_r$ and azimuthal $V_\theta$. The filled ellipses represent $1.5\sigma$ confidence contour and the Gaussians represent the mean and the standard deviation in the $V_\phi$ components of the member objects of these groups.}
\label{fig:Fig_AMR_vel}
\end{figure*}

These streams-\GSE\ associations are reported here for the first time. Unlike \cite{Bonaca2021}, we do not find streams Ophiuchus and Fimbulthul as members of this group. As for the globular cluster-- \GSE\ associations, our list contains half of those $10$ clusters that were previously associated with this group by \cite{Myeong_2018_SausageGCs}. However, more recent studies have attributed a large number of globular clusters to the \GSE\ merger. For instance, \cite{Massari_2019} associated $\approx 32$ globular clusters to this merger; although some of their associations were tentative. They found these association by making hard cuts in the ($J_\phi,E,L_\perp$) space that were previously used by \cite{Helmi2018} to select the \GSE\ stellar debris. \cite{Massari_2019} further supported their associations by arguing that the resulting globular clusters show a tight age–metallicity relation (AMR). We show the AMR of our \GSE\ globular clusters in Figure~\ref{fig:Fig_AMR_vel} that (visually) appears to be tighter than Figure~4 of \cite{Massari_2019}. The study of \cite{Forbes_2020}, that is based on the analysis of \cite{Massari_2019}, found $28$ globular cluster-\GSE\ associations. We find that some of these additional globular clusters, that have recently been linked with \GSE\ by other studies, likely belong to a different merger group (see Section~\ref{subsec:Pontus}). 

The halo objects associated with the \GSE\ merger span a very wide range in [Fe/H] from $-2.4$~dex to $-1.1$~dex, with the median of the MDF located at [Fe/H]=$-1.6$. This large spread in MDF supports the scenario that \GSE\ was a massive galaxy. We estimate the mass of the progenitor galaxy as $M_{\rm halo}\sim 10\times10^{10}\msun$ and $M_{*}\sim 50\times10^7\msun$. This mass estimate is consistent with those found by previous studies from chemical evolution models (e.g., \citealt{Helmi2018, Fernndez-Alvar_2018}), counts of metal-poor and highly eccentric stars (e.g., \citealt{Mackereth_2020}) and the mass-metallicity relation (e.g., \citealt{Naidu2020}).

\subsection{Group~4: The \asi\ merger}\label{subsec:asi}

This group is highly-retrograde and its member objects possess dynamical properties in the range $E\sim[ -1.27 , -1.02 ]\times10^5\km2s2$, $J_R\sim [ 20 , 1190 ]\kms\kpc$, $J_\phi\sim [ 1880 , 2955 ]\kms\kpc$, $J_z\sim [ 230 , 940 ]\kms\kpc$, $L_\perp\sim [ 980 , 2530 ]\kms\kpc$, eccentricity$\sim [ 0.1 , 0.6 ]$, $ r_{\rm peri}\sim [ 5 , 14 ]\kpc$, $ r_{\rm apo}\sim [ 15 , 37 ]\kpc$, $ \phi\sim [ 25 \deg, 46 \deg]$. We refer to this group as the \asi\ group, because it is likely comprised of objects that actually resulted from $3$ independent mergers: Sequoia \citep{Myeong2019}, Arjuna and I'itoi \citep{Naidu2020}. This understanding comes from \cite{Naidu2020}, who performed chemo-dynamical analysis of stars and proposed that at this $(E,J_\phi)$ location, there exist three different (but somewhat overlapping) stellar populations: a metal-rich population whose MDF peaks at [Fe/H]$\approx -1.2$~dex (namely Arjuna), another one whose MDF peaks at [Fe/H]$\approx -1.6$~dex (namely Sequoia) and the most metal poor among these whose MDF peaks at [Fe/H]$\approx -2$~dex (namely I'itoi). Here, we analyse this detected group as a single merger, because our detected grouping contains only a handful of objects and therefore it is difficult to detect any plausible sub-groups within this \asi\ group. 

We find that $9$ objects belong to this group: $7$ streams (namely, Phlegethon, Gaia-9, NGC~3201, Gj\"oll, GD-1, Kshir, Ylgr) and $2$ globular clusters (namely, NGC~6101, NGC~3201). These two globular clusters were previously associated with Sequoia by \cite{Myeong2019}, although they associated $5$ additional clusters to this group that we do not identify here. To discover the Sequoia group, \cite{Myeong2019} applied a `Friends-of-Friends' grouping algorithm to the projected action space containing only globular clusters (essentially, they applied their algorithm to the \Gaia\ DR2 version of the top-left panel shown in Figure~\ref{fig:Fig_EJ_of_objects}). With this, they found a group of globular clusters (that they named Sequoia) whose combined dynamical properties ranged from $E=[-2.2, -0.97]\times10^5\km2s2$, $J_R\sim [ 54 , 1400 ]\kms\kpc$, $J_\phi=[250,3210]\kms\kpc$ and $J_z=[66,800]\kms\kpc$ (they used the same Galactic potential model as ours). This dynamical range is larger than the range we infer for the \asi\ group, especially in $J_\phi$ and $E$. Moreover, it could be due to their wide $E$ selection that even the low-energy cluster NGC~6401 ends up in their Sequoia group; we note that NGC~6401 possesses such a low energy \citep{Massari_2019} that it likely belongs to the Galactic bulge (see below). On the other hand, our two \asi\ globular clusters were previously associated with both Sequoia and \GSE\ by \cite{Massari_2019}; we note that their ($J_\phi,E$) selection is motivated by the results of \cite{Myeong2019}. But \cite{Massari_2019} also found $5$ additional member clusters for Sequoia and many of these are not present in \cite{Myeong2019} selection. The Sequoia group found by \cite{Forbes_2020} is very similar to that of \cite{Massari_2019}, likely because the selection of the former study is based on the latter. As for the stream-\asi\ associations, \cite{Myeong2019} analysed \EJ\ of only GD-1 and argued against its association. However, \cite{Bonaca2021} favoured an association of GD-1 with \asi, along with those of Phlegethon, Gj\"oll and Ylgr.

The MDF of this group spans a wide range from $-2.24$~dex to $-1.56$~dex, with the median of [Fe/H]$\sim-1.78$~dex. Interestingly, this [Fe/H] median is similar to the [Fe/H] of the Kshir stream \citep{Malhan_Kshir2019}. Kshir is broad stream that moves in the Milky Way along very similar orbit as that of GD-1, and this observation encouraged \cite{Malhan_Kshir2019} to propose that Kshir is likely the stellar stream produced from the tidal stripping of the merging galaxy that brought in GD-1. If true, this indicates that Kshir is likely the stream of the progenitor \asi\ galaxy; perhaps that of the Sequoia galaxy (given the similarity in their [Fe/H]). 

Using only the member globular clusters of \asi, and not the streams, we estimate the mass of the progenitor galaxy as $M_{\rm halo}\sim 1.2\times10^{10}\msun$ and $M_{*}\sim 1.5\times10^7\msun$. Note that the actual masses should be higher than these computed masses, since this group contains several streams (as compared to globular clusters) whose masses we could not account for. Interestingly, these mass values are similar to those derived by \cite{Myeong2019} for the Sequoia merger, using similar techniques; even though we could not identify many of their globular clusters as members of our \asi\ group.

\subsection{Group~5: The \lms1\ merger}\label{subsec:lms1}

This group has a slight prograde motion and its member objects are very tightly clumped in \EJ\ space. It possesses dynamical properties in the range $E\sim[ -1.41 , -1.19 ]\times10^5\km2s2$, $J_R\sim [ 100 , 605 ]\kms\kpc$, $J_\phi\sim [ -1560 , -210 ]\kms\kpc$, $J_z\sim [ 875 , 2710 ]\kms\kpc$, $L_\perp\sim [ 1400 , 3085 ]\kms\kpc$, eccentricity$\sim [ 0.2 , 0.5 ]$, $ r_{\rm peri}\sim [ 5 , 13 ]\kpc$, $ r_{\rm apo}\sim [ 15 , 25 ]\kpc$, $ \phi\sim [ 58 \deg, 85 \deg]$. This polar group corresponds to the {\it Low-mass-stream-1 (LMS-1)/Wukong} merger \citep{Yuan2020, Naidu2020, Malhan_2021_LMS1}.

We find that $11$ objects belong to this group: $7$ streams (LMS-1 itself, Phoenix, Pal~5, C-19, Indus, Sylgr, Jhelum) and $4$ globular clusters (namely NGC~5272/M~3, NGC~5053, Pal~5, NGC~5024/M~53). In regard to the stream-\lms1\ associations, Phoenix, Indus, Jhelum and Sylgr were tentatively associated with this merger by \cite{Bonaca2021}. The association of Indus was also favoured by \cite{Malhan_2021_LMS1}, however, this study had argued against Jhelum's association. Our result here could be different from \cite{Malhan_2021_LMS1} because here we are using different data for streams; that consequentially results in different \EJ\ solutions. Furthermore, since Indus and Jhelum are tidal debris of dwarf galaxies \citep{Li_2021_12streams}, this indicates that they were likely the satellite dwarf galaxies of the progenitor \lms1\ galaxy (also see \citealt{Malhan_2021_LMS1}). As for the globular cluster-\lms1\ associations, \cite{Koppelman_2019_Helmi, Massari_2019} associated NGC~5024, NGC~5053 and NGC~5272 with the \Helmi\ substructure \citep{Helmi_1999}. \cite{Koppelman_2019_Helmi}, in particular, supported the association of four additional clusters with the \Helmi\ substructure, but we find many of these clusters as part of the \GSE\ group. As for \cite{Massari_2019}, they could only associate the clusters with those merger groups that were known at that time, but \lms1\ was detected after their study by \cite{Yuan2020, Naidu2020}. On the other hand, recent studies have shown that there indeed exists a strong association of NGC~5024 and NGC~5053 with the \lms1\ group \citep{Yuan2020, Naidu2020, Malhan_2021_LMS1}, based on the fact that these clusters lie within the phase-space distribution of the LMS-1 stream (e.g., \citealt{Yuan2020, Malhan_2021_LMS1}). Another recent study by \cite{Wan_2020_Phoenix} advocates for a dynamical connection between Phoenix, Pal~5 and NGC~5053. In summary, our analysis supports these recent studies and makes a stronger case that all of these objects are associated with the \lms1\ merger. 

We find that \lms1\ is the most metal-poor merger of the Milky Way, because this group contains the three most metal-poor streams of our Galaxy, namely C-19, Sylgr and Phoenix (see their [Fe/H] values in Table~\ref{tab:table_stream_EJ}). Overall, this group has a wide MDF ranging from $-3.38$~dex to $-1.41$~dex with the median of [Fe/H]$\sim-2$~dex. We note that this median is similar to the metallicity of the LMS-1 stream \citep{Malhan_2021_LMS1}. Using the masses of the globular clusters, we estimate the progenitor galaxy's mass as $M_{\rm halo}\sim 2.7\times10^{10}\msun$ and $M_{*}\sim 5.5\times10^7\msun$.  These mass values are higher than those reported in \cite{Malhan_2021_LMS1} because here we find higher number of globular cluster-\lms1 associations. 

\subsection{Group~6: Discovery of the \Pontus\ merger}\label{subsec:Pontus}

We detect a new group that possesses low energy and is slightly retrograde. Its dynamical properties are in the range $E\sim[ -1.72 , -1.56 ]\times10^5\km2s2$, $J_R\sim [ 245 , 725 ]\kms\kpc$, $J_\phi\sim [ -5 , 470 ]\kms\kpc$, $J_z\sim [ 115 , 545 ]\kms\kpc$, $L_\perp\sim [ 390 , 865 ]\kms\kpc$, eccentricity$\sim [ 0.5 , 0.8 ]$, $ r_{\rm peri}\sim [ 1 , 3 ]\kpc$, $ r_{\rm apo}\sim [ 8 , 13 ]\kpc$, $ \phi\sim [ 54 \deg, 89 \deg]$. We refer to this group as \Pontus\footnote{In Greek mythology, ``\Pontus'' (meaning ``the Sea'') is the name of one of the first children of the Gaia deity.}. We find that $8$ objects belong in this group: $1$ stream (namely, M~92) and and $7$ clusters (namely, NGC~288, NGC~5286, NGC~7099/M~30, NGC~6205/M~13, NGC~6341/M~92, NGC~6779/M~56, NGC~362). There exist two additional objects close to this group, namely the globular cluster NGC~6864/M~75 and the stream NGC~7089, but their association was not very strong (because of their slightly higher $J_R$ and slightly lower $J_\phi$ values). 

\Pontus\ lies close to \GSE\ in ($J_\phi,E$) space (although the two groups possess very different $J_R$ values) and essentially all the \Pontus's globular clusters (that we mentioned above) have been previously associated with \GSE\ \citep{Massari_2019}. Given this potential overlapping between the two groups, it is natural to ask: do these groups in fact represent different merging events or is it that our procedure has fragmented the large \GSE\ group into two pieces? We argue that fragmentation can not be the reason, otherwise the neighbouring \Sgr\ and \Cetus\ groups should also be regarded as a single group; as these latter groups are much closer to each other in \EJ\ space compared to the former groups. Similarly, even the neighbouring \lms1\ and \GSE\ groups could be distinguished by our detection procedure. To understand the nature of \Pontus\ and \GSE\ groups, we use their member objects and analyse their dynamical properties and the age-metallicity relationship. 

First, we find that the objects belonging to these two groups possess different dynamical properties. The average eccentricity of \Pontus\ objects is smaller than that of the \GSE\ objects (see Figure~\ref{fig:Fig_rper_rapo}); while we note that the eccentricity range of our \GSE\ group is similar to that of \cite{Myeong_2018_SausageGCs}. This implies that the orbits of \GSE\ objects are more radial than those of the \Pontus\ objects (this can also be discerned by comparing their $J_R$ values). Also, the average $r_{\rm apo}$ of \Pontus\ objects is smaller than that of the \GSE\ objects. Furthermore, we also compare the velocity behaviour of their member objects in spherical polar coordinates, namely radial $V_r$ and azimuthal $V_\phi$ (see Figure~\ref{fig:Fig_AMR_vel}). The motivation for adopting this particular coordinate system comes from \cite{Belokurov2018}, who used a similar system to originally identify the ``sausage'' structure. From Figure~\ref{fig:Fig_AMR_vel}, we note that both the \Pontus\ and \GSE\ distributions are stretched along the $V_r$ direction (implying radial orbits), although their $V_\phi$ components (on average) differ by $\approx 60\kms$. This implies, as noted above, that \Pontus\ objects are more retrograde than the \GSE\ objects. Also, \GSE\ objects possess larger dispersion in $V_\phi$ as compared to \Pontus\ objects. 

Moreover, in Figure~\ref{fig:Fig_AMR_vel}, the AMR for the globular clusters belonging to these two groups also appear quite different; especially the age difference of their metal-poor clusters ($\simlt-1.5$~dex) is $\simgt 1\Gyr$. In view of this investigation, we conclude that \Pontus\ and \GSE\ represent two distinct and independent merging events: \GSE\ comprising of slightly younger globular clusters than those present in \Pontus\, and \GSE's objects possessing overall different dynamical properties compared to \Pontus's objects.

Similarly, we argue that the \Pontus\ group is also different than the {\it Thamnos} substructures identified by \cite{Koppelman2019}. \cite{Koppelman2019} suggested that at the location $(E,J_\phi) \sim (-1.65\times10^5\km2s2, 1500\kms\kpc)$ and $\sim  (-1.75\times10^5\km2s2, 900\kms\kpc)$, there lies two substructures, namely Thamnos~1 and Thamnos~2 (see their Figure~2). Motivated by their selection, \cite{Naidu2020} selected Thamnos stars around a small region of $(E,J_\phi) \sim (-1.75\times10^5\km2s2, 500\kms\kpc)$ (see their Figure~23). Given that these $(E,J_\phi)$ locations for {\it Thamnos} are different than that of \Pontus, we argue that \Pontus\ is independent of {\it Thamnos}. Moreover, the metallicity of Thamnos~2 members is different from that of \Pontus\ members. This we argue by inspecting Figure~2 of \cite{Koppelman2019} that shows that the metallicity of Thamnos~2 stars range from [Fe/H]$\sim-1.4$~dex to $-1.1$~dex, and this is different from the metallicity range of \Pontus\ (see below). We also note that a few of the \Pontus\ member clusters were previously tentatively associated with the {\it Canis Major} progenitor galaxy \citep{Kruijssen_2019}.

The MDF of \Pontus\ spans a range from $-2.3$~dex to $-1.3$~dex with a median of [Fe/H]$=-1.7$~dex. We estimate the mass of the progenitor \Pontus\ galaxy as $M_{\rm halo}\sim 5\times10^{10}\msun$ and $M_{*}\sim 15\times10^7\msun$.

\begin{table*}
\scriptsize
\centering
\caption{Various groups detected in our study along with their member globular clusters, stellar streams and satellite galaxies. This table is based on the associations that are visible in Figure~\ref{fig:Fig_probability_cut}. The detailed properties of these groups are described in Section~\ref{sec:analyse_mergers}.}
\label{tab:merger_summary}
\begin{tabular}{|l|c|l|l|l|}
\hline
\hline
Merger/        & No. of & Member            & Member  & Member\\
in-situ group  & members   & globular clusters & stellar streams & satellite galaxies \\
\hline
\hline

\Sgr\                      &    8   & Pal~12, Whiting~1, Terzan~7, Terzan~8,   & Elqui                        & Sagittarius dSph\\
(Section~\ref{subsec:Sgr}) &        & NGC~6715/M~54, Arp~2                     &                              &                 \\
                           &        &                                          &                              &                  \\

\hline
\Cetus\                      &  6-8  & NGC~5824                              & Cetus [stream of \Cetus],        & Willman~1\\
(Section~\ref{subsec:Cetus}) &       &                                       & Slidr, Atlas, AliqaUma           &          \\
                             &       &                                       &                                  &          \\
                             &       & tentative:                            &                                  &          \\
                             &       & NGC~4590/M~68 [stream:Fj\"orm]        & Fj\"orm [stream of NGC~4590/M~68]&          \\
                             &       &                                       &                                  &\\

\hline
\GSE\                        & 16-18  & NGC~7492, NGC~6229, NGC~6584,         & C-7, Hr\`id,                    &\\
(Section~\ref{subsec:GSE})   &        & NGC~5634, IC~1257,  NGC~1851,         & M~5 [stream of NGC~5904/M~5]    &\\
                             &        & NGC~2298, NGC~4147, NGC~1261,         &                                 &\\
                             &        & NGC~6981/M~72, NGC~1904/M~79,         &                                 &\\
                             &        & NGC~7089/M~2 [stream:NGC~7089],       &                                 &\\
                             &        & NGC~5904/M~5 [stream: M~5]            &                                  &\\
                             &       &                                        &                                  & \\
                             &       & tentative:                             &                                  & \\
                             &       & NGC~6864/M~75                          &    NGC~7089 [stream of NGC~7089/M~2] & \\
                             &       &                                        &                                   &\\

\hline
\asi\                        &   9   & NGC~6101,                               & GD-1, Phlegethon, Gaia-9, Kshir,   & \\
(Section~\ref{subsec:asi})   &       & NGC~3201 [streams: NGC~3201,Gj\"oll]    & NGC~3201 [stream of NGC~3201],     & \\
                             &       &                                         & Gj\"oll [stream of NGC~3201], Ylgr &\\
                             &        &                                        &                                     &\\

\hline
\lms1\                       &  11   & NGC~5272/M~3, NGC~5053,                 & LMS-1 [stream of \lms1],         &\\
(Section~\ref{subsec:lms1})  &       & NGC~5024/M~53,                          & C-19, Sylgr, Phoenix, Indus,     &\\
                             &       & Pal~5 [stream: Pal~5]                   &  Jhelum, Pal~5 [stream of Pal~5] &\\
                             &       &                                         &                                  &\\

\hline
\Pontus\                      & 8-10 & NGC~288, NGC~5286, NGC~7099/M~30        & M~92 [stream of NGC~6341/M~92] &\\
(Section~\ref{subsec:Pontus}) &      & NGC~6205/M~13, NGC~6779/M~56,           & &\\
                              &      & NGC~6341/M~92 [stream:M~92], NGC~362    & &\\
                              &      &                                         & &          \\
                              &      & tentative:                              &    &          \\
                              &      & NGC~6864/M~75                           &    NGC~7089 [stream of NGC~7089/M~2]   &          \\
                              &      &                                           &&\\

\hline
{\it Candidate merger} & 5    &  NGC~5466 [stream: NGC~5466],        & Gaia-10,                           & Tucana~III\\
(not detected, but     &      &   NGC~7492                           & NGC~5466 [stream of NGC~5466]      & \\
 selected, Section~\ref{sec:merger_candidate})    &      &                                      &                                    &\\

\hline
Galactic disk                   & 6-7 & Pal~10, NGC~6838/M~71, NGC~6356,        &&\\
(Section~\ref{subsec:insitu}) &     & IC~1276/Pal~7, Pal~11, NGC~104/47Tuc     &&\\
                                &     & tentative:                             &                             &          \\
                                &     & NGC~7078/M~15                           &&\\
                                &     &                                           &&\\

\hline
Galactic Bulge                 & 28 & Terzan~2/HP~3, 1636-283/ESO452, Gran~1,          &&\\
(Section~\ref{subsec:insitu})  &    & Djorg~2/ESO~456, NGC~6453, NGC~6401,      &&\\
                               &    & NGC~6304,  NGC~6256, NGC~6325, Pal~6,             &&\\
                               &    & Terzan~6/HP~5, Terzan~1/HP~2, NGC~6528,               &&\\
                               &    & NGC~6522, NGC~6626/M~28, Terzan~9,        &&\\
                               &    & Terzan~5~11, NGC~6355, NGC~6638 ,       &&\\ 
                               &    & NGC~6624, NGC~6266/M~62, NGC~6642,          &&\\
                               &    & NGC~6380/Ton1, NGC~6717/Pal9, NGC~6558,            &&\\
                               &    & NGC~6342, HP~1/BH~229, NGC~6637/M~69  &&\\ 
                               &    &                                           &&\\

\hline
Galactic Bulge/               & 11  & Terzan~3, NGC~6569, NGC~6366, NGC~6139,    &&\\ 
disk/ low-energy              &     & BH~261/AL~3, NGC~6171/M~107, Pal~8,        &&\\
(Section~\ref{subsec:insitu}) &     & Lynga~7/BH~184, NGC~6316, FSR~1716,  &&\\
                              &     & NGC~6441                                    &&\\
                              &     &                                           &&\\

\hline
\hline
\end{tabular}
\tablecomments{From left to right the columns provide name of the group, total number of halo objects that are members of this group, names of the member globular clusters, names of the member streams, names of the member satellite galaxies. In case of globular clusters, we provide in brackets the names of their streams (if any present in our sample). Likewise, in case of stellar streams, we provide in brackets the names of their parent globular clusters (in case the parent globular cluster of the stream is known).}
\end{table*}
\subsection{The in-situ Groups~7, 8 and 9}\label{subsec:insitu}

We detect three additional groups, however, their locations in the ($E,J_\phi$) space indicates that they do not represent any merger, but actually belong to the in-situ population of the Milky Way -- the population of the Galactic disk and the Galactic bulge. This we infer based on the fact that the member objects of these groups possess low $E$, low $r_{\rm apo}$ and high [Fe/H] -- as expected from the in-situ globular cluster population (see Figure~\ref{fig:Fig_orbits_FeH}). 

The first of these groups possess dynamical properties in the range $E\sim[ -1.77 , -1.66 ]\times10^5\km2s2$, $J_R\sim [ 15 , 135 ]\kms\kpc$, $J_\phi\sim [ -1315 , -960 ]\kms\kpc$,  $J_z\sim [ 5 , 235 ]\kms\kpc$, $L_\perp\sim [ 115 , 730 ]\kms\kpc$, eccentricity$\sim [ 0.1 , 0.4 ]$, $ r_{\rm peri}\sim [ 3 , 6 ]\kpc$, $ r_{\rm apo}\sim [ 7 , 9 ]\kpc$, $ \phi\sim [ 5 \deg, 36 \deg]$. Given these dynamical properties, especially low eccentricity (implying circular orbits), low $\phi$ value (implying that the objects orbit close to the Galactic plane) and the values of $r_{\rm peri}$ and $ r_{\rm apo}$ being similar to stars in the Galactic disk, we interpret this as the disk group. This group contains $6$ globular clusters (their names are provided in Table~\ref{tab:merger_summary}). There exists one additional cluster, NGC~7078/M~15, that lies close to this group in \EJ\ space but we do not identify it as a strong associate. The member globular clusters are metal rich and the corresponding MDF ranges from $-0.8$~dex to $-0.1$~dex with a median of [Fe/H]$\sim-0.65$~dex (see Figure~\ref{fig:Fig_MDF}). It is interesting to note that this MDF minimum is consistent with the results of \cite{Zinn_1985}, who found [Fe/H]$\geq-0.8$ as the threshold between disk and halo clusters (they inferred this simply on the basis of the bi-modality of the [Fe/H] distribution of the globular clusters). All of our globular clusters, including the very metal-poor NGC~7078/M~15, were previously associated with the disk by \cite{Massari_2019}; although they associated a total of $26$ clusters to the disk but we could not identify all of these objects. 

The second group possesses the lowest energy among all the detected groups, implying that its member objects orbit deep in the potential of the Milky Way -- close to the Galactic centre. The members of this group possess the dynamical properties in the range $E\sim[ -2.55 , -2.22 ]\times10^5\km2s2$, $J_R\sim [ 5 , 140 ]\kms\kpc$, $J_\phi\sim [ -380 , 150 ]\kms\kpc$, $J_z\sim [ 0 , 275 ]\kms\kpc$, $L_\perp\sim [ 10 , 405 ]\kms\kpc$, eccentricity$\sim [ 0.1 , 0.8 ]$, $ r_{\rm peri}\sim [ 0 , 2 ]\kpc$, $ r_{\rm apo}\sim [ 1 , 4 ]\kpc$, $ \phi\sim [ 1 \deg, 89 \deg]$. This group is comprised of $28$ globular clusters (their names are provided in Table~\ref{tab:merger_summary}). Given these dynamical properties, especially very low values of $E$, $r_{\rm peri}$ and $r_{\rm apo}$ and that the objects are spherically distributed (as we note from the range of $\phi$ parameter), we interpret this as the Galactic bulge group. The member objects span a wide range in [Fe/H], ranging from $-1.5$~dex to $-0.1$~dex with a median at [Fe/H]$\sim -1.0$~dex. We confirm that several of these objects have been associated with the Galactic bulge by \cite{Massari_2019}; although they associated a total of $36$ globular clusters to the bulge. A few of our member objects were interpreted by \cite{Massari_2019} as the ``unassociated objects with low-Energy'', but \cite{Forbes_2020} interpreted these objects as those accreted inside the {\it Koala} progenitor galaxy. We further note that some of our clusters have also been interpreted as the bulge objects by \cite{Horta_2020} on the basis of their high alpha-element abundances and high [Fe/H] values. 

We detect a third group that is slightly prograde and possesses $E$ values between that of the bulge and disk groups (see Figure~\ref{fig:Fig_probability_cut}). Its dynamical properties lie in the range $E\sim[ -2.17 , -1.92 ]\times10^5\km2s2$, $J_R\sim [ 10 , 110 ]\kms\kpc$, $J_\phi\sim [ -630 , -250 ]\kms\kpc$, $J_z\sim [ 55 , 190 ]\kms\kpc$, $L_\perp\sim [ 215 , 485 ]\kms\kpc$, eccentricity$\sim [ 0.2 , 0.4 ]$, $ r_{\rm peri}\sim [ 1 , 3 ]\kpc$, $ r_{\rm apo}\sim [ 3 , 6 ]\kpc$, $ \phi\sim [ 18 \deg, 56 \deg]$. This group contains $11$ globular clusters (these are listed in Table~\ref{tab:merger_summary}). The metallicity of these objects range from [Fe/H]$\sim-1.65$~dex to $-0.4$~dex with a median of [Fe/H]$=-0.7$~dex. For this group, while its dynamical properties appear consistent with that of the disk (e.g., low eccentricity, the $r_{\rm apo}$ range and low $\phi$ values), its relatively lower [Fe/H] value appears more consistent wit that of the bulge. This makes it challenging to associate these objects with either disk or bulge. Perhaps \cite{Massari_2019} were also in a similar conundrum that they interpreted some of these objects as the bulge-clusters, some as the disk-clusters and others simply as ``low energy objects''. Among our member objects, NGC~6441 was tentatively associated with the {\it Kraken} merger by \cite{Kruijssen_2020_Kraken}. We argue that our objects likely do not belong to {\it Kraken} because our objects possess slightly negative $J_\phi$ values (on average), while {\it Kraken} objects have an average $J_\phi \sim 0$ (that we observed from Figure~\ref{fig:Fig_literature} in Appendix~\ref{appendix:literature}). Also, a few of our objects have been interpreted as either bulge or simply low-energy clusters by \cite{Horta_2020} on the basis of their chemical compositions.

\begin{figure*}
\begin{center}
\vspace{-0.1cm}
\includegraphics[width=0.95\hsize]{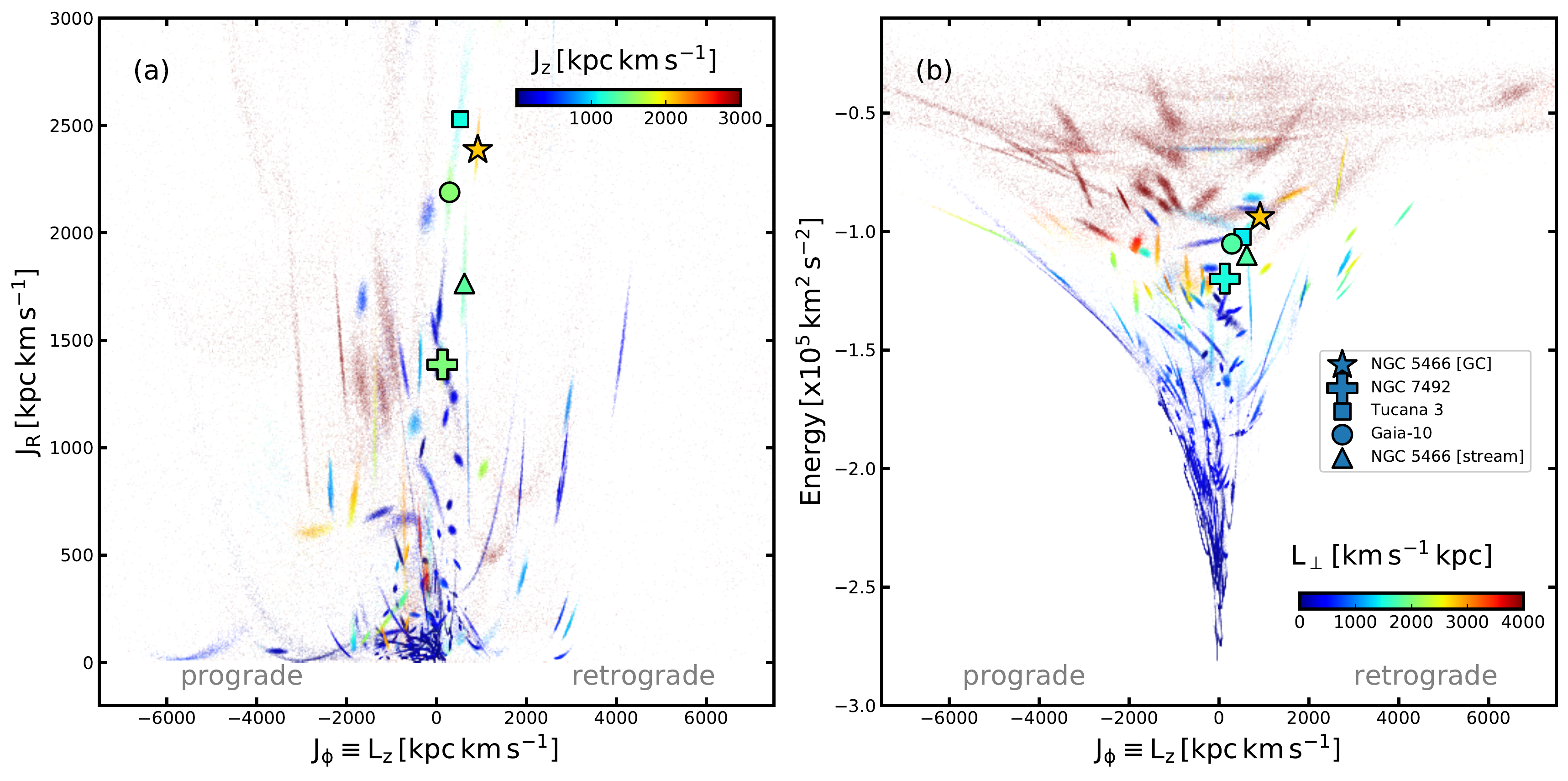}
\end{center}
\vspace{-0.5cm}
\caption{\EJ\ distribution of the objects belonging to the candidate merger (see Section~\ref{sec:merger_candidate}). These objects are shown along with all the other objects in our sample. Similar to Figure~\ref{fig:Fig_EJ_of_objects}, each object is shown as a `cloud' of $1000$ Monte Carlo representations of its orbit.}
\label{fig:Fig_EJ_candidate_merger}
\end{figure*}
\begin{figure}
\begin{center}
\includegraphics[width=0.95\hsize]{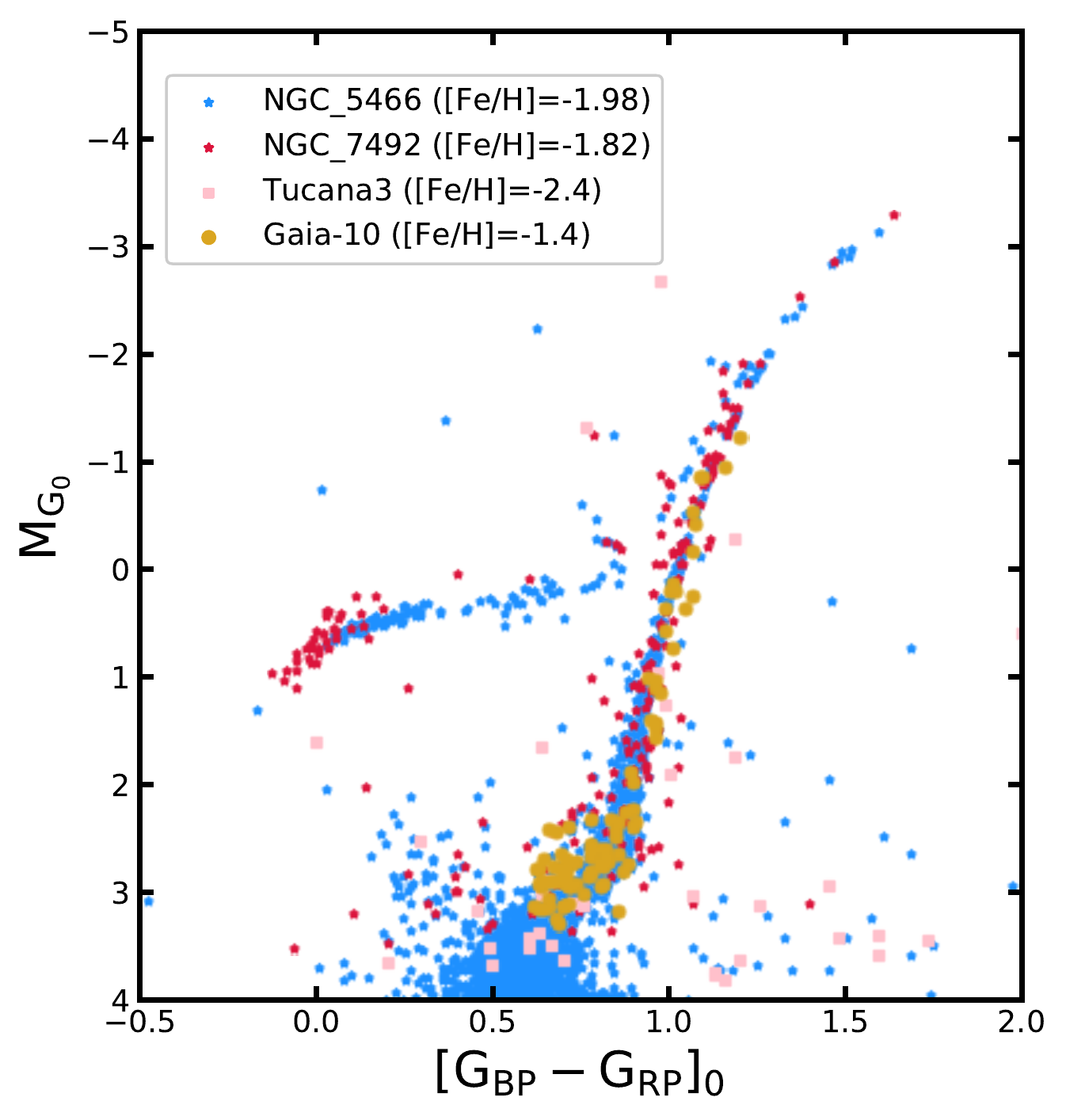}
\end{center}
\vspace{-0.5cm}
\caption{Comparing the color-magnitude distribution of the objects belonging to the candidate merger  (see Section~\ref{sec:merger_candidate}). We use \Gaia\ EDR3 photometry that is corrected for extinction using the \citealt{Schlafly_2014} dust-extinction map. The absolute magnitude ($M_{G_0}$) is obtained by correcting the de-reddened magnitudes for the distance of the objects. The distances of globular clusters are taken from \cite{Vasiliev_GCs_2021}, of Tucana~III is taken from \cite{McConnachie_2020_GaiaEDR3} and for the stream from Table~\ref{tab:table_stream_helio}. The quoted [Fe/H] values correspond to the spectroscopic measurements of these objects (see text).}
\label{fig:Fig_CMD}
\end{figure}
\section{A candidate merger}\label{sec:merger_candidate}

All the above mentioned groups were detected at $\geq2\sigma$ significance following the \enlink\ procedure described in Section~\ref{sec:ENLINK}. However, during our multiple \enlink\ runs (while we were initially experimenting with different parameters), we noticed a particular group that comprised $5$ objects whose $\Pg$ values were fluctuating close to the the detection threshold. This is likely due to the \enlink\ parameter $\texttt{min\_cluster\_size}$ that we set to $5$ (see Section~\ref{subsec:app_enlink}), thus making it difficult for \enlink\ to detect groups containing $\leq5$ objects. Motivated by the possibility that this group may represent an actual merger, we discuss its properties below. 

This group possesses slight retrograde motion and relatively high energy (as shown in Figure~\ref{fig:Fig_EJ_candidate_merger}). The dynamical properties of this group lie in the range $E\sim[ -1.2 , -0.94 ]\times10^5\km2s2$, $J_R\sim [ 1385 , 2525 ]\kms\kpc$, $J_\phi\sim [ 130 , 915 ]\kms\kpc$,  $J_z\sim [ 1125 , 2095 ]\kms\kpc$, $L_\perp\sim [ 1435 , 2795 ]\kms\kpc$,  eccentricity$\sim [ 0.7 , 0.8 ]$, $ r_{\rm peri}\sim [ 3 , 7 ]\kpc$, $ r_{\rm apo}\sim [ 27 , 47 ]\kpc$, $ \phi\sim [ 69 \deg, 85 \deg]$. The group comprises $2$ globular clusters (namely, NGC~5466 and NGC~7492), $2$ stellar streams (NGC~5466 and Gaia-10) and 1 dwarf galaxy (Tucana~III). The $\phi$ parameter indicates that the member objects have very `polar' orbits. Particularly for Tucana~III, Gaia-10 and NGC~5466, we note that their orbital planes are very similar; this further lends credence to their possible association. The MDF of this group ranges from [Fe/H]=$-2.4$~dex to $-1.4$~dex. These minima and maxima are set by Tucana~III \citep{Simon_2017} and Gaia-10, respectively. NGC~5466 has [Fe/H]$\sim-1.98$~dex \citep{Lamb_2015} and NGC~7492 has [Fe/H]$\sim-1.8$~dex \citep{Cohen_2005}. We further compare the stellar population of these objects in terms of their color-magnitude distributions (CMDs), and this is shown in Figure~\ref{fig:Fig_CMD}. These objects possess strikingly similar CMDs, despite their differences in [Fe/H]\footnote{The reason that Tucana~III's CMD appears scattered is because we construct the CMD using the photometry from \Gaia\ EDR3, and \Gaia\ has a limiting magnitude at $G\sim21$.}. In summary, the similarities in the stellar population of these objects, together with their co-incidence in the \EJ\ space, indicates that these objects were perhaps born at the same time inside the same progenitor galaxy.

Previous studies have associated NGC~5466 and NGC~7492 with the Sequoia and \GSE\ groups, respectively \citep{Massari_2019, Forbes_2020}. On the other hand, Gaia-10 (that is being analysed here for the first time) and Tucana~III have not been previously associated with any merger group.

From Figure~\ref{fig:Fig_EJ_candidate_merger}, one notices that the globular cluster NGC~5466 and its stream have different values of \EJ, although they should ideally have very similar orbits. Upon closer inspection we find that the computed orbit of the NGC~5466 stream fits the phase-space data nicely, however, its orbital distance solution has an offset of $\sim 1.5$~kpc from that of the globular cluster. This offset in distance changes the resulting \EJ\ solutions. This offset arises because we are constraining the distance of the streams using only the parallaxes (see Section~\ref{subsec:compute_EJ}), but the uncertainties on parallaxes are too large to properly constrain the strong distance gradient that is present particularly in this stream \citep{Jensen_2021}.

\section{Physical connections between streams and other objects}\label{sec:Streams_progs}

Until this point of the analysis, the question that we have tried to answer is: which set of globular clusters, streams and satellites were accreted inside which merger galaxies? In this section, our objective is slightly different than the main objective of this paper. Here, we want to investigate: which stream was produced by the tidal stripping of which parent system (i.e., which globular cluster or satellite galaxy)? At this juncture of our analysis, exploring this question is relatively straightforward because we possess the orbital information of all the halo objects. Therefore, all we require to do is to compare the orbits of streams with those of globular clusters and satellites to find plausible physical connections amongst them\footnote{This is a very simple method of finding parent systems of streams, and it does not guarantee that the resulting connections represent the reality (at least not in all the cases). For confirmation, such methods should be complemented with the [Fe/H] comparison of the stream and also proper N-body modeling; however, these are beyond the scope of this paper. It is for these reasons that we proceeded in Section~\ref{sec:ENLINK} with both the globular cluster and stellar stream counterparts, as we deemed it less biased.}. Below, we first briefly describe the motivation behind this investigation and then explain our method of identifying these physical connections. 

A stellar stream is generally pictured as two tidal tails emerging from its parent system (be it a globular cluster or a satellite galaxy, e.g., \citealt{Dehnen_2004}). While this is true for a few streams in our sample (e.g., Pal~5, Fj\"orm, NGC~3201, NGC~5466), a majority of them are observed without their parent systems ($\approx 75\%$ of the streams considered in this study). The two main reasons for this could be: (1) if the parent system of a stream has been completely dissolved due to tidal stripping, then it cannot be observed at the present day, and (2) if the parent system is spatially separated from the stream, then it becomes difficult to recognize any physical connection between the two. The latter scenario is especially possible for those streams that merged inside their progenitor galaxies, since mergers can deposit their stellar content in disparate regions of phase-space (e.g., \citealt{Jean-Baptiste_2017}). It is this second scenario that we want to explore here. Note that this knowledge of a stream's parent object is also crucial for the stream's N-body dynamical modeling (e.g, \citealt{Thomas_2016_Pal5, Bonaca_2019_GD1}). 

Our strategy to find parent objects of streams is straightforward. We compare \EJ\ and [Fe/H] values of streams with those of globular clusters and satellites. Thus, a stream is considered to be physically associated with an object if 1) their \EJ\ values differ by $<2\sigma$, and 2) their [Fe/H] values are similar. Through this \EJ-[Fe/H] comparison, we identify $4$ stream-globular cluster pairs and $1$ stream-satellite pair. The former pairs include Gj\"oll--NGC~3201 (originally noted in \citealt{Hansen_2020_3201_Gjoll, Palau_2021_NGC3201_Gjoll}); Fimbulthul-- NGC~5139/$\omega$~Centauri \citep{Ibata_2019_wcen}; Fj\"orm--NGC~4590/M~68 \citep{Palau2019_M68_Fjorm} and Ophiuchus--Kim~3 (we note that the large dispersion in \EJ\ values of Kim~3 renders this association tentative). The latter pair includes Cetus--Willman~1. The details of these parent objects are provided in Table~\ref{tab:table_GC_EJ} and these can be compared with the parameters of streams in Table~\ref{tab:table_stream_EJ}. 

We find additional associations based only on the \EJ\ values, since a majority of streams lack [Fe/H] measurements. In this case, we consider only those associations for which the \EJ\ values differ by $<1\sigma$. These pairs include C~3--NGC~288. 

We also examine if there are any streams that are possibly connected to other streams; implying that these physically separated structures actually represent different segments of the same stream. To this end, we consider only those associations for which the \EJ\ values differ by $<1\sigma$ and find the following associations: Atlas-AliqaUma (originally noted in \citealt{Li_2021_AAS}); C~3--Gaia~6; Fimbulthul--Gunnthr\`a. This last association suggests that Gunnthr\`a is the leading tidal arm of the $\omega$~Centauri cluster and Fimbulthul is the trailing arm; in a way we complete the picture proposed by \cite{Ibata_2019_wcen}.

\begin{table*}
\centering
\caption{Actions, energies, orbital parameters and [Fe/H] of those globular clusters/satellite galaxies that are physically connected to stellar streams (see Section~\ref{sec:Streams_progs}).}
\label{tab:table_GC_EJ}
\begin{tabular}{|l|l|c|c|c|c|c|c|}

\hline
\hline
Globular cluster/ & $(J_R,J_\phi,J_z)$ & Energy & $r_{\rm peri}$ & $r_{\rm apo}$ & $z_{\rm max}$ & eccentricity & [Fe/H]\\
satellite galaxy  & [$\kpc\kms$] & [$\km2s2$] & [kpc] & [kpc] & [kpc] & & [dex]\\
\hline
\hline
& & & & & & &\\

NGC~3201  & $( 787 ^{+ 29 }_{- 27 }, 2724 ^{+ 21 }_{- 18 }, 276 ^{+ 5 }_{- 4 })$  & $ -115824 ^{+ 752 }_{- 724 }$  & $ 8.3 ^{+ 0.0 }_{- 0.0 }$  & $ 27.1 ^{+ 0.5 }_{- 0.5 }$  & $ 10.8 ^{+ 0.2 }_{- 0.2 }$  & $ 0.53 ^{+ 0.01 }_{- 0.01 }$  & $ -1.59 $ \\

NGC~4590  & $( 765 ^{+ 23 }_{- 25 }, -2359 ^{+ 14 }_{- 15 }, 824 ^{+ 12 }_{- 12 })$  & $ -113703 ^{+ 611 }_{- 668 }$  & $ 9.1 ^{+ 0.1 }_{- 0.0 }$  & $ 28.1 ^{+ 0.4 }_{- 0.5 }$  & $ 18.5 ^{+ 0.3 }_{- 0.3 }$  & $ 0.51 ^{+ 0.0 }_{- 0.01 }$  & $ -2.23 $ \\

NGC~5139  & $( 157 ^{+ 5 }_{- 5 }, 540 ^{+ 10 }_{- 10 }, 136 ^{+ 6 }_{- 6 })$  & $ -185068 ^{+ 253 }_{- 258 }$  & $ 2.7 ^{+ 0.1 }_{- 0.1 }$  & $ 6.9 ^{+ 0.0 }_{- 0.0 }$  & $ 3.0 ^{+ 0.1 }_{- 0.1 }$  & $ 0.44 ^{+ 0.01 }_{- 0.01 }$  & $ -1.53 $ \\

Kim~3  & $( 623 ^{+ 4925 }_{- 526 }, -569 ^{+ 800 }_{- 1117 }, 2095 ^{+ 3421 }_{- 1872 })$  & $ -117699 ^{+ 58531 }_{- 40035 }$  & $ 9.4 ^{+ 5.7 }_{- 3.7 }$  & $ 24.5 ^{+ 90.9 }_{- 14.0 }$  & $ 21.8 ^{+ 91.7 }_{- 16.6 }$  & $ 0.5 ^{+ 0.29 }_{- 0.24 }$  & $ -1.6^{+0.45}_{-0.30} $ \\

Willman~1  & $( 837 ^{+ 306 }_{- 262 }, -2432 ^{+ 758 }_{- 490 }, 2726 ^{+ 1775 }_{- 892 })$  & $ -93587 ^{+ 8055 }_{- 7180 }$  & $ 16.0 ^{+ 4.4 }_{- 2.6 }$  & $ 42.4 ^{+ 7.1 }_{- 6.9 }$  & $ 37.3 ^{+ 10.1 }_{- 8.6 }$  & $ 0.44 ^{+ 0.07 }_{- 0.08 }$  & $ -2.1 $ \\


& & & & & & & \\
\hline
\hline
\end{tabular}
\tablecomments{The orbital parameters are derived in this study and the values represent the medians of the sampled posterior distributions and the corresponding uncertainties reflect their $16$ and $84$ percentiles. The [Fe/H] measurements are taken from the literature. We take [Fe/H] of all the globular clusters from the \cite{Harris2010} catalog, except for Kim~3 (we use \citealt{Kim_2016_GC}). [Fe/H] value of satellite galaxies are taken from the \cite{McConnachie_2020_GaiaEDR3} catalog.
}
\end{table*}
\section{Discussion and Conclusion}\label{sec:Disc_Conc}

We implemented an objective search strategy and detected $N=\nmergers$ apparently independent mergers that our Galaxy has suffered during its long history. We achieve this by searching for statistically significant groups of halo objects in \EJ\ space. Our data set comprises $n=\nobjects$ objects (namely, \nGCs\ globular clusters, \nss\ stellar streams and \nDGs\ satellite galaxies). The \EJ\ values of these objects are derived using \Gaia\ EDR3 based catalogs. To detect the mergers in the \EJ\ space, we used the \enlink\ group-finding software, coupled with our statistical procedure that accounts for the uncertainties on the derived \EJ\ values of independent objects (that arises due to the measurement uncertainties on the phase-space properties of the objects). All of the mergers are detected at $\geq2\sigma$ confidence, and together they comprise $\nMergerObjects$ objects ($\fMergerObjects$ of the total sample size), including $35$ globular clusters, $25$ streams and $2$ satellite galaxies. 

We successfully recovered many of the previously known mergers (namely \Sgr, \Cetus, \GSE, \asi, \lms1)\footnote{The mergers that we could not detect are \Helmi\ \citep{Helmi_1999},  {\it Thamnos~1 and 2} \citep{Koppelman2019}, \Kraken\ \citep{Kruijssen_2020_Kraken, Forbes_2020}.}. Moreover, we discovered a new merger (that we call \Pontus) and a possible merger candidate. Section~\ref{sec:analyse_mergers} details the overall properties of these mergers -- their member objects (Table~\ref{tab:merger_summary}), their \EJ\ distribution (Figure~\ref{fig:Fig_probability_cut}), their orbital properties as a function of [Fe/H] of the member objects (Figure~\ref{fig:Fig_orbits_FeH}), their [Fe/H] distribution function (Figure~\ref{fig:Fig_MDF}), their other orbital properties (Figure~\ref{fig:Fig_rper_rapo}) and the masses of their progenitor galaxies. 

Below, we first discuss some of the key results of our study, and then consider how our analysis advances our knowledge about the formation of the Milky Way halo, and then finally explore what information this study provides to galaxy formation models. 

\subsection{Key results}\label{subsec:key_results}

First, we report the detection of a new merger \Pontus\ (see Section~\ref{subsec:Pontus}). This is a retrograde merger (i.e., $J_\phi>0$) that possesses relatively low-energy (see Figure~\ref{fig:Fig_probability_cut}). Its low-energy could be indicating an early accretion, but this is hard to confirm at this stage. We note that some of the globular clusters in \Pontus\ are quite metal poor and old (Figure~\ref{fig:Fig_AMR_vel}). While \Pontus\ is detected by our analysis, we find another group through visual inspection of the \EJ\ data, and this represents a new candidate merger (Section~\ref{sec:merger_candidate}). 

We find that the \lms1\ merger \citep{Yuan2020, Naidu2020} represents the most metal poor merger of the Milky Way, with its MDF having a minimum of [Fe/H]$\approx -3.4$~dex (Figure~\ref{fig:Fig_MDF}, Section~\ref{subsec:lms1}). The three most metal poor streams -- C-19 ([Fe/H]$=-3.38\pm0.06$~dex, \citealt{Martin2022_C19}), Sylgr ([Fe/H]$=-2.92\pm0.06$, \citealt{Roederer_2019}) and Phoenix ([Fe/H]$=-2.70\pm0.06$, \citealt{Wan_2020_Phoenix}) -- belong to this merger. Thus the progenitor of \lms1\ was probably an ancient and extremely metal-poor proto-galaxy that formed in the early universe. Furthermore, most of the member objects of \lms1 possess low-mass. The exceptions are the globular clusters NGC~5024 ($M\sim 4.5\times10^5\msun$) and NGC~5272/M~3 ($M\sim 4\times10^5\msun$, \citealt{Baumgardt2019}), but we note that both of them are Type I clusters\footnote{Type I clusters correspond to those that exhibit a simpler stellar population, and where first and second generations of stars are not easily distinguishable (likely because they host only one generation of stars, \citealt{Milone_2017}).} and they are also metal-poor (\citealt{Harris2010, Boberg_2016}). Moreover, NGC~5024 has also been confirmed to possess a stellar population similar to that of the LMS-1 stream \citep{Malhan_2021_LMS1}. These results, at some level, lend credence to the possibility that massive globular clusters can form inside metal-poor proto-galaxies. In future, it will be interesting to combine the chemical properties of the member clusters, and also streams, of the \lms1\ group (e.g., their [Fe/H], alpha-abundances, etc) to explore their formation inside the progenitor \lms1\ galaxy. In a broader context, this will also be useful to understand the formation of the most metal deficient globular clusters inside those ancient proto-galaxies that formed in the early universe. 

Furthermore, in the context of the hierarchical build up of the Milky Way halo, the distribution function of these accreted mergers can be qualitatively constrained by examining Figure~\ref{fig:Fig_probability_cut} (and also Figure~\ref{fig:Fig_EJ_candidate_merger}). For instance, there is an even distribution of mergers along the prograde and retrograde direction. A second interesting feature that we note in Figure~\ref{fig:Fig_probability_cut}a is that all the prograde mergers possess higher $J_z$ values and all the retrograde mergers possess lower $J_z$ values. Moreover, three of these mergers have their orbits inclined at $\phi \sim 90\deg$ to the Galactic plane; i.e., they possess polar orbits structures in the Milky Way halo. These mergers include \Sgr\ (Section~\ref{subsec:Sgr}), \Cetus\ (Section~\ref{subsec:Cetus}) and \lms1\ (see Section~\ref{subsec:lms1}). The presence of these `polar' mergers could be indicating that the polar orbits in the Milky Way are indeed stable (see \citealt{Penarrubia_2002}), and therefore, those mergers that would have accreted along polar orbits would remain polar (although their orbits may still precess). We also note a lack of mergers that orbit close to the Galactic plane. A possible reason could be that objects accreted along such orbits are quickly phase-mixed, and this renders their detection rather difficult with present methods. 

Our analysis also allows us to (roughly) constrain the accreted stellar mass of the Milky Way halo. This parameter depends critically on the most massive mergers, such as those we detect here. To compute this parameter we add up the stellar mass values of the independent mergers, which yields a Galactic stellar halo mass of $M_{\rm *, MW\, halo}\sim10^9\msun$. This value is smaller than that obtained by some recent studies (e.g., $\sim1.5\times10^9\msun$ from \citealt{Deason_2019}). But we caution that our value is based only on the member globular clusters in each merger and not the member stellar streams; and the inclusion of streams can change this result. 

We also find various physical associations between stellar streams and other objects (i.e., globular clusters and satellite galaxies, see Section~\ref{sec:Streams_progs}). Most of these associations are based on the similarities in their \EJ\ and [Fe/H] values, however, a few of them are proposed based only on \EJ\ comparisons. To confirm these latter connections, future [Fe/H] and stellar population comparisons are necessary.

\begin{figure*}
\begin{center}
\vspace{-0.1cm}
\includegraphics[width=\hsize]{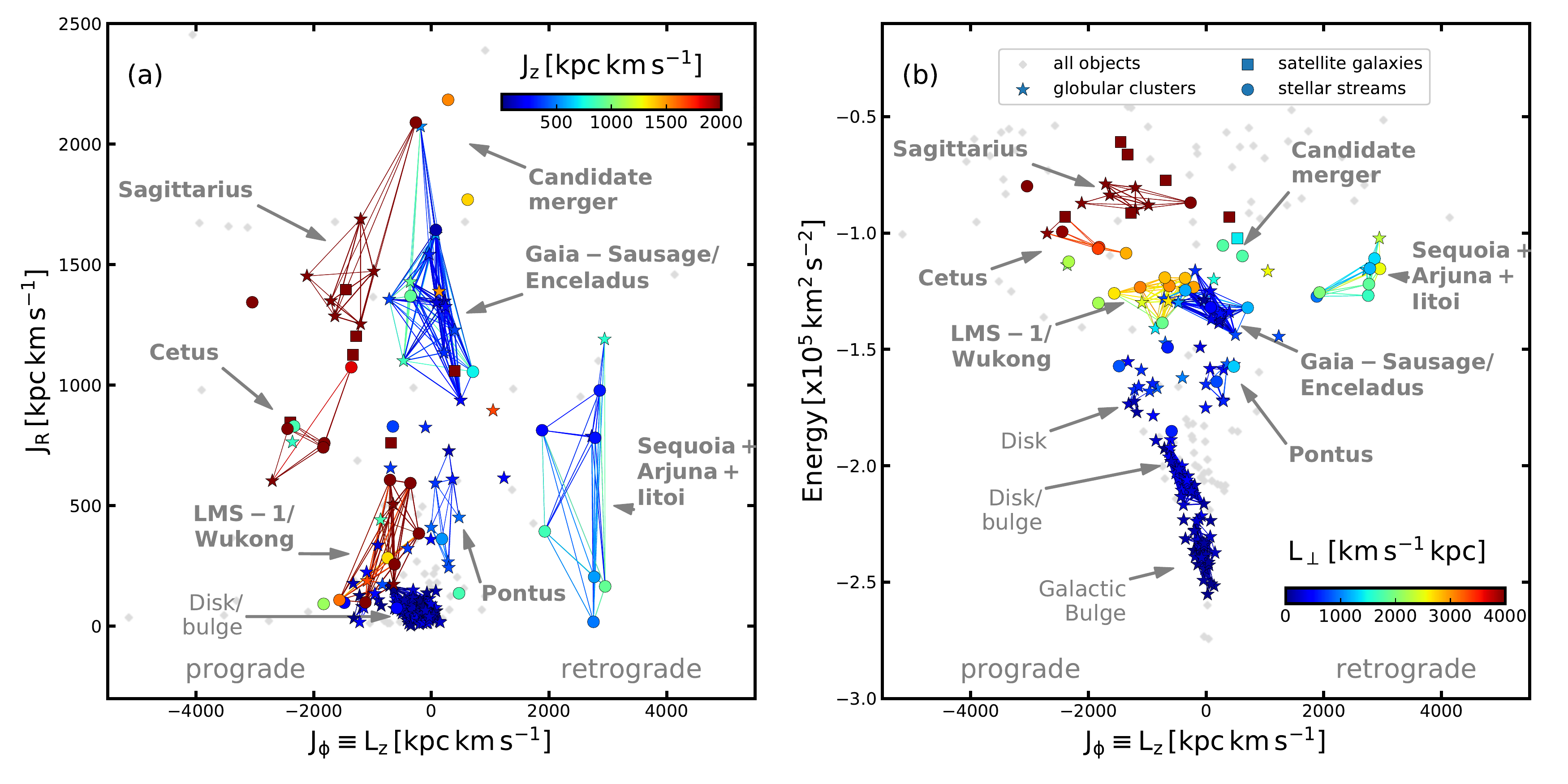}
\end{center}
\vspace{-0.5cm}
\caption{Same as Figure~\ref{fig:Fig_probability_cut}, but lowering the probability threshold value $\Pth$ to $0.2$ (see Section~\ref{subsec:remaining_objs}).}
\label{fig:Fig_probability_cut_lowPth}
\end{figure*}
\subsection{What about the remaining halo objects, that are not identified as members of any detected group?}\label{subsec:remaining_objs}

A large fraction of objects in our sample are not identified as members of any detected groups -- be it the merger groups or the in-situ groups (i.e., Galactic disk and bulge). Quantitatively, this is $58\%$ of the sample of $\nobjects$ objects at a $2\sigma$ threshold. We consider below some possible interpretations of the low-significance groupings.

In regard to those objects that possess low $\Pg$ value and high-energy (with say $E\simgt -1.6\times10^5\km2s2$), there are three possible scenarios: (1) Some of these objects were accreted inside the same mergers that we detected here, but our method could not confidently associate them with these mergers. This could owe either to the inefficiency of our method or large phase-space uncertainties of these objects; or both. (2) Some of these objects were accreted inside other mergers that we could not detect here (e.g., {\it Helmi} structure, \citealt{Helmi_1999}). A possible reason that we did not detect these mergers could be that these are highly phase-mixed substructures and their member objects are largely dispersed in \EJ\ space. (3) A majority of these objects may have accreted inside a multitude of low-mass mergers, that perhaps had masses in the range $M_{\rm halo}\simlt10^{7-9}\msun$. Since such low-mass galaxies typically harbour only $\sim 1-2$ globular clusters (e.g., \citealt{Forbes_2018}), their detection is not possible using our method. 

To explore these possible scenarios, we consider Figure~\ref{fig:Fig_probability_cut_lowPth} that shows all those objects with $\Pg \geq0.20$. This time we find $78$ objects ($30\%$ of our total sample) represent potential merger objects. Lowering this threshold also reveals the candidate merger that we described in Section~\ref{sec:merger_candidate}. But it is clearly difficult to interpret the reality of these low-significance objects. 

In regard to those objects that possess low $\Pg$ value and low-energy (say, with $E\simlt -1.6\times10^5\km2s2$), there are two possible scenarios: (1) All of these objects indeed represent the in-situ population (i.e., disk-objects or bulge-objects), however, we could only identify those objects that are very tightly clumped in the \EJ\ space. (2) Some of these objects were probably accreted into the Milky Way but we could not identify them as members of high-significance groups. This may include the \Kraken\ group \citep{Kruijssen_2020_Kraken, Forbes_2020}, as shown in Figure~\ref{fig:Fig_literature}. 

In particular, most of the satellite galaxies did not form part of any group according to our analysis. This could be because these satellites have only recently arrived into the Milky Way halo (as indicated by their high $E$ values in Figure~\ref{fig:Fig_EJ_of_objects}) and themselves are undergoing merging (see \citealt{Hammer_2021}). Naturally, each satellite independently represents a (minor) merger event.

\subsection{Caveats and Limitations}\label{subsec:limitations}

Our method can not detect those mergers that initially only possessed a population of stars, and no globular clusters or streams (e.g., low-mass galaxies with $M_{\rm halo}\simlt10^{7-9}\msun$). Perhaps this is the reason that we could not detect the {\it Thamnos} substructure \citep{Koppelman2019} and/or the \Helmi\ substructure \citep{Helmi_1999}\footnote{Although \Helmi\ substructure was previously (unambiguously) associated with $5$ globular clusters by \cite{Massari_2019}, we find all of these clusters to be associated with one of the detected mergers.}. Such mergers can only be identified using the phase-space information of the stars (e.g., \citealt{Naidu2020}). Hence, our study is capable of only putting a lower limit on the number of Milky Way mergers (in this case $N=\nmergers$)

To detect mergers using \enlink, we had to supplement the \J\ information with the (redundant) $E$ information. The additional information of $E$ acts as a `weight', thus enhancing the contrast between different detected mergers. As we noted above, this is because the relative uncertainties on $E$ are smaller than those on \J, and thus \enlink\ uses this additional (precise) $E$ information to detect groups. Future \Gaia\ datasets will deliver much more precise phase-space measurements of the halo objects, and this in turn will reduce the uncertainties on the derived \J\ quantities, and it may eventually become possible to detect mergers by using only the \J\ information. 

Our strategy of detecting mergers relies on the derived \EJ\ values of the halo objects, and these parameters depend on the choice of the Galactic potential. Admittedly, we tried only one potential model \citep{McMillan2017}, but this is a realistic, static and axisymmetric model. On the other hand, there is now mounting evidence that the potential of the Milky Way halo is time dependent and asymmetric. This is due to the ongoing accretion of the massive LMC system that has perturbed the Galactic halo  (especially in the outer halo regions $r_{\rm gal}\simgt30\kpc$, e.g., \citealt{Garavito-Camargo_2021_LMCperturb, Petersen_2021}). Studies indicate that this perturbation is significant enough to alter the orbits of halo objects. In future, it will be important to also try time dependent potential models to integrate the orbits of the halo objects (e.g., \citealt{CorreaMagnus_2021}), compute \J\ values along these orbital paths and then detect the mergers. 

Moreover, we also neglected the effect of the bar – together with all the other (almost) un-knowable time-dependent asymmetries that the  Galaxy has possessed over its long history  – as well as the effect of dynamical friction. The effect of the bar should not have a significant effect on the detected mergers, first of all because its quadrupole potential drops off quickly with distance (as $r^{-3}$), and it will therefore solely affect (weakly) groups that have small pericenters, like \GSE\ and \Pontus. Resonances with the bar should not be an issue for the halo orbits under consideration, and since the streams have a relatively narrow range of energies, whilst the stream path itself will not be the same in the barred potential as in an axisymmetric one, the stream will not be  disrupted. In any case, objects that remained close to each other in the \EJ\ space, as computed within the axisymmetric background potential, can certainly still be associated to each other. A likely much more important issue that we also neglected is dynamical friction. For instance, \citet{Villalobos} simulated the accretion of a merger with a mass of $10^{11}\msun$, and showed that the mass of the merger, which remains bound, decreases mostly during pericentric passages of the merger. For this reason, we cannot in principle formally exclude that \GSE\ and \Pontus\ represent different stages of disruption of one single massive accretion event. However, \Pontus\ has lower orbital energy than \GSE\ -- meaning it should then have been stripped later from the common progenitor galaxy, and while low-mass mergers are known to circularize under dynamical friction, high-mass ones tend to radialize \citep{Vasiliev_2021}. Therefore, with a progenitor mass of $\sim 5 \times 10^{10}\msun$, \Pontus\ is safely in the radialization regime, which would therefore make its {\it lower} eccentricity than \GSE\ an argument against the scenario of these two groups being different stages of disruption of one single massive accretion event. Moreover, we recall that the two groups have a different age-metallicity relationship, which reinforces the hypothesis that these are in fact the remnants of {\it distinct} accretion events.

\vspace{\baselineskip}
The dynamical atlas of the Milky Way mergers that we present here provides a global view of the Galaxy formation in action. Thus, our study contributes to the initial steps of unravelling the full hierarchical build-up of our Galaxy, and also understanding the origin of the globular clusters and stellar streams of the Milky Way halo. This endeavour of detecting mergers using the \EJ\ quantities is only possible due to the amazingly rich phase-space information that ESA/\Gaia\ mission has provided, and this places us in a very exciting position to disentangle the merging events in the Milky Way halo. The community now looks forward to future \Gaia\ data releases and upcoming spectroscopic surveys (e.g., WEAVE, \citealt{Dalton_2014}, 4MOST, \citealt{deJong_2019}, SDSS-V), as the combination of these datasets will provide a gold mine of information in terms of phase-space, metallicity, elemental abundances and ages of stars. With such a wealth of information, we will be able to explore the `temporal' aspect of Galactic archaeology (e.g., \citealt{Feuillet_2021}) by building an understanding of the `chronological' merging history of the Milky Way.

%
\section*{Acknowledgements}
We thank the referee for helpful comments and suggestions. KM acknowledges support from the Alexander von Humboldt Foundation at Max-Planck-Institut f\"ur Astronomie, Heidelberg. KM is also grateful to the IAU's Gruber Foundation Fellowship Programme for their finanacial support. RI, BF, ZY, NFM acknowledge funding from the Agence Nationale de la Recherche (ANR project ANR-18-CE31-0006, ANR-18-CE31-0017 and ANR-19-CE31-0017), from CNRS/INSU through the Programme National Galaxies et Cosmologie, and from the European Research Council (ERC) under the European Unions Horizon 2020 research and innovation programme (grant agreement No. 834148). MB acknowledges the financial support to this research by INAF, through the Mainstream Grant 1.05.01.86.22 assigned to the project “Chemo-dynamics of globular clusters: the Gaia revolution” (P.I. E. Pancino). GT acknowledges support from the Agencia Estatal de Investigaci\'on (AEI) of the Ministerio de Ciencia e Innovaci\'on (MCINN) under grant FJC2018-037323-I and from the financial support through the grant (AEI/FEDER, UE) AYA2017-89076-P, as well as by the Ministerio de Ciencia, Innovaci\'on y Universidades (MCIU), through the State Budget and by the Consejer\'ia de Econom\'ia, Industria, Comercio y Conocimiento of the Canary Islands Autonomous Community, through the Regional Budget.

This work has made use of data from the European Space Agency (ESA) mission
{\it Gaia} (\url{https://www.cosmos.esa.int/gaia}), processed by the {\it Gaia}
Data Processing and Analysis Consortium (DPAC,
\url{https://www.cosmos.esa.int/web/gaia/dpac/consortium}). Funding for the DPAC
has been provided by national institutions, in particular the institutions
participating in the {\it Gaia} Multilateral Agreement.

Funding for SDSS-III has been provided by the Alfred P. Sloan Foundation, the Participating Institutions, the National Science Foundation, and the U.S. Department of Energy Office of Science. The SDSS-III web site is http://www.sdss3.org/.

SDSS-III is managed by the Astrophysical Research Consortium for the Participating Institutions of the SDSS-III Collaboration including the University of Arizona, the Brazilian Participation Group, Brookhaven National Laboratory, Carnegie Mellon University, University of Florida, the French Participation Group, the German Participation Group, Harvard University, the Instituto de Astrofisica de Canarias, the Michigan State/Notre Dame/JINA Participation Group, Johns Hopkins University, Lawrence Berkeley National Laboratory, Max Planck Institute for Astrophysics, Max Planck Institute for Extraterrestrial Physics, New Mexico State University, New York University, Ohio State University, Pennsylvania State University, University of Portsmouth, Princeton University, the Spanish Participation Group, University of Tokyo, University of Utah, Vanderbilt University, University of Virginia, University of Washington, and Yale University.

Guoshoujing Telescope (the Large Sky Area Multi-Object Fiber Spectroscopic Telescope LAMOST) is a National Major Scientific Project built by the Chinese Academy of Sciences. Funding for the project has been provided by the National Development and Reform Commission. LAMOST is operated and managed by the National Astronomical Observatories, Chinese Academy of Sciences. 
%

\appendix

\section{Comparing the orbital phase and eccentricity of the halo objects}\label{appendix:phase}

\begin{figure}
\begin{center}
\vspace{-0.3cm}
\includegraphics[width=\hsize]{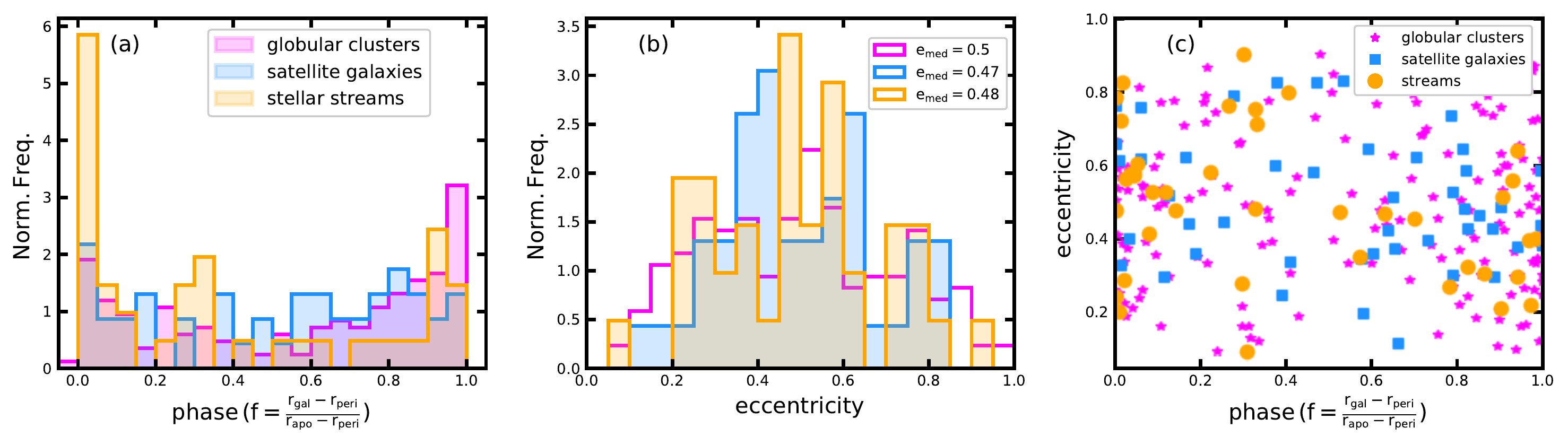}
\end{center}
\vspace{-0.7cm}
\caption{Analysing the orbital phase ($f$) and eccentricity ($e$) of globular clusters, stellar streams and satellite galaxies. Panel `a' compares the $f$ distribution of various objects, panel `b' compares the $e$ distribution (where medians of different distributions are also quoted), and panel `c' shows $f$ vs. $e$.}
\label{fig:Fig_phase}
\end{figure}

As a passing analysis in Section~\ref{subsec:analyse_orbits}, we compared the distribution of orbital phase ($f$) and eccentricity ($e$) of globular clusters, stellar streams and satellite galaxies. This comparison is shown in Figure~\ref{fig:Fig_phase}.

Figure~\ref{fig:Fig_phase}a compares the $f$ distribution of different types of objects. We define the orbital phase of an object (in the radial direction) as $f=\frac{r_{\rm gal}-r_{\rm peri}}{r_{\rm apo}-r_{\rm peri}}$, where $r_{\rm gal}$ is the present day galactocentric distance of the object, and $r_{\rm peri}$ and $r_{\rm apo}$ represent its pericenter and apocenter distances, respectively. This definition of $f$ comes from \cite{Fritz_2018, Li_2021_12streams}. As per this definition, $f=0$ implies that the object is at its pericenter and $f=1$ implies that the object is at its apocentre. The $f$ distribution in Figure~\ref{fig:Fig_phase}a shows piling-up of objects at both the pericenter and the apocenter. This effect is prevalent for globular clusters and stellar streams, but not so much for satellite galaxies. Particularly, in the case of streams, we note that more objects are piled-up at the pericenter than at the apocenter. This effect points towards our inefficiency in detecting those streams that may currently be located at larger distances ($D_{\odot}\simgt30\kpc$) close to their their apocenters. This inefficiency, in part, could be due to \Gaia's limiting magnitude at $G\sim21$. We note that this inference on the $f$ distrbution of streams is different than that of \cite{Li_2021_12streams}, who found that more streams are piled-up at the apocentre (likely due to a different selection strategy). 

Figure~\ref{fig:Fig_phase}b compares the $e$ distribution of different types of objects. The medians of the $e$ distribution for globular clusters, streams and satellites are provided in panel `b'. The median of the eccentricity distribution lies at $e\approx0.5$, which is true for all the three types of objects. This implies that, in general, very radial orbits (with $e\approx1$) or very circular orbits (with $e\approx0$) are rare occurrences. This inference, in particular for streams, is consistent with that of \cite{Li_2021_12streams}. 

Figure~\ref{fig:Fig_phase}c shows $e$ as a function of $f$, and we infer no obvious trend from this plot.

\begin{figure*}
\begin{center}
\vspace{-0.25cm}
\vbox{
\includegraphics[width=0.88\hsize]{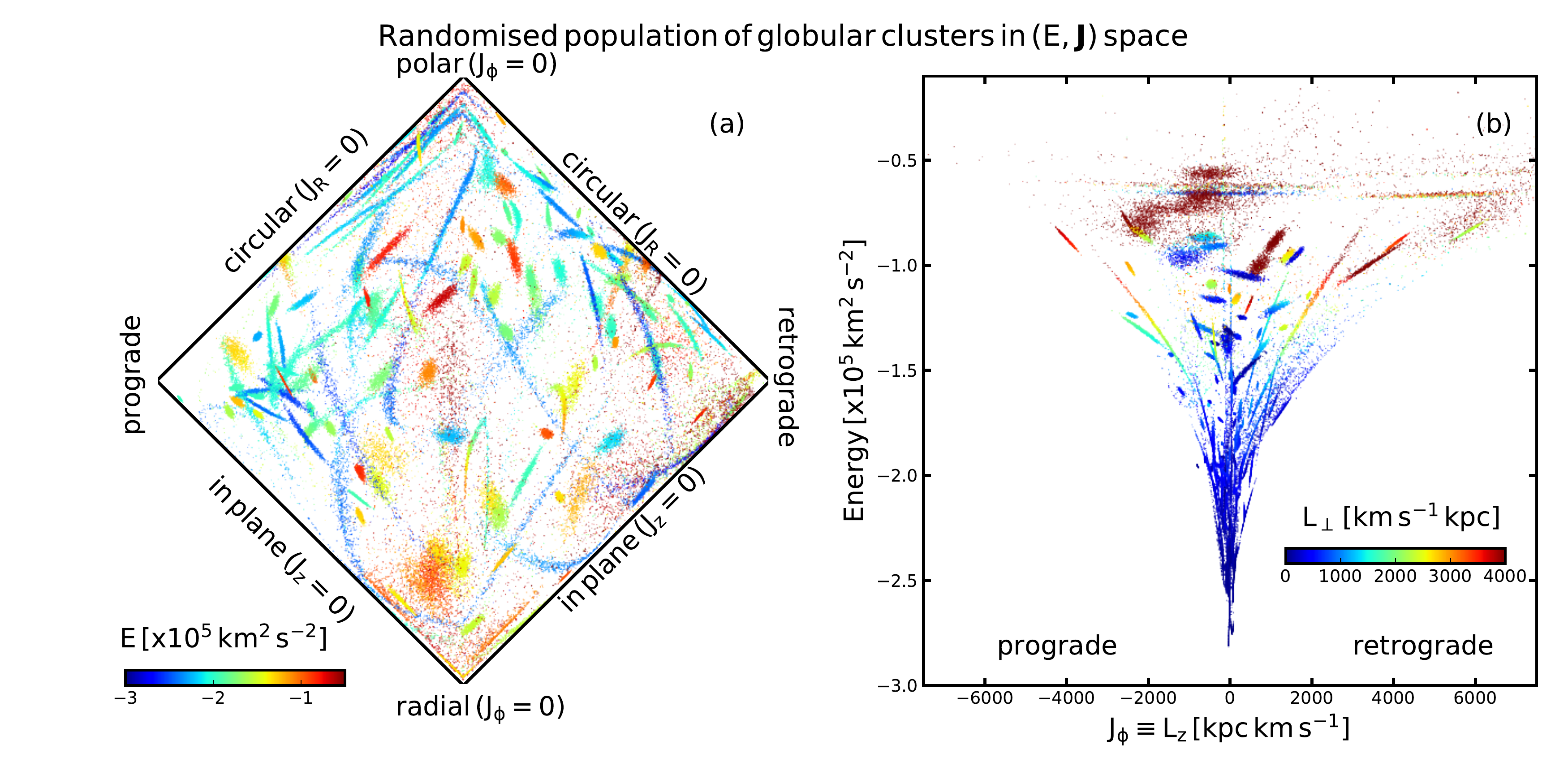}
\includegraphics[width=0.88\hsize]{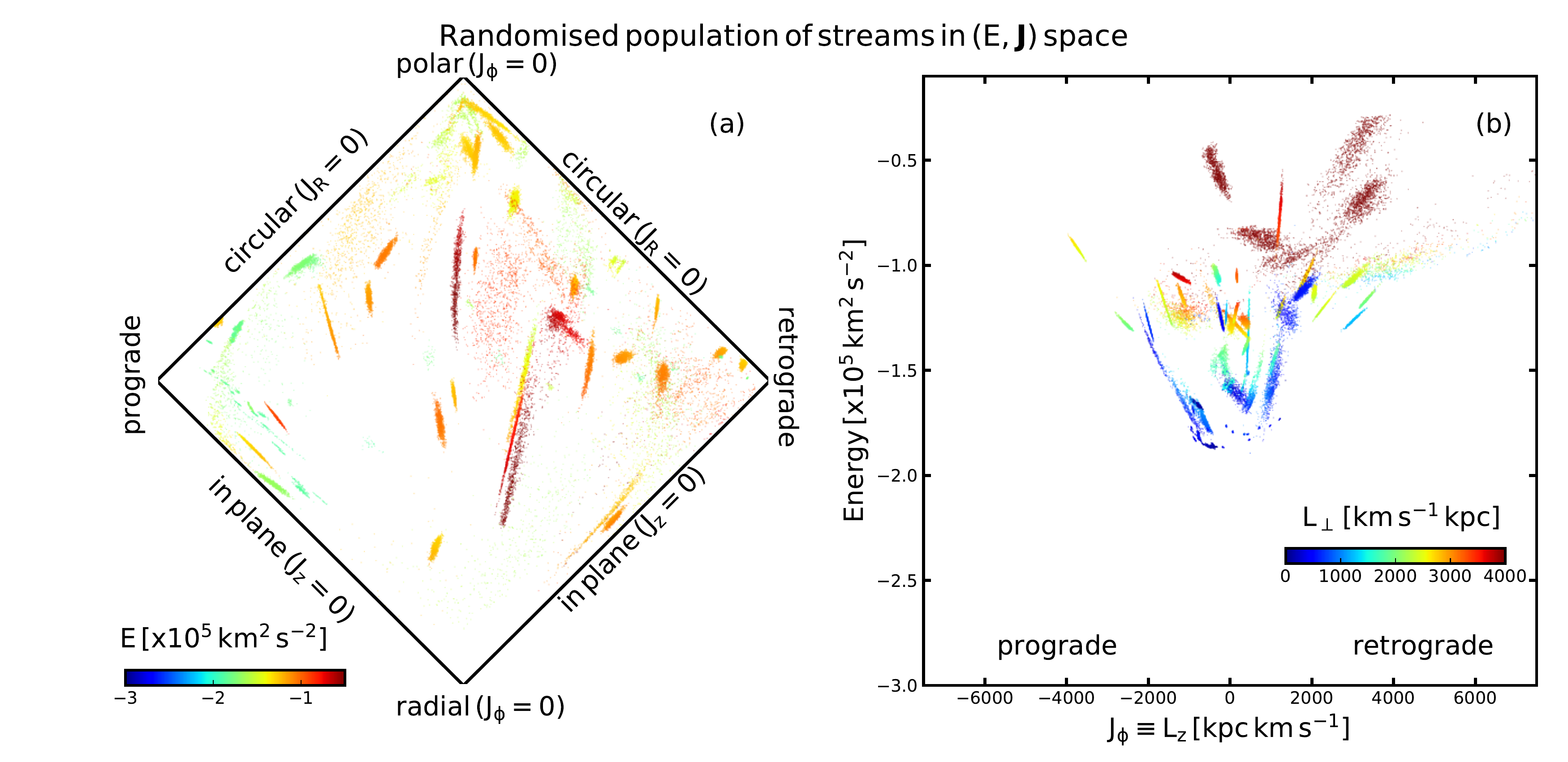}
\includegraphics[width=0.88\hsize]{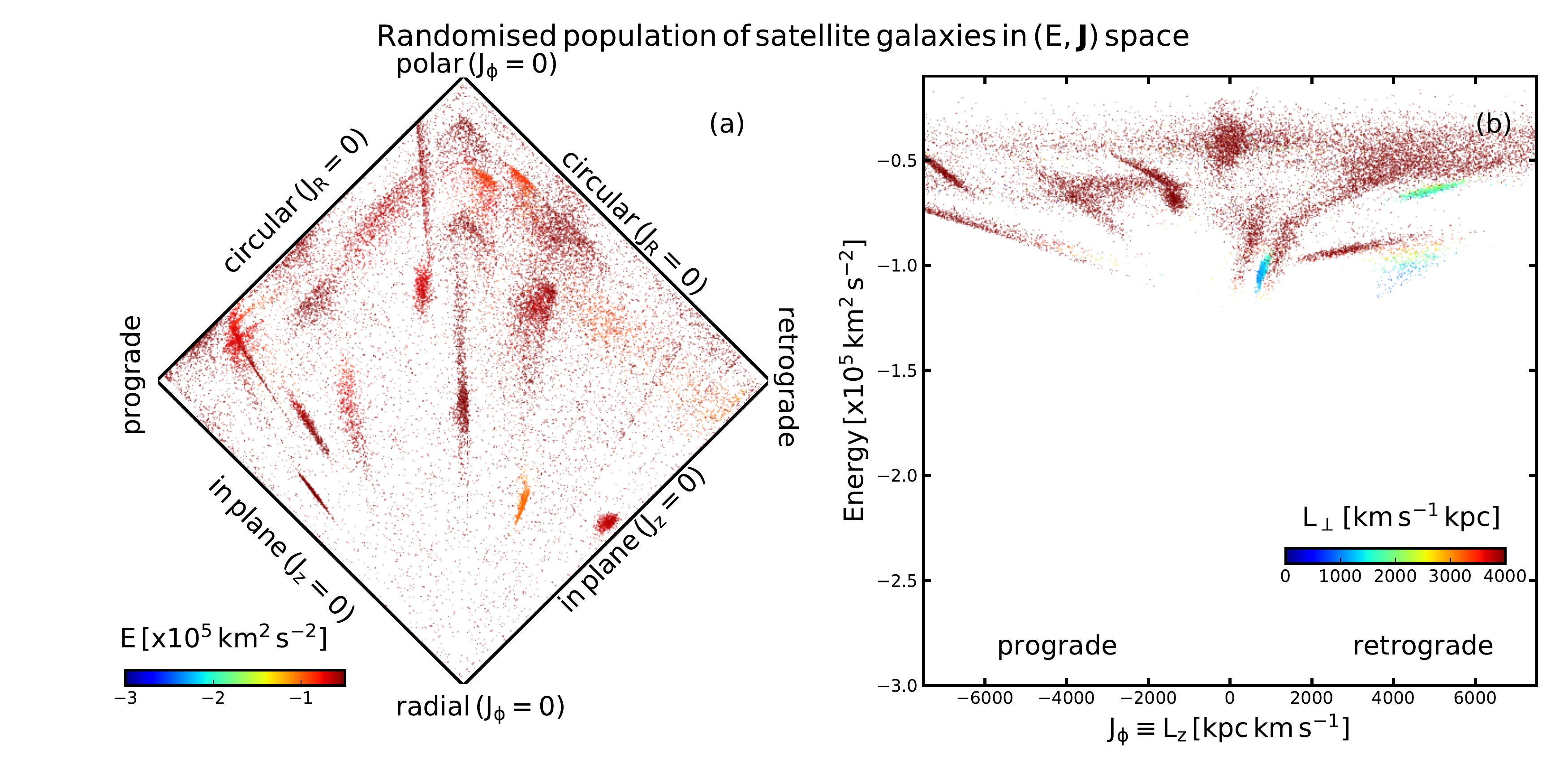}
}
\end{center}
\vspace{-0.5cm}
\caption{Same as Figure~\ref{fig:Fig_EJ_of_objects}, but for the artificially randomised \EJ\ distribution drawn from the sample.}
\label{fig:Fig_EJ_of_random_objects}
\end{figure*}
\begin{figure*}
\begin{center}
\vspace{-0.3cm}
\includegraphics[width=0.80\hsize]{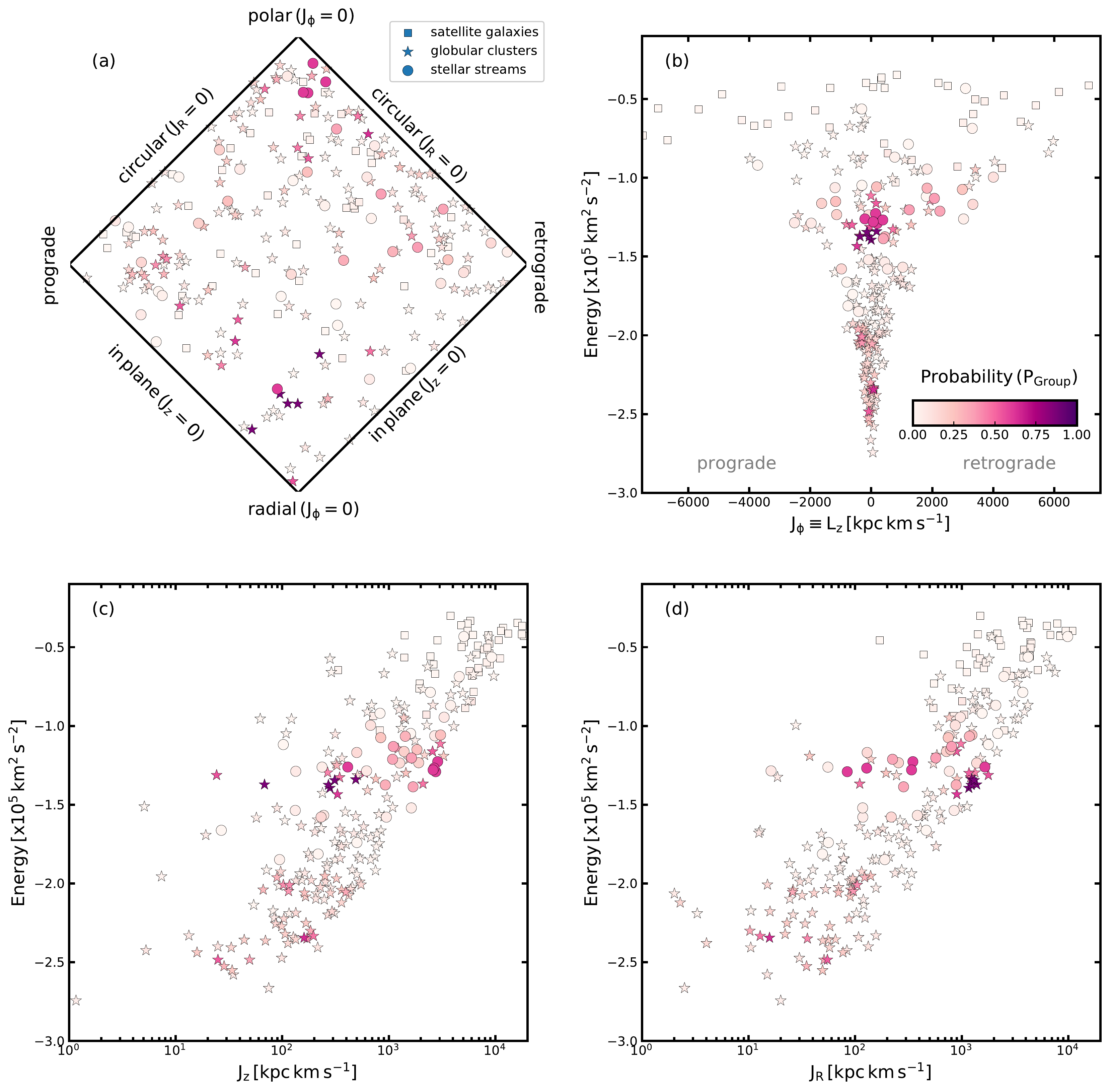}
\end{center}
\vspace{-0.5cm}
\caption{Same as Figure~\ref{fig:Fig_EJ_probability}, but for the artificially randomised \EJ\ distribution.}
\label{fig:Fig_prob_of_random_objects}
\end{figure*}
\begin{figure}
\begin{center}
\vspace{-0.1cm}
\includegraphics[width=0.39\hsize]{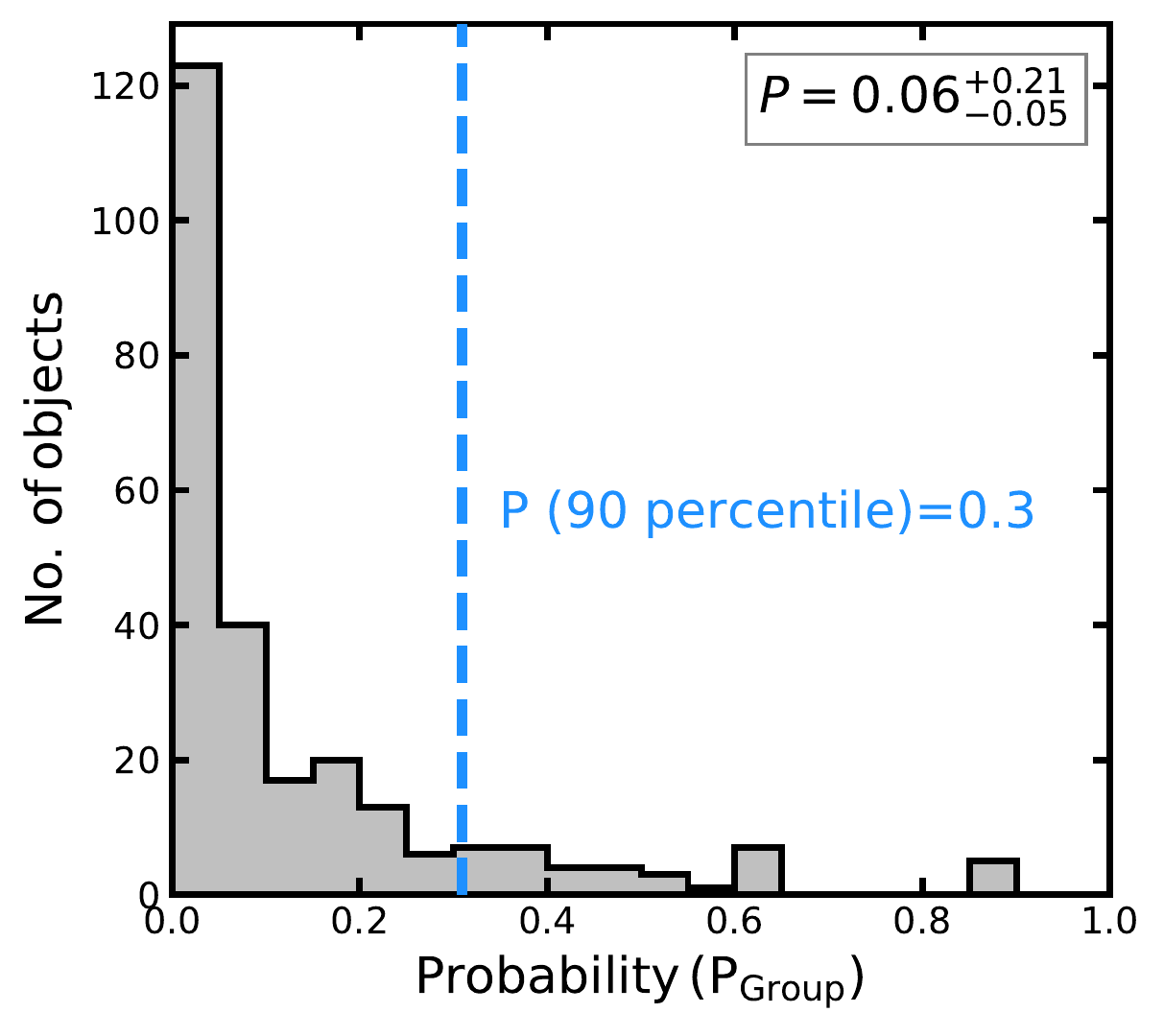}
\end{center}
\vspace{-0.5cm}
\caption{Same as Figure~\ref{fig:Fig_ProbDF}, but for the artificially randomised \EJ\ distribution of objects. The vertical line shows the $90$ percentile of this PDF and we use the corresponding value to set the desired $\Pth$ value. The other quoted value is the median of the distribution and the corresponding uncertainties reflect its $16$ and $84$ percentiles.}
\label{fig:Fig_PDFrandom}
\end{figure}
\begin{figure}
\begin{center}
\vspace{-0.1cm}
\includegraphics[width=\hsize]{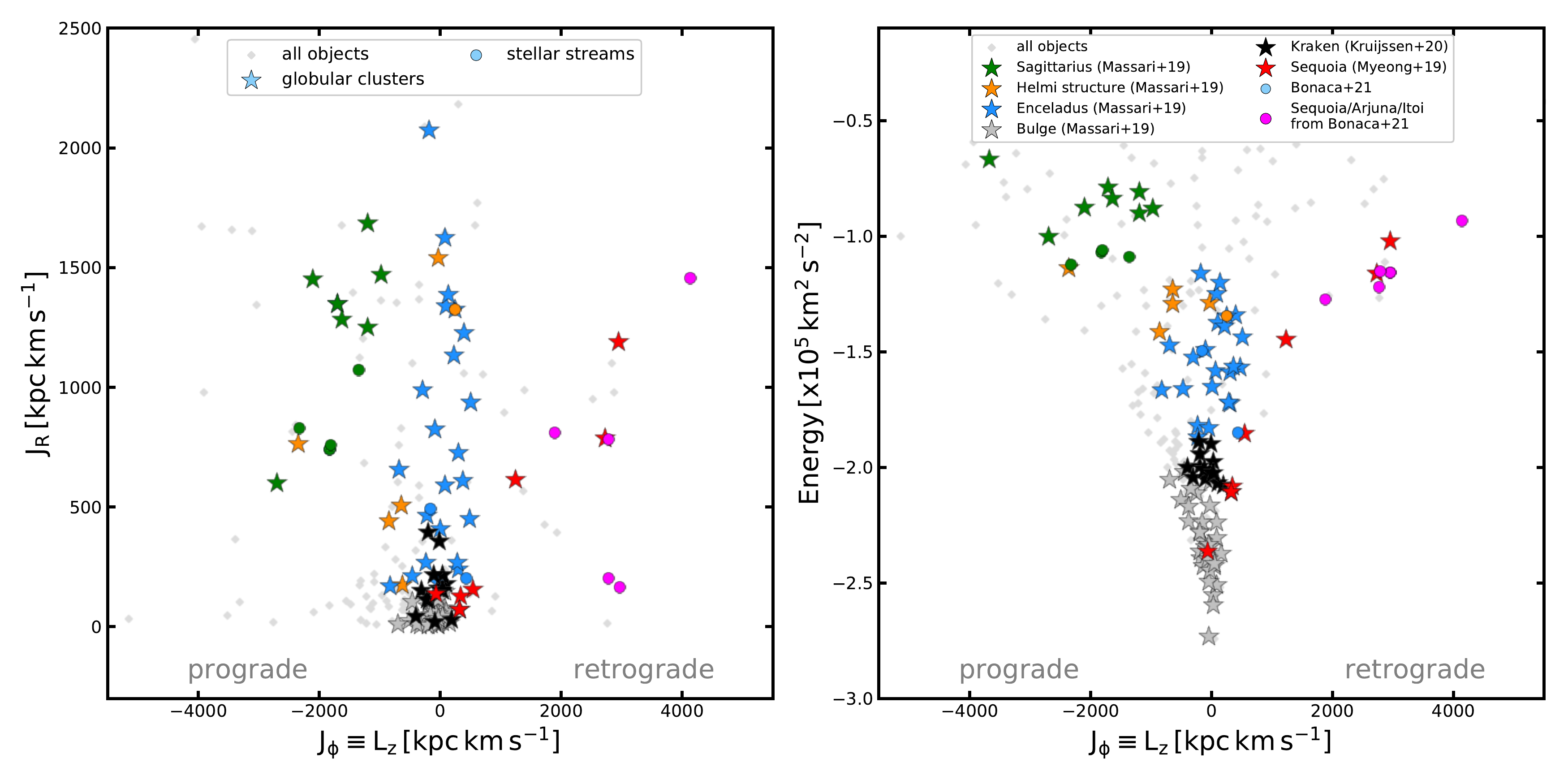}
\end{center}
\vspace{-0.5cm}
\caption{\EJ\ space showing groupings of objects as found by previous studies. Different groups are shown with different colors, the globular clusters are shown with the `star' markers and stellar streams are shown with the `circle' markers.}
\label{fig:Fig_literature}
\end{figure}
\section{Determining the $\Pth$ value}\label{appendix:EJ_random}

For Section~\ref{sec:ENLINK}, we required a suitable probability threshold value $\Pth$, such that all the halo objects with $\Pg\geq\Pth$ can be considered to belong to high-significance groups. To set this $\Pth$ value, we artificially construct a `randomised' version of the real \EJ\ data as follows.

For a given halo object, we first randomly set its orbital pole and then recompute its \EJ\ corresponding to this new orbit (taking the real \EJ\ uncertainties into account). This is repeated for all the halo objects. This randomisation erases any real correlations between the objects in  \EJ\ space. For a given halo object, this transformation from the real coordinate system to this new coordinate system is done as follows. From the $3D$ position vector and $3D$ velocity vector $({\bf x, v})$ we draw (isotropically) a random value $({\bf x', v'})$. After performing this transformation on all the objects, we compute the new \EJ\ values. This new randomised \EJ\ distribution for all the objects is shown in Figure~\ref{fig:Fig_EJ_of_random_objects}. We note that, as a consequence of this transformation, the $E$ parameter of a given object only slightly changes, however, the \J\ parameters are significantly altered. 

Next, we find groups in this randomised \EJ\ distribution using exactly the same procedure as described in Section~\ref{sec:ENLINK}. Figure~\ref{fig:Fig_prob_of_random_objects} shows the probabilities $\Pg$ for all these randomised objects in the \EJ\ space. Initially, we were expecting that all the objects would receive very low $\Pg$ values (close to zero), since all the correlations must be erased and hence no groups would be found. While this is true for most of the objects in Figure~\ref{fig:Fig_prob_of_random_objects}, however, a few objects located at $(E,J_\phi)\sim(-1.25\times10^5\km2s2, 0 \kpc\kms)$ possess strikingly high $\Pg$ values. In other words, \enlink\ has detected one significant group at this $(E,J_\phi)$ location. Intriguingly, this $(E,J_\phi)$ location coincides with the dynamical location of the \GSE\ merger. We repeated the above \EJ\ randomisation procedure multiple times, however, a group was always detected at this $(E,L_z)$ location. We thus tentatively suggest that this compromises the reality of those objects that have been linked with the \GSE\ merger (both in our study and in previous studies). We defer the exploration of this problem to future work. 

Figure~\ref{fig:Fig_PDFrandom} shows the PDF of the computed probabilities $\Pg$ of all the randomised objects, and this serves as the background model for our main analysis. From this PDF, we use its $90$ percentile to set the threshold probability at $\Pth=0.3$ (rounded off for $0.295$) for a $2\sigma$ detection.

\section{Groupings of previous studies}\label{appendix:literature}

We construct Figure~\ref{fig:Fig_literature} that shows groupings of objects in the \EJ\ space from previous studies. We do this so that the previous results can be compared with our result shown in Figure~\ref{fig:Fig_probability_cut}. 

To construct Figure~\ref{fig:Fig_literature}, we use our derived \EJ\ values of globular clusters and streams, but use the associations that are described by previous studies. To this end, the globular cluster groupings are adopted from Table~1 of \cite{Massari_2019} for the groups \Sgr, \Helmi, \GSE\ and Bulge; Table~1 of \cite{Myeong2019} for {\it Sequoia} and Table~1 of \cite{Kruijssen_2020_Kraken} for {\it Kraken}. The stream groupings are adopted from Table~1 of \cite{Bonaca2021}. From these previous studies, we specifically highlight only those objects that have been unambiguously linked to just one group and avoid highlighting those objects that have been linked to multiple groups; the only exception is the \asi\ group.

\bibliography{sample631}
\bibliographystyle{aasjournal}

\end{document}